\documentclass{article}

\usepackage{PRIMEarxiv}

\usepackage[utf8]{inputenc} 
\usepackage[T1]{fontenc}    
\usepackage{hyperref}       
\usepackage{url}            
\usepackage{booktabs}       
\usepackage{amsfonts}       
\usepackage{nicefrac}       
\usepackage{microtype}      
\usepackage{lipsum}
\usepackage{fancyhdr}       
\usepackage{graphicx}       
\usepackage{epstopdf}
\usepackage{ragged2e}
\usepackage{algorithm}
\usepackage{algpseudocode}
\usepackage{enumerate}
\usepackage{subfigure}
\usepackage{multirow}
\usepackage{amsmath}
\usepackage{bm}
\usepackage{stfloats}
\usepackage{color}
\usepackage{pifont}
\graphicspath{{media/}}     
\usepackage{makecell}
\usepackage{authblk}
\title{A Review of Machine Learning for Cavitation Intensity Recognition in Complex Industrial Systems}

\author[1,9]{Yu Sha}
\author[3]{Ningtao Liu}
\author[5]{Haofeng Liu}
\author[1,4]{Junqi Tao}
\author[6]{Zhenxing Niu}
\author[1]{Guojun Huang}
\author[7]{Yao Yao}
\author[8]{Jiaqi Liang}
\author[1]{Moxian Qian}
\author[9,10,11]{Horst Stoecker}
\author[12]{Domagoj Vnucec}
\author[12]{Andreas Widl}
\author[1,2,9,$^\dagger$]{Kai Zhou}

\affil[1]{School of Science and Engineering, The Chinese University of Hong Kong, Shenzhen 518172, China}
\affil[2]{School of Artificial Intelligence, The Chinese University of Hong Kong, Shenzhen 518172, China}
\affil[3]{School of Computer, Luoyang Institute of Science and Technology, Luoyang 471023, China}
\affil[4]{Key Laboratory of Quark \& Lepton Physics (MOE) and Institute of Particle Physics, Central China Normal University, Wuhan 430079, China}
\affil[5]{School of Artificial Intelligence, Xidian University, Xi'an 710126, China}
\affil[6]{School of Computer Science and Technology, Xidian University, Xi'an 710126, China}
\affil[7]{School of Information Engineering, HangZhou Polytechnic University, HangZhou 310018, China}
\affil[8]{School of Business, China University of Political Science and Law, Beijing 102249, China}
\affil[9]{Frankfurt Institute for Advanced Studies, Frankfurt am Main 60438, Germany}
\affil[10]{Institut f{\"u}r Theoretische Physik, Goethe Universit{\"a}t Frankfurt, Frankfurt am Main 60438, Germany}
\affil[11]{GSI Helmholtzzentrum f{\"u}r Schwerionenforschung GmbH, Darmstadt 64291, Germany}
\affil[12]{SAMSON AG, Frankfurt am Main 60314, Germany}

\begin{document}
\renewcommand{\thefootnote}{}
\footnotetext{Emails: \{yusha, zhoukai\}@cuhk.edu.cn}
\footnotetext{$^\dagger$Corresponding author: Kai Zhou}
\renewcommand{\thefootnote}{\arabic{footnote}}
\maketitle
\begin{abstract}
Cavitation intensity recognition (CIR) is a critical technology for detecting and evaluating cavitation phenomena in hydraulic machinery, with significant implications for operational safety, performance optimization, and maintenance cost reduction in complex industrial systems. Despite substantial research progress, a comprehensive review that systematically traces the development trajectory and provides explicit guidance for future research is still lacking. To bridge this gap, this paper presents a thorough review and analysis of hundreds of publications on intelligent CIR across various types of mechanical equipment from 2002 to 2025, summarizing its technological evolution and offering insights for future development. The early stages are dominated by traditional machine learning approaches that relied on manually engineered features under the guidance of domain expert knowledge. The advent of deep learning has driven the development of end-to-end models capable of automatically extracting features from multi-source signals, thereby significantly improving recognition performance and robustness. Recently, physical informed diagnostic models have been proposed to embed domain knowledge into deep learning models, which can enhance interpretability and cross-condition generalization. In the future, transfer learning, multi-modal fusion, lightweight network architectures, and the deployment of industrial agents are expected to propel CIR technology into a new stage, addressing challenges in multi-source data acquisition, standardized evaluation, and industrial implementation. The paper aims to systematically outline the evolution of CIR technology and highlight the emerging trend of integrating deep learning with physical knowledge. This provides a significant reference for researchers and practitioners in the field of intelligent cavitation diagnosis in complex industrial systems.
\end{abstract}


\section{Introduction}
\label{sec: 1_introduction}
Cavitation intensity recognition (CIR) is a critical technology for quantitatively evaluating cavitation phenomena that occur during the operation of hydraulic machinery \cite{wu2022acoustic}. Cavitation not only generates intense noise and vibration, but can also cause erosion of component, performance degradation and structural failure, posing severe threats to the safety, stability and cost-effectiveness of industrial systems \cite{kang2025research}. The high performance of CIR enables the timely implementation of control measures, optimization of operating conditions, extension of equipment service life and reduction of maintenance and downtime costs. Therefore, CIR holds significant engineering importance and practical relevance across sectors such as energy, shipping, aerospace and the chemical industries, where pumps, turbines, propellers and valves serve as critical components \cite{caupin2006cavitation}. With advances in sensing technology and improvements in data acquisition capabilities, multi-source signals (e.g., acoustic, vibration, pressure, high-speed imaging, etc.) provide abundant diagnostic information for CIR, which has attracted increasing attention from both academia and industry over the past two decades.

Intelligent cavitation intensity recognition (ICIR) refers to the application of machine learning and deep learning techniques from modern artificial intelligence to cavitation intensity recognition \cite{peres2020industrial}. Unlike traditional diagnostic approaches that rely on manual feature extraction and expert experience, ICIR can adaptively learn from complex multi-source sensor data. It automatically captures cavitation features and establishes mappings between signals and cavitation intensity, significantly reducing dependence on human intervention \cite{jan2023artificial}. Firstly, ICIR can process multi-modal information, enabling comprehensive cross-signal-type analysis and improving diagnostic performance and robustness. Secondly, it is capable of handling nonlinear, multi-scale and high-noise signals, allowing it meet recognition requirements across various equipment types and operating conditions. Thirdly, it realizes an automated workflow from raw data to recognition results via end-to-end models, reducing diagnostic time and enhancing real-time monitoring capabilities. Fourthly, it can incorporate embedded physical knowledge and constraint-based loss functions to improve model interpretability and cross-condition generalization, providing technical support for sustainable monitoring and predictive maintenance in industrial scenarios.

\begin{figure*}
    \centering
    \includegraphics[width=0.9\textwidth,height=200mm]{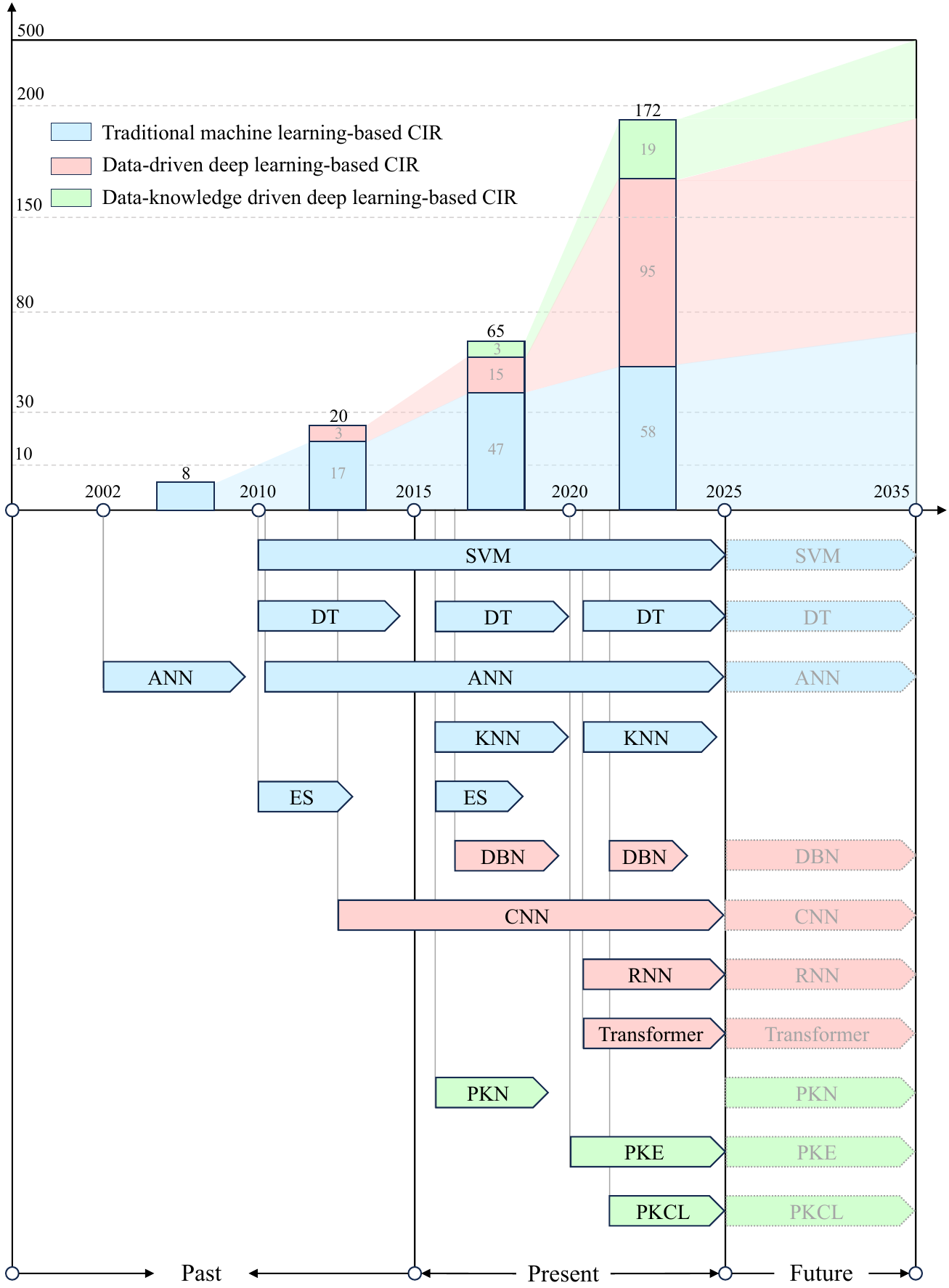}
    \caption{Development path and significant milestones of cavitation intensity recognition using machine learning.}
    \label{fig: Introduction}
\end{figure*}

In recent years, ICIR has attracted increasing attention from academia and industry, a trend that is closely related to the rapid development of machine learning technologies. To gain a comprehensive understanding of the technological evolution and research dynamics in this field, we conducted a systematic statistical analysis of several hundred relevant publications from 2004 to 2025, as shown in Figure \ref{fig: Introduction}. This analysis aims to reveal the principal technical characteristics and developmental trends at each stage, which provides a basis for delineating subsequent research phases. Based on the observed changes in publication activity and corresponding technological advances, we roughly divide the research on CIR into the following three periods.

In the past, traditional machine learning (TML) methods have served as the primary technical approach in CIR. These methods typically rely on manually engineered features extracted from acoustic, vibration or pressure signals by domain experts. The extracted features are then classified using classical algorithms such as support vector machines (SVM) \cite{cortes1995support}, decision trees (DT) \cite{quinlan1986induction}, artificial neural networks (ANN) \cite{rumelhart1986learning}, k-Nearest neighbors (KNN) \cite{fukunaga2013introduction} and expert systems (ES) \cite{jackson1986introduction}. This stage of development reduced reliance on purely manual judgment. However, the construction of features remains time-consuming and labor-intensive and heavily dependent on expert experience, which limits the generalization capability across different equipment types and operating conditions.

At present, the rise of deep learning has ushered ICIR into an era of end-to-end modeling. Deep learning models can automatically learn cavitation features directly from multi-source raw signals (e.g., acoustic, vibration, pressure, high-speed imaging, etc.) and perform multi-modal information fusion to enhance recognition performance and robustness. The application of deep belief networks (DBN) \cite{hinton2006fast}, convolutional neural networks (CNN) \cite{lecun2002gradient}, residual neural networks (ResNet) \cite{he2016deep}, densely connected convolutional networks (DenseNet) \cite{huang2017densely}, MobileNet \cite{howard2017mobilenets}, ShuffleNet \cite{zhang2018shufflenet}, recurrent neural networks (RNN) \cite{zaremba2014recurrent}, long short-term memory (LSTM) \cite{hochreiter1997long}, gated recurrent units (GRU) \cite{cho2014learning} and Transformer \cite{vaswani2017attention} architectures has significantly improved the real-time performance and stability of CIR under complex operating conditions.

In the future, the development of ICIR will depend on standardized high-quality multi-source data acquisition and the design of physical informed deep learning diagnostic models. Through incorporating physical mechanisms, physical knowledge embedding and physical constraint losses into end-to-end data-driven deep learning, which can enhance the physical consistency of the model \cite{karniadakis2021physics}. This approach not only helps improve the model's generalization capabilities under small-sample, high-noise and cross-operating conditions, but also significantly enhances the interpretability and engineering credibility of the diagnostic results \cite{cuomo2022scientific,moseley2022physics}. Compared to purely data-driven methods, physical-informed diagnostic models can better capture the non-linear, multi-scale characteristics of cavitation processes, ensuring diagnostic outputs are closely aligned with practical physical mechanisms \cite{ganga2024exploring}. In addition, in future industrial applications, the combination of real-time data acquisition by the industrial internet of things and edge computing deployment will enable physical informed deep learning diagnostic models to provide robust support for ICIR across equipment and operating conditions \cite{hua2023edge,sisinni2018industtationrial}.

To summarize the research of CIR, Wu at al. \cite{wu2022acoustic} discussed various methods for characterizing cavitation intensity and emphasizes three acoustic approaches. Liu et al. \cite{liu2023review} analyzed pump cavitation detection methods based various signals and discussed future trends with intelligent algorithms and advanced sensing. Adil et al. \cite{adil2025detect} examined experimental methods and control systems for diagnosing and managing cavitation in centrifugal pumps, and proposed future improvements using more sensors, advanced data processing and artificial intelligence for early detection to enhance reliability and extend service life. Fern{\'a}ndez et al. \cite{fernandez2024review} outlined acoustic emission techniques for cavitation and fracture detection in hydraulic turbines and emphasized artificial intelligent as a key future direction for condition monitoring. Mousmoulis et al. \cite{mousmoulis2019review} discussed cavitation monitoring tools for centrifugal pumps and advocated for developing reliable, low-cost and easy-to-install multi-sensor setups. Folden et al. \cite{folden2023classification} categorized cavitation models, analyzed key examples and outlined future improvements in flow simulation. Existing reviews on CIR have the following shortcomings. Firstly, the previous reviews just focus on a single technical approach or specific equipment type (e.g. pumps, turbines, acoustic detection, modeling methods, etc.) and lacks a systematic and integrated framework across different methods and devices, which makes it difficult to comprehensively compare their feasibility and cost-effectiveness. Secondly, these reviews typically cover only specific stage of CIR development, lacking a systematic overview that integrates past, present and future progress. Thirdly, the current reviews generally do not provide a roadmap for the future development of CIR, while potential research trends over the next five to ten years are of great interest to review readers. Fourthly, the present reviews provide limited coverage of emerging technologies such as deep learning and physical informed deep learning. 

In order to overcome the above shortcomings, this paper systematically examines hundreds of studies across various hydraulic machinery from 2002 to 2025 and outlines a technological roadmap for this field. The contributions of this review are summarized as follows:
\begin{itemize}
    \item The development of CIR is organized into three stages: past traditional machine learning, present deep learning and future physical informed deep learning, which helps systematically trace the technological evolution, clarify the key characteristics of each stage and provide guidance for future research directions.
    \item A future roadmap for CIR is proposed, highlighting physical-informed diagnostic models, multi-modal fusion, lightweight network architectures and industrial agents deployment as promising directions to overcome challenges in multi-source data acquisition, standardized evaluation and industrial implementation.
    \item A systemic view of CIR's technological evolution is presented and the trend of integrating deep learning with physical knowledge is emphasized, providing a valuable reference for intelligent cavitation diagnosis in complex industrial systems.
\end{itemize}

The rest of this review is organized as follows. Section \ref{sec: 2_TML} examines the past development of CIR, focusing on the applications of traditional machine learning theories. Section \ref{sec: 3_DL} reviews the applications of deep learning theories, representing the present period in the development of CIR. Section \ref{sec: 4_DKDL} argues applications of physical informed deep learning to CIR. In Section \ref{sec: 5_Dissusion}, a roadmap is outlined in conjunction with the challenges of CIR. Conclusions are provided in Section \ref{sec: conclusions}.

\section{Traditional Machine Learning}
\label{sec: 2_TML}
\begin{figure*}
    \centering
    \includegraphics[width=\textwidth,height=80mm]{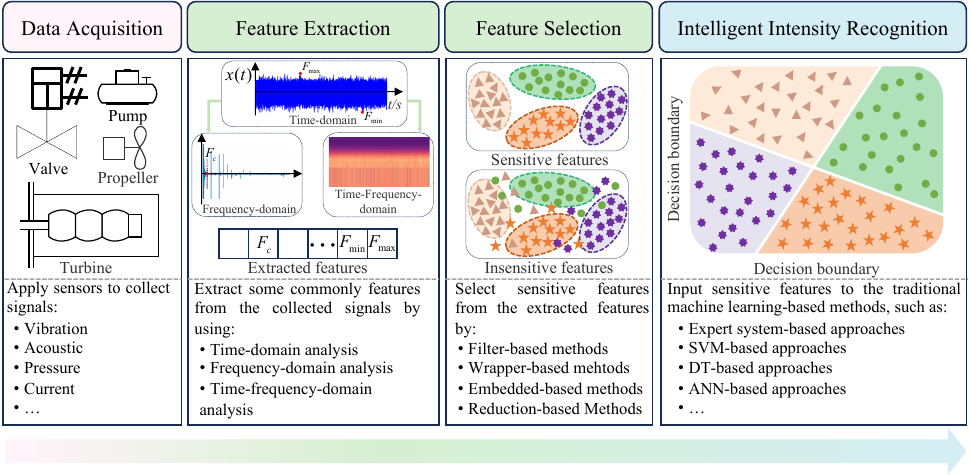}
    \caption{Diagnosis procedure of cavitation intensity recognition using traditional machine learning methods, which consists of data acquisition, feature extraction, feature selection and intelligent intensity recognition.}
    \label{fig: TML}
\end{figure*}

Traditionally, cavitation intensity recognition (CIR) has primarily relied on operators' experience-based judgment of equipment operating conditions and on-site inspections \cite{liu2023review,mousmoulis2019review}. It not only increase maintenance workload but also affects the performance and consistency of diagnostic results. In recent years, with the continuous development of sensor technology and signal processing methods, researchers have gradually introduced traditional machine learning (TML) into CIR tasks, achieving automated prediction and recognition of cavitation states \cite{2017Cavitation,kumar2010study}. A typical TML-based CIR includes four key steps: data acquisition, feature extraction, feature selection and intelligent intensity recognition \cite{hasanpour2024pump,dutta2018centrifugal}, see Figure \ref{fig: TML}. Each step will be discussed in the following subsections.

\subsection{Data Acquisition}
\label{subsec: Data Collection}
During the data acquisition phase, sensors are typically installed at critical locations on equipment (e.g. valves \cite{mahdavi2014performance,jazi2009detecting}, pipes \cite{wen2002time,de2015monitoring}, pumps \cite{tash2009application,sakthivel2012automatic}, turbines \cite{escaler2006detection,escaler2004cavitation} and propellers \cite{jamalpassive,grenie1990acoustic}) to enable continuous monitoring of operational conditions. Researchers employ various types of sensors to effectively capture the formation, development and associated physical changes of cavitation bubbles. Different types of sensors have their own advantages in terms of adaptability, sensitivity, resolution and anti-interference capability \cite{mousmoulis2019review,escaler2002field}, see Table \ref{tab: TML_DataAcquisition}. In general, pumps, valves, pipes and turbines commonly utilize vibration, acoustic and pressure sensors \cite{kumar2021identification,sha2022acoustic,zhang2015applicability,mahmoodi2009determination,feng2024cavitation,kirschner2023cavitation}, while propellers mainly use acoustic, pressure, current and high-speed sensors \cite{pereira2004experimental,tsai2021multi}. By appropriately selecting and integrating multi-source sensor data, the performance and robustness of CIR can be significantly improved.
\begin{table*}[htbp]
\centering
\caption{Comparison of different sensors in data acquisition stage of cavitation intensity recognition.}
\label{tab: TML_DataAcquisition}
\small
\setlength{\tabcolsep}{1mm}{
\begin{tabular}{c|c|c}
\toprule 
Type              & Principle               & Characteristics                                                  \\ 
\midrule
Ultrasonic Sensor & Analyzes variations in high-frequency               & High resolution, sensitive to medium                             \\
High-speed Camera & Captures bubble flashing frequency                                & High temporal resolution, requires transparent medium, high cost \\
Energy Sensor     & Measurement of energy changes   & Quantitative, requires high-precision devices, high cost         \\
Pressure Sensor   & Detects local pressure pulsations            & High sensitivity, real-time, vulnerable to corrosion and impact  \\
Vibration Sensor  & Monitors surface micro-vibrations           & Flexible installation, low cost, source interference possible    \\
Acoustic Sensor   & Monitors sound wave changes            & Remote monitoring, strong penetration, needs noise suppression   \\ 
\bottomrule
\end{tabular}}
\end{table*}

\subsection{Feature Extraction}
\label{subsec: Feature Extraction}
Feature extraction serves as a crucial bridge between raw sensor signals and TML models in CIR. By analysing the time-domain, frequency-domain and time-frequency-domain characteristics of the signals, multi-dimensional features that effectively represent the cavitation state can be extracted \cite{samanipour2017cavitation,al2020detection,zhou2024multi}, as detailed below.
\begin{itemize}
    \item Time-domain features: These are statistical quantities directly derived from the raw signal, including mean, variance, kurtosis, skewness, maximum value, minimum value, shape factor, peak factor, clearance factor, impulse factor, etc., reflecting the overall fluctuation characteristics of the signal \cite{shervani2018cavitation,panda2018prediction}.
    \item Frequency-domain features: These are obtained from the frequency spectrum, including mean frequency, frequency center,  frequency standard deviation, root mean square frequency, spectral entropy, energy spectral density, etc., which help capture frequency variations caused by cavitation \cite{liu2024cavitation}.
    \item Time-frequency-domain features: These are suitable for non-stationary signal analysis, common methods include wavelet transform, wavelet packet transform, short-time Fourier transform and empirical model decomposition \cite{dong2019cavitation,li2024intelligent,kang2017analysis}. These approaches enable the characteristics of the dynamic evolution of cavitation across different time scales.
\end{itemize}

\subsection{Feature Selection}
\label{subsec: Feature Selection}
Feature selection aims to filter out a discriminative and sensitive feature subset from high-dimensional features, reducing computational complexity, alleviating the dimension diaster and improving recognition performance \cite{li2017feature,jovic2015review}. For CIR task, commonly used feature selection methods include filter-based, wrapper-based, embedded-based and reduction-based approaches, each of which has its own advantages and is suitable for different data scenarios and modeling requirements. 

\subsubsection{Filter-based Methods}
\label{subsubsec: Filter-based Methods}
Filter-based methods evaluate the correlation or statistical properties between each feature and the target variable, performing feature selection independently before the model training \cite{porkodi2014comparison}. The following briefly introduces several commonly used filters.
\begin{itemize}
    \item Variance threshold \cite{wang2010comparative}: Remove features with low variance and low information content to enhance generalization ability of the model.
    \item Correlation coefficient analysis \cite{hall1999correlation}: Calculate the linear or non-linear correlation between features and target variable (e.g. Pearson, Spearman or Kendall coefficients, etc.)
    \item Mutual information \cite{vergara2014review}: Measure the non-linear dependency between features and target variable, which is suitable for feature selection under complex distribution structures.
    \item Distance Evaluation \cite{patel2020euclidean}: Based on the principle of minimizing intra-class distance and maximizing inter-class distance, features with strong class discrimination capabilities are selected.
    \item Information gain and gain ratio \cite{karegowda2010comparative}: Evaluate the effective amount of information carried by a feature. Features with higher information gain and gain ratio are generally more advantageous in enhancing the performance of diagnostic models.
    \item Minimum redundancy maximum relevance \cite{peng2005feature}: Select the most representative feature subset by maximizing the relevance between features and the target variable and minimizing redundancy among features.
    \item Relevant features \cite{rudnicki2014all}: Construct relevance metrics to evaluate the sensitivity of features to the target variable, and then select key features that are highly correlated with the target variable and exhibit significant class discrimination capabilities.
    \item Chi-square test \cite{mochammad2021stable}: Evaluate the independence between features and the target variable to determine the most relevant and sensitive features for the diagnostic model.
\end{itemize}

\subsubsection{Wrapper-based Methods}
\label{subsubsec: Wrapper-based Methods}
Wrapper-based methods select optimal feature subsets by relying on the performance of TML models. The core idea of these methods is to convert the feature selection issue into a feature subset search task, i.e. guiding the selection of the feature subset based on the model's performance \cite{jovic2015review}. The following are common wrapper-based methods.
\begin{itemize}
    \item Sequential selection \cite{bermejo2009incremental}: Features are gradually added or remove features via forward selection or backward elimination to optimize model performance, which is suitable for high-dimensional feature spaces.
    \item Recursive feature elimination \cite{jeon2020hybrid}: Iteratively remove features with lower contributions based on the model's assessment of feature importance, ultimately retaining the subset of features that most significantly improves model performance.
    \item Heuristic search \cite{sadeghian2025review}: Utilise heuristic strategies (e.g. genetic algorithms, simulated annealing, ant colony algorithms, etc.) to perform a global search in the feature space to find the optimal or near-optimal feature subset. This approach is effective for complex feature combinations and large search spaces.
\end{itemize}

\subsubsection{Embedded-based Methods}
\label{subsubsec: Embedded-based Methods}
Embedded-based methods integrate the feature selection process into the model training, leveraging the model's own parameter learning and optimization mechanisms to automatically identify and retain sensitive features for prediction results \cite{siham2021feature}. The following are common embedded-based methods.
\begin{itemize}
    \item Regularization \cite{moslemi2023tutorial}: Introduce ${\mathcal{L}_1}$ or ${\mathcal{L}_2}$ regularization terms during model training to constrain feature coefficients, achieving automatically selecting and retaining the most influential sensitive features for prediction results.
    \item Tree models \cite{manikandan2021feature}: Evaluate information gain, gini impurity reduction or purity improvement during node splitting to quantify the contribution of each feature to model decisions (e.g. decision tree, random forest, gradient boosting decision tree, etc.) and identify highly important and sensitive features.
\end{itemize}

\subsubsection{Reduction-based Methods}
\label{subsubsec: Reduction-based Methods}
Reduction-based methods transform the original high dimensional feature space into a low dimensional representation through mathematical mapping, effectively compressing the number of features and eliminating redundancy while preserving as much critical information as possible \cite{masaeli2010transformation}. The following are common reduction-based methods.
\begin{itemize}
    \item Linear reduction \cite{sumithra2015review}: Utilize linear techniques (e.g. rthogonal transformations or matrix decomposition, etc.) to project high-dimensional feature space onto low dimensional linear subspaces. This approach can effectively eliminate features linear correlations while retaining the global structure of the original feature (e.g. principal component analysis, linear discriminant analysis, non-negative matrix factorization, etc.).
    \item Non-linear reduction \cite{fodor2002survey}: Employ non-linear transformation techniques (e.g. manifold learning, kernel methods, etc.) to embed features into a low dimensional non-linear space. This method can effectively capture the local neighborhood relationships and manifold structures in the original feature space (e.g. kernel principal component analysis, t-distributed stochastic neighbor embedding, isometric mapping etc.).  
\end{itemize}

\subsection{Intelligent Intensity Recognition}
\label{subsec: Intelligent Intensity Recognition}
Intelligent intensity recognition establishes a mapping between selected features and cavitation intensity through a recognition model based on TML. To achieve this goal, the recognition model needs to be trained and learned on labeled cavitation samples. According to current research progress, we briefly introduce five commonly used TML methods for CIR in the following sections.

\subsubsection{SVM-based Approaches}
\label{subsubsec: SVM-based Approaches}
\begin{figure}[htbp]
    \centering
    \includegraphics[width=0.45\textwidth,height=60mm]{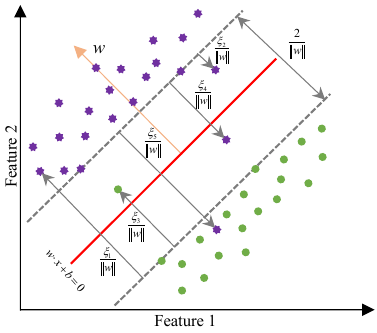}
    \caption{Schematic diagram of a SVM. The red solid line and the gray dashed line represent the separating hyperplane and the margin boundaries, respectively.}
    \label{fig: SVM}
\end{figure}

Support vector machine (SVM) is a supervised learning method based on statistical learning theory, which is widely used in classification, regression and anomaly detection tasks \cite{cortes1995support}. Its fundamental idea is to find an optimal hyperplane in the feature space to maximize the interclass margin, achieving effective separation of samples. Given a training dataset $\{ \mathcal{X},\mathcal{Y}\}  = ({x_i},{y_i})$ with $N$ samples, ${x_i} \in {\mathbb{R}^d}$ represents the feature vector of the $i$-th sample and ${y_i} \in \{  + 1, - 1\}$ denotes its corresponding class label, the basic optimization objective of SVM can be expressed as:
\begin{equation}
\label{eq: SVM Objective}
\begin{array}{c}
\mathop {\min }\limits_{w,b} \frac{1}{2}{\left\| w \right\|^2}\\
{\rm{s.t.}}\;\;{y_i}({w^{\rm{T}}}{x_i} + b) \ge 1,\;\;i = 1,2, \ldots ,N
\end{array},
\end{equation}
where $w$ is the normal vector of the hyperplane and $b$ is the bias term. In practical applications, due to noise interference and overlapping class boundaries, it is difficult for samples to satisfy the strict linear separability assumption. Therefore, SVM introduces slack variable ${\xi _i}$ to construct a soft-margin classifier, as shown in Figure \ref{fig: SVM}. The optimization objective is adjusted as follows:
\begin{equation}
\begin{array}{c}
\mathop {\min }\limits_{w,b,\xi} \frac{1}{2}{\left\| w \right\|^2} + C\sum\limits_{i = 1}^N {{\xi _i}} \\
{\rm{s.t.}}\;\;{y_i}({w^{\rm{T}}}{x_i} + b) \ge 1 - {\xi _i},\;\;{\xi _i} \ge 0
\end{array},
\end{equation}
where $C$ is the penalty factor used to control the balance between margin maximization and classification error minimization. In addition, SVM also introduces kernel methods to implicitly map the original features to a high-dimensional feature space. The typical kernel functions include radial basis kernel, polynomial kernel and sigmoid kernel, which are crucial role in accurately modeling complex patterns. 

The application of SVM to CIR are systematically summarized in Table \ref{tab: SVM Methods}. According to the research results, the SVM have exhibited excellent performance in CIR for equipment such as pumps, turbine and pipes. For the above diagnostic objects, the SVM-based methods need to able to recognize multiple flow cavitation states, not only limited to the binary classification of cavitation and non-cavitation. Therefore, one-versus-rest (OVR) and one-versus-one (OVO) strategies have become solutions for SVM \cite{han2024off}. In addition, the researchers improved the diagnostic accuracy of modifying the SVM and optimizing the SVM parameters. For the former, researchers applied the modified SVM to CIR, such as least square SVM \cite{de2015cavitation}, proximal SVM \cite{shervani2018cavitation}, wavelet SVM \cite{de2015cavitation} and one-class SVM \cite{jeong2012surface}, which all achieved better diagnostic performance than the traditional SVM.
For algorithm optimization, researchers focused on simplifying the complex solution process and optimizing parameter selection of SVM, such as particle swarm optimization-based SVM \cite{li2024intelligent}, cuckoo search algorithm-based SVM \cite{chen2022csa} and butterfly optimization algorithm-based SVM \cite{zhou2024multi}.
\begin{table*}[htbp]
\centering
\caption{Summary of applications of SVM-based methods in cavitation intensity recognition.}
\label{tab: SVM Methods}
\small
\setlength{\tabcolsep}{5mm}{
\begin{tabular}{llc}
\toprule 
Objects & References & Signals \\ 
\midrule
\multirow{4}{*}{Pump} 
            &  \cite{hasanpour2024pump}, 
               \cite{zhou2024multi},
               \cite{shervani2018cavitation},
               \cite{panda2018prediction},
               \cite{li2024intelligent}, 
               \cite{han2024off}, 
               \cite{he2022intelligent},
               \cite{dutta2020comparative},          
               \cite{rapur2019multifault},
               \cite{casoli2019multi},
               & \multirow{4}{*}{Vibration, Pressure, Current, Acoustic} \\
               &
               \cite{qiu2019experimental},
               \cite{tingfeng2011ls},
               \cite{stephen2025evaluation},
               \cite{hu2025cavitation}, 
               \cite{ranawat2025blockage}, 
               \cite{dias2025edge},
               \cite{araste2023fault},
               \cite{huang2023condition},
               \cite{xiao2022research},
               \cite{al4615504incipient},
               &                         \\
               
              & 
               \cite{kamiel2022cavitation},
               \cite{bordoloi2017identification}, 
               \cite{dutta2023svm},
               \cite{araste2020support},
               \cite{youlin2017wavelet},
               \cite{xue2014intelligent},
               \cite{wang2019hydraulic},
               \cite{muralidharan2014fault},
               \cite{sakthivel2010application},
               
               &                         \\

                & \cite{matloobi2018identification}, 
                \cite{al2021faults},
                \cite{altobi2019fault},
                \cite{orru2020machine},
                \cite{song2024centrifugal}
                 &                         \\

Turbine      & \cite{stephen2025evaluation},
               \cite{gregg2018method},
               \cite{de2019intelligent},
               \cite{liu2014underwater},
               \cite{kang2022cavitation}
             & Vibration, Acoustic         \\ 
             
Pipe         & \cite{de2015cavitation},
               \cite{chen2022csa},
               \cite{bagherzadeh2025prediction},
               \cite{liu2025machine},
               \cite{hong2025improving},
               \cite{dias2024soft}
             & Ultrasonic, Pressure, Acoustic       \\ 
             
Propeller    & \cite{jeong2012surface},
               \cite{gypa2023propeller}
             & Vibration         \\

\bottomrule
\end{tabular}}
\end{table*}

\subsubsection{DT-based Approaches}
\label{subsubsec: DT-based Approaches}
\begin{figure}[htbp]
    \centering
    \includegraphics[width=0.45\textwidth,height=40mm]{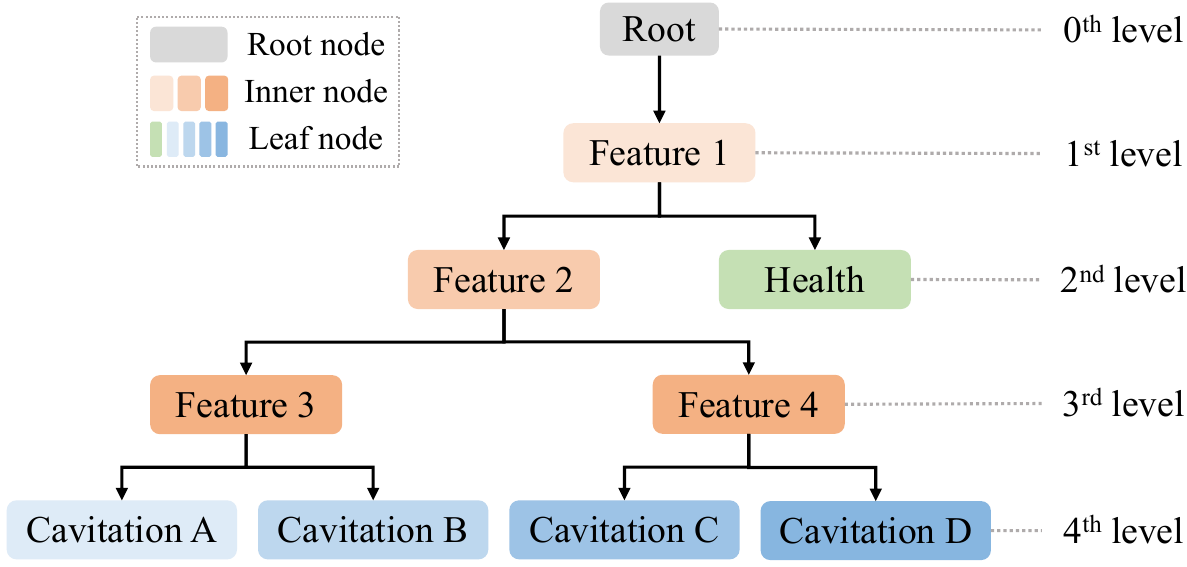}
    \caption{Schematic diagram of a DT. In general, the root node represents the entire sample set.}
    \label{fig: DT}
\end{figure}

Decision tree (DT) is a supervised learning method for classification and regression tasks, which makes decisions at each internal node by partitioning the feature space into distinct subregions based on feature values \cite{quinlan1986induction}. The decision at each node is made by evaluating a specific feature and splitting the data into subsets according to a threshold value. Given a training dataset $\{ \mathcal{X},\mathcal{Y}\}  = ({x_i},{y_i})$ with $N$ samples, ${x_i} \in {\mathbb{R}^d}$ represents the feature vector of the $i$-th sample and ${y_i} $ denotes its corresponding class label, the objective of DT is to recursively split the data at each node in order to create a tree structure that minimizes a measure of impurity, as shown in Figure \ref{fig: DT}. The basic optimization objective of DT is as follows:
\begin{equation}
\label{eq: DT Objective}
\begin{array}{c}
\mathop {\min }\limits_{S} \sum\limits_{i = 1}^N {I(y_i|S)}\\
{\rm{s.t.}}\left\{ \begin{array}{l}
G(S)=1- \sum\limits_{c=1}^C p(c|S)\\ 
E(S)=-\sum\limits_{c=1}^C p(c|S)logp(c|S)
\end{array} \right.,
\end{array}
\end{equation}
where $I(y_i|S)$ is a measure of impurity, e.g. Gini impurity $G(S)$ or entropy $E(S)$, $S$ represents the current subset of the data at given node and $p(c|S)$ is the proportion of class $c$ sample in subset $S$. To prevent overfitting, the DT are often pruned by setting a maximum depth or by requiring a minimum number of samples in each leaf node. In addition, the DT can be enhanced by ensemble methods, such as random forests (RF) \cite{breiman2001random}, gradient boosting machines (GBM) \cite{friedman2001greedy}, extreme gradient boosting (XGBoost) \cite{chen2016xgboost}, light gradient boosting machine (LightGBM) \cite{ke2017lightgbm} and extremely randomized trees (ERT) \cite{geurts2006extremely}, which combine multiple DTs to improve performance and reduce overfitting.

The application of DT and its derived methods to CIR are comprehensively summarized in Table \ref{tab: DT Methods}. Based on the research results, the DT and its its derivatives have also been widely applied in CIR for equipment such as pumps, turbines, pipelines and valves. For the above diagnostic objects, DT-based methods are able to deal with both binary classification of cavitation and non-cavitation and  multi-class recognition of different cavitation states. In addition, researchers have proposed a variety of improved DT methods, such as RF \cite{casoli2019multi,chang2024random} and XGBoost \cite{sha2022acoustic}, which effectively overcome the shortcomings of traditional DTs by ensemble learning and gradient boosting mechanisms. Furthermore, Kamali et al. \cite{kamali2025characterization} employed a combination of bayesian optimization with DT to automatically select hyperparameters.

Compared to SVM-based approaches, DT-based methods have natural interpretability and can automatically generate explicit diagnostic rules without relying on additional expert interpretation. In addition, DT-based methods are able to perform effective diagnosis even in the presence of missing data, which exhibits strong robustness. However, DT-based methods are often prone to suffer from overfitting, leading to weak generalization and compromising their effectiveness in complex diagnostic tasks. It is also worth noting that the design of many DT-based models still depends on expert knowledge, which restricts their universality and flexibility in certain application scenarios.
\begin{table*}[htbp]
\centering
\caption{Summary of applications of DT-based methods in cavitation intensity recognition.}
\label{tab: DT Methods}
\small
\setlength{\tabcolsep}{5mm}{
\begin{tabular}{llc}
\toprule 
Objects & References & Signals \\ 
\midrule
\multirow{2}{*}{Pump} 
                &  \cite{dutta2018centrifugal},
                   \cite{han2024off},
                   \cite{casoli2019multi},
                   \cite{chang2024random},
                   \cite{kamali2025characterization},
                   \cite{zhang2022anti},
                   \cite{gatica2025classifying},
                   \cite{orhan2024machine},

                & \multirow{2}{*}{Vibration, Current, Acoustic, Ultrasonic, Pressure} \\
                & \cite{karagiovanidis2023early},
                \cite{sakthivel2010vibration}, 
                \cite{muralidharan2013feature},
                  \cite{farokhzad2013fault},
                  \cite{sakthivel2011use},
                  \cite{stephen2024prediction},
                  \cite{manikandan2023vibration}          &                         \\

Turbine      & \cite{de2019intelligent},
               \cite{kang2020stacked},
               \cite{gruberdetection},
               \cite{gruber2014detection},
               \cite{kang2022analysis},
               \cite{amihai2018industrial},
               \cite{saari2018selection}

             &  Acoustic, Vibration, Pressure, Ultrasonic  \\ 
             
Pipe         & \cite{sha2022acoustic},
               \cite{liu2025machine},
               \cite{xu2022leakage},
               \cite{shen2022tree}
             & Acoustic, Vibration    \\ 
             
Valve        & \cite{sha2022acoustic},
              \cite{ehemann2023ai},
              \cite{yakupov2023application}
             & Acoustic, Vibration     \\

\bottomrule
\end{tabular}}
\end{table*}

\subsubsection{ANN-based Approaches}
\label{subsubsec: ANN-based Approaches}
\begin{figure}[htbp]
    \centering
    \includegraphics[width=0.4\textwidth,height=70mm]{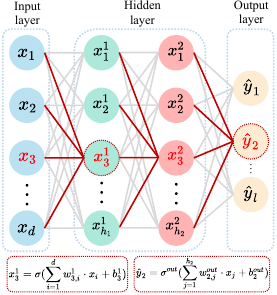}
    \caption{Schematic diagram of a ANN with two hidden layers.}
    \label{fig: ANN}
\end{figure}

Artificial neural network (ANN) is a traditional supervised learning method based on multilayer perceptrons (MLP), which is commonly applied to classification and regression tasks \cite{rumelhart1986learning}. Its basic structure consists of an input layer, several hidden layers and an output layer, which achieves feature mapping through forward propagation and parameter updating using back propagation, as shown in Figure \ref{fig: ANN}. Given a training dataset $\{ \mathcal{X},\mathcal{Y}\}  = ({x_i},{y_i})$ with $N$ samples, ${x_i} \in {\mathbb{R}^d}$ represents the feature vector of the $i$-th sample and ${y_i} \in {\mathbb{R}^l}$ denotes its corresponding class label. During the forward propagation process, the input data ${x_i}$ are sequentially transmitted through each layer. Each layer receives the output of the previous layer as input, first calculates the weighted sum of that input, then performs a nonlinear transformation and passes the result to the next layer. The specific process is as follows:
\begin{equation}
\label{eq: forward propagation}
{(x_i^h)_j} = \sigma (\sum\limits_{i = 1}^{{n_{h - 1}}} {w_j^h}  \cdot x_i^{h - 1} + b_j^h),
\end{equation}
where ${(x_i^h)_j}$ is the output of the $j$-th neuron in the $h$-th hidden layer, ${w_j^h}$ is the weight between the neurons in the previous layer, $b_j^h$ is the bias of the $j$-th neuron in the $h$-th hidden layer, $\sigma$ represents the activation function (e.g. Sigmoid, ReLU, Tanh, etc.). The predicted output of ANN is:
\begin{equation}
\label{eq: output layer}
{({{\hat y}_i})_k} = {\sigma ^{out}}(\sum\limits_{i = 1}^{{n_H}} {w_j^{out}}  \cdot x_i^H + b_j^{out}),
\end{equation}
where ${({{\hat y}_i})_k}$ is the predicted output of the $k$-th neuron in the output layer, $\sigma ^{out}$ represents the activation function (Sigmoid and Softmax), $w_j^{out}$ and $b_j^{out}$ are the weights and bias of the output layer, respectively. The optimization objective of ANN minimizes the error between the predicted output and the groundtruth:
\begin{equation}
\label{eq: ANN Objective}
\mathop {\min }\limits_{\theta} \sum_{i=1}^{N} \mathcal{L}(y_i,{{\hat y}_i}),
\end{equation}
To achieve the above objective, the training parameters $w$ and $b$ are updated by gradient descent during the back propagation, as detailed below:
\begin{equation}
\label{eq: parameter update}    
\begin{array}{cc}
w^*\leftarrow w-\eta \cdot \frac{\partial \mathcal{L}}{\partial w},\;\;\;\;\;\;b^*\leftarrow b-\eta \cdot \frac{\partial \mathcal{L}}{\partial b},
\end{array}
\end{equation}
where $\eta$ is the learning rate, ${\partial \mathcal{L}}/{\partial w}$ and ${\partial \mathcal{L}}/{\partial b}$ are the gradients with respect to parameter $w$ and $b$, respectively. In general, ANN consists of MLP, radial basis function neural network (RBFNN) \cite{orr1996introduction}, self-organizing map (SOM) \cite{kohonen2012self} and hopfield neural network (HNN) \cite{hopfield1982neural}, adaptive resonance theory network (ARTN) and so on \cite{haykin2009neural}.

The application of ANN to CIR are systematically reviewed in Table \ref{tab: ANN Methods}. As shown in Table \ref{tab: ANN Methods}, ANN-based methods have been widely applied to CIR in various hydraulic machinery and fluid systems, such as pumps, turbines, pipes and propeller. In particular, pump-related research focused on the variety of signal types, which reflects their comprehensive employment of multi-source information in cavitation monitoring. For pipelines and valves, acoustic, pressure and vibration measurements are commonly used for effective detection of local cavitation. In general, ANN-based methods is widely used on different objects and multiple signal sources, which demonstrates the universality and adaptability of ANN to CIR task.
\begin{table*}[htbp]
\centering
\caption{Summary of applications of ANN-based methods in cavitation intensity recognition.}
\label{tab: ANN Methods}
\small
\setlength{\tabcolsep}{0.7mm}{
\begin{tabular}{llc}
\toprule 
Objects & References & Signals \\ 
\midrule
\multirow{4}{*}{Pump} 
                & \cite{kumar2021identification},
                  \cite{liu2024cavitation},
                  \cite{dong2019cavitation},
                  \cite{casoli2019multi},
                  \cite{stephen2025evaluation},
                  \cite{matloobi2018identification},
                  \cite{al2021faults},
                  \cite{altobi2019fault},
                  \cite{orru2020machine},
                  \cite{song2024centrifugal},

                & \multirow{4}{*}{Current, Pressure, Vibration, Acoustic, High-speed images} \\
                & \cite{de2019intelligent},
                \cite{lavretsky2002health},
                 \cite{bernardini2009expert},
                  \cite{rathinasabapathydevelopment},
                  \cite{kane2016application},
                  \cite{hussain2016condition},
                  \cite{babikir2019noise},
                  \cite{maradey2020methodology},
                  \cite{sayed2020classification},

                &  \\

                & 
                \cite{adeodu2020adaptive},
                  \cite{matloobi2021identification},
                  \cite{lan2022experimental},
                  \cite{tong2023cavitation},
                  \cite{nasiri2011vibration},
                  \cite{siano2018diagnostic},
                  \cite{farokhzad2012897},
                  \cite{zhang2016effect},
                  \cite{al2022using},

                &  \\

                & 
                
                  \cite{zouari2004fault},
                  \cite{tan2023fault},
                  \cite{wang2025optimization},
                  \cite{azizi2017improving},
                  \cite{arendra2020investigating},
                  \cite{xu2021multi}
                &  \\

Turbine      & \cite{stephen2025evaluation},
               \cite{hovcevar2005prediction},
               \cite{saeed20133d},
               \cite{zhao2020use},
               \cite{souza2024application},
               \cite{de2012neural},
               \cite{sun2019prediction},
               \cite{benabdesselam2025multi}

             & Vibration, Pressure, Ultrasonic, High-speed images  \\ 
             
Pipe         & \cite{de2015cavitation},
               \cite{yang2008leak},
               \cite{wu2023neural},
               \cite{you2024cavitation}
             & Acoustic, Pressure    \\ 
             
Valve        & \cite{salah2019automated},
               \cite{gao2019cavitation}
             & Acoustic, Vibration     \\ 

Propeller    & \cite{schizas2002artificial}
             & Vibration     \\

\bottomrule
\end{tabular}}
\end{table*}

Compared to SVM and DT-based methods, ANN-based CIR methods can effectively capture potential features in multi-source monitoring signals to achieve high performance of different equipment cavitation states. However, ANN-based method still faces two challenges in engineering applications. First, the parameter size and complexity of the model dramatically increase with the dimension of the input data, which reduces the training efficiency and result in overfitting. Second, the ANN-based models show a typical black-box nature without theoretical foundation, making its diagnostic results difficult to support operation and maintenance decisions. 

\subsubsection{Other Approaches}
\label{subsubsec: Other Approaches}
In addition, other methods are also widely employed in FIR, such as k-nearest neighbours (KNN), expert system (ES). We will briefly introduce them in this subsection.

\begin{figure}[htbp]
    \centering
    \includegraphics[width=0.45\textwidth,height=50mm]{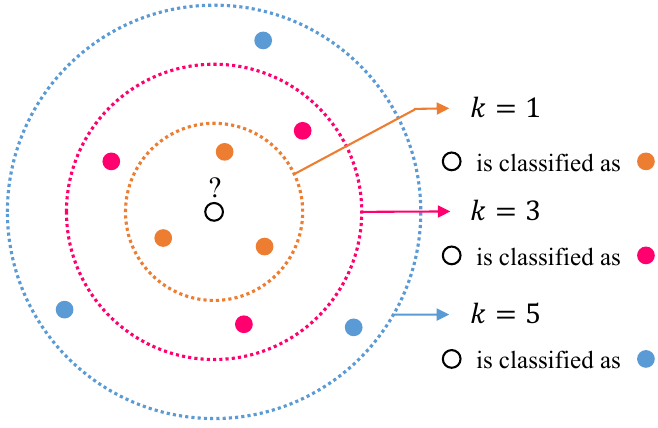}
    \caption{Illustration of a KNN algorithm.}
    \label{fig: KNN}
\end{figure}

\noindent\textbf{KNN-based Approaches.} K-nearest neighbours (KNN) is a distance-based supervised learning method, which is typically used for classification and regression tasks \cite{fukunaga2013introduction}. The core idea is that given a predicted sample, the distance is firstly calculated between this sample and all labelled samples in the taring set. Then, the $k$ nearest neighbours are found. Finally, the output of the sample is predicted based on the class or value of these $k$ neighbours, as shown in Figure \ref{fig: KNN}. 

KNN has received much attention in intelligent CIR research, especially in pumps \cite{han2024off,dutta2020comparative,kamiel2015vibration}, turbines \cite{de2019intelligent} and valves \cite{ehemann2023ai}. Han et al. \cite{han2024off} proposed a KNN-based method for detection cavitation in centrifugal pumps using vibration and current signals. Paolo et al. \cite{casoli2019multi} attempted a KNN-based model based on vibration signal for a hydraulic axial piston pump.  Sousa et al. \cite{de2019intelligent} developed a method for identifying incipient cavitation in wind turbines by extracting statical features of vibration signals and applying a KNN-based model. In addition, some researchers investigated the modified KNN models to enhance cavitation intensity diagnostic effects.  Kermani et al. \cite{fadaei2018cavitation} presented a fuzzy KNN-based method for recognizing the cavitation intensity of dam spillway utilising pressure signals. Zeng et al . \cite{zeng2024multi} proposed a method integrating dual-tree complex wavelet transform with variational mode decomposition and Bayesian optimized locally weighted KNN algorithm for intelligent recoignition of multi-state cavitation in vortex pumps based on current signals.

Although KNN has simplicity and effectiveness, it has several limitations in practical applications. First, the model is very sensitive to the choice of distance metric and parameter $k$, and inappropriate settings can significantly affect the final performance. Second, it has high computational complexity during the inference stage since it needs to calculate the distance between the test sample and all the training samples, resulting in extremely low efficiency in large-scale datasets. Third, it is very sensitive to irrelevant or redundant features. Finally, the basic KNN connot automatically perform feature weighting or learn the importance of features, causing it to depend on feature preprocessing to obtain the desired performance.

\begin{figure}[htbp]
    \centering
    \includegraphics[width=0.45\textwidth,height=55mm]{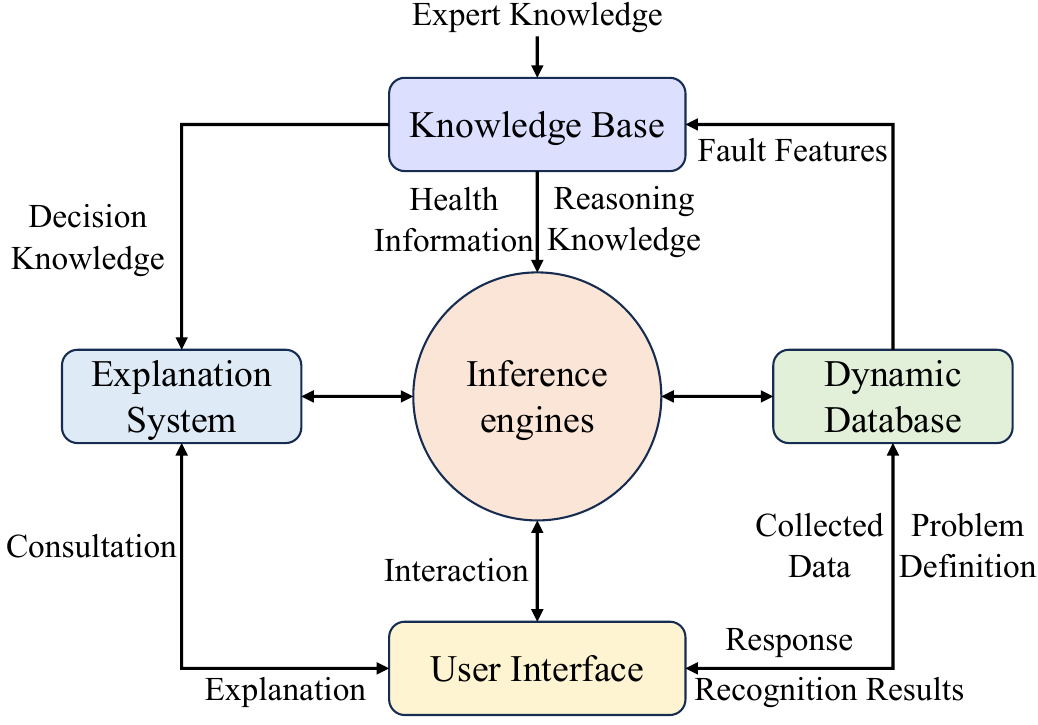}
    \caption{Schematic diagram of an ES.}
    \label{fig: ES}
\end{figure}

\noindent\textbf{ES-based Approaches.} Expert system (ES) is a rule-based artificial intelligence models that emulate the decision making process of human experts by encoding domain knowledge into a knowledge base and inference engine, which is widely used in condition monitoring of fluid machinery \cite{jackson1986introduction}. The core of ES is to utilise predefined logic rules extracted from practical experience and expert knowledge to parse the sensor signals and identify different cavitation states, as shown in Figure \ref{fig: ES}. Sakthivel et al. \cite{sakthivel2012automatic} proposed a combination of rough set rule extraction and a fuzzy inference system to achieve cavitation detection and intensity diagnosis in centrifugal pump using vibration signals. Saeid et al. \cite{farokhzad2013vibration} presented a vibration signal diagnosis method based on the combination of FFT feature extraction and adaptive fuzzy inference system for intelligent fault diagnosis of centrifugal pumps. Azadeh et al. \cite{azadeh2010fuzzy} developed a fuzzy rule-based inference system approach to realise timely diagnosis of cavitation states in centrifugal pumps.

Although ES has a strong interpretability, it faces several limitations in practical applications. First, the construction of ES is heavily relies on a priori knowledge of domain experts, resulting in time-consuming and labor-intensive process of knowledge acquisition. Second, ES cannot update knowledge base or optimize inference rules online, making it difficult to adapt to new fault patterns. Third, ES typically relys on deterministic logical reference, resulting in weakly handling noise and incomplete information in signals. Finally, ES adds new fault types or modifies new rules to require redesigning the rule base, resulting in system inflexibility.

\section{Data Driven Deep Learning}
\label{sec: 3_DL}
\begin{figure*}
    \centering
    \includegraphics[width=0.85\textwidth,height=80mm]{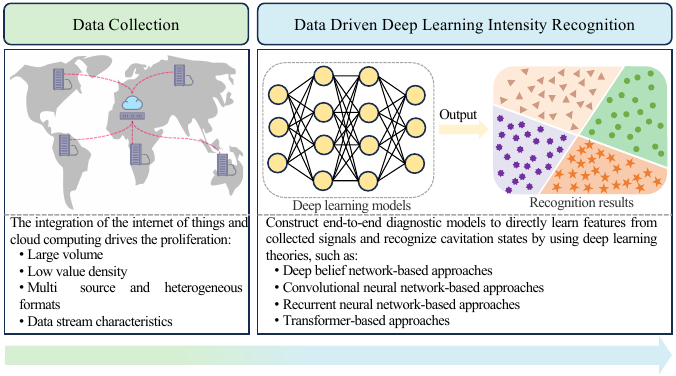}
    \caption{Diagnosis procedure of cavitation intensity recognition using deep learning methods, which consists of data acquisition and data-driven deep learning intensity recognition.}
    \label{fig: DL}
\end{figure*}

In contrast to TML approaches, data-driven deep learning (DL) methods have gained attention in CIR since their powerful end-to-end feature learning and representation capabilities. Instead of relying on manually extracted features, DL models can automatically extract multi-level features directly from original sensor signals, which can reduce the dependence on expert experience and complex signal preprocessing \cite{hoang2019survey}. A typical DL-based CIR pipeline normally consists of data collection and end-to-end intensity recognition, as shown in Figure \ref{fig: DL}, which will be detailed in the following subsections.

\subsection{Data Collection}
\label{subsec: Data Collection}
In modern industrial monitoring systems, data collection is the basis for CIR and other equipment health diagnostics. The process typically relies on a variety of sensors and control system to continuously acquire signals from the equipment under a wide range of operating conditions. The collected data not only presents the classical characteristics of volume, velocity, variety and veracity, but also exhibits domain-specific attributes determined by industrial environment \cite{mayer2013big}.
\begin{itemize}
    \item Large volume: Over long-term operation, monitoring systems for equipment continuously record multiple channel signals. High-frequency measurement (e.g. vibration signals from gearboxes, acoustic emission signals from cavitation-prone pipelines segments) can significantly increase the data volume.
    \item Low value density: Although a large amount of sensor data is collected, only a small portion is directly related to cavitation events or fault states. The vast majority of the records correspond to the normal operating cditions of the equipments and some of data may be of poor quality due to  transmission interruptions, environmental disturbances or sensor failures.
    \item Multi-source and heterogeneous formats: Multiple sensors (e.g. vibration, pressure, acoustic, etc.) are often integrated in industrial monitoring, resulting in heterogeneous data in both structure and sampling rate.
    \item Data stream characteristics: The development of technologies (e.g. internet of things, high-speed networks, GPU computing, edge computing, etc.) make it possible to acquire real-time, high-frequency monitoring data streams, enabling immediate access to machine health information.
\end{itemize}

Cavitation-related monitoring data are collected from the target system across different operating conditions. The acquisition devices include acoustic emission sensors, energy sensors, pressure sensors, vibration accelerometers, ultrasonic sensors and high-speed camera (more detailed see Table \ref{tab: TML_DataAcquisition}), which are placed at key locations to capture the entire "occurrence-development-collapse" process of cavitation. All signals are recorded simultaneously by the industrial data acquisition unit and stored in original waveforms and lightly preprocessed forms, ensuring the data covers a wide range of cavitation intensities, which satisfies the training requirements of DL models.

\subsection{End-to-end intensity recognition}
\label{subsec: Deep learning-based Approaches}
The DL model directly performs feature learning and cavitation intensity prediction simultaneously from raw monitoring data, eliminating manual feature extraction and feature selection. In general, its architecture includes an automatic feature learning module and a prediction module. The DL model training is based on end-to-end error backpropagation, while optimizing the parameters of both the feature and prediction layers to highly match the learned representation with the recognized target. According to current research progress, we briefly introduce four typical DL structures for CIR in the following sections.

\subsubsection{DBN-based Approaches}
\label{subsubsec: DBN-based Approaches}
Deep belief network (DBN) is a class of generative graphical models stacked with multiple layers of restricted boltzmann machine (RBM) \cite{hinton2006fast}. It is able to learn hierarchical abstract features of the input data in an unsupervised learning manner and achieve classification or regression by supervised fine-tuning, as shown in Figure \ref{fig: DBN}. The RBM is an undirected bipartite graph consisting of a visible layer $v$ and a hidden layer $h$, which are fully connected between the layers and unconnected within the same layer. Its energy function is:
\begin{figure}[htbp]
    \centering
    \includegraphics[width=0.45\textwidth,height=55mm]{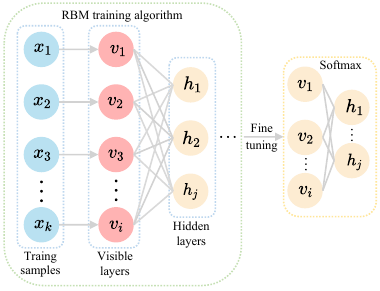}
    \caption{Schematic diagram of a DBN.}
    \label{fig: DBN}
\end{figure}
\begin{equation}
\label{eq: RBF energy function}
E(v,h) =  - \sum\limits_{i = 1}^m {{a_i}{v_i}}  - \sum\limits_{j = 1}^n {{b_i}{h_j}}  - \sum\limits_{i = 1}^m {\sum\limits_{j = 1}^n {{v_i}{w_{ij}}{h_j}} } ,
\end{equation}
where $w_{ij}$ is the weight matrix, $a$ and $b$ are the bias terms of the visible and hidden layers, respectively. Based on the Equation \ref{eq: RBF energy function}, the marginal distribution of the visible units can be calculated as:
\begin{equation}
\label{eq: marginal distribution RBM}
P(v,h) = \frac{1}{Z}{e^{ - E(v,h)}},
\end{equation}
where $Z = \sum\limits_{v,h} {{e^{ - E(v,h)}}} $ represents partition function. The activation conditions for visible and hidden units are defined as follows:
\begin{equation}
\label{eq: activation conditions}
\left\{ \begin{array}{l}
P({v_i} = 1|h) = \sigma (\sum\limits_{j = 1}^m {{w_{ij}}{h_j} + {a_i}} )\\
P({h_j} = 1|v) = \sigma (\sum\limits_{i = 1}^n {{w_{ij}}{v_i} + {b_j}} )
\end{array} \right.,
\end{equation}
where $\sigma $ denotes the Sigmoid function. Based on contrastive divergence (CD), the weight matrix $w_{ij}$ is updated as follows:
\begin{equation}
\label{eq: weight matrix update}
\Delta {w_{ij}} = \eta ( < {v_i}{h_j}{ > _{data}} -  < {v_i}{h_j}{ > _{recon}}),
\end{equation}
where, $\eta$ is the learning rete, $< {v_i}{h_j}{ > _{data}}$ and $< {v_i}{h_j}{ > _{recon}}$ represent the expected values of the product $v_i h_j$ under the data distribution and reconstructed distributions, respectively. 

DBN has been widely applied in CIR due to its strong capability in learning hierarchical and discriminative features from raw, high-dimensional and noisy monitoring signals. Wang et al. \cite{wang2017hydraulic} proposed a fast and effective cavitation diagnosis approach for hydraulic system vibration signals based on a DBN with sliding window spectrum feature. Huang et al. \cite{huang2019analysis} presented an automatic weak cavitation recognition method for hydraulic systems based on the combination of multi-scale entropy feature extraction of fault-sensitive intrinsic mode function and DBN. Liu et al. \cite{liu2023optimization} developed a method to improve the cavitation diagnosis performance of gear vibration signals by optimizing the number of hidden layer neurons and learning rate of DBN. Yan et al. \cite{yan2019fault} devised an improved data preprocessing method combined with a genetic algorithm optimized DBN to enhance cavitation diagnosis performance for actual gas turbine.

Although DBN exhibits strong capability in automatic feature learning, it also encounters several limitations in practical applications. First, the performance of DBN is highly dependent on the setting of hyperparameters, which typically require repeated tuning for different datasets. Second, DBN training involves layer-wise pretraining and fine-tuning, leading to relatively high computational cost and long training time.

\subsubsection{CNN-based Approaches}
\label{subsubsec: CNN-based Approaches}
Convolutional neural network (CNN) is a type of deep feedforward neural network inspired by the receptive field mechanism of the visual cortex, which is widely applied in computer vision \cite{wu2023towards,huang2022feature,bao2023lightweight,lu2022understanding}, speech recognition \cite{palaz2015analysis,abdel2014convolutional,abdel2013exploring}, natural language processing \cite{li2025encoder,lavanya2021deep,huang2025median,fu2025pair}, time-frequency signal analysis \cite{qian2025diffusion,li2022study,jiao2024self,wu2024novel} and so on \cite{gou2025dynamic,yuan2025prioritization,pang2018equation,lu2024generic,xiang2022phase}. Its basic structure typically consists of an input layer, several convolutional layers, pooling layers, fully connected layers and an output layer. The CNN achieves hierarchical automatic feature extraction through convolution and pooling operations, as shown in Figure \ref{fig: Conv}, followed by nonlinear transformations via activation functions and finally performs classification or regression through fully connected layers. Given a training dataset $\{ \mathcal{X},\mathcal{Y}\}  = ({x_i},{y_i})$ with $N$ samples, ${x_i} \in {\mathbb{R}^d}$ represents $i$-th sample and ${y_i} \in \{  + 1, - 1\}$ is the corresponding label. For a one dimensional convolutional network, the operation for the $k$-th feature map in the $h$-th layer can be expressed as:
\begin{figure*}
    \centering
    \includegraphics[width=0.8\textwidth,height=60mm]{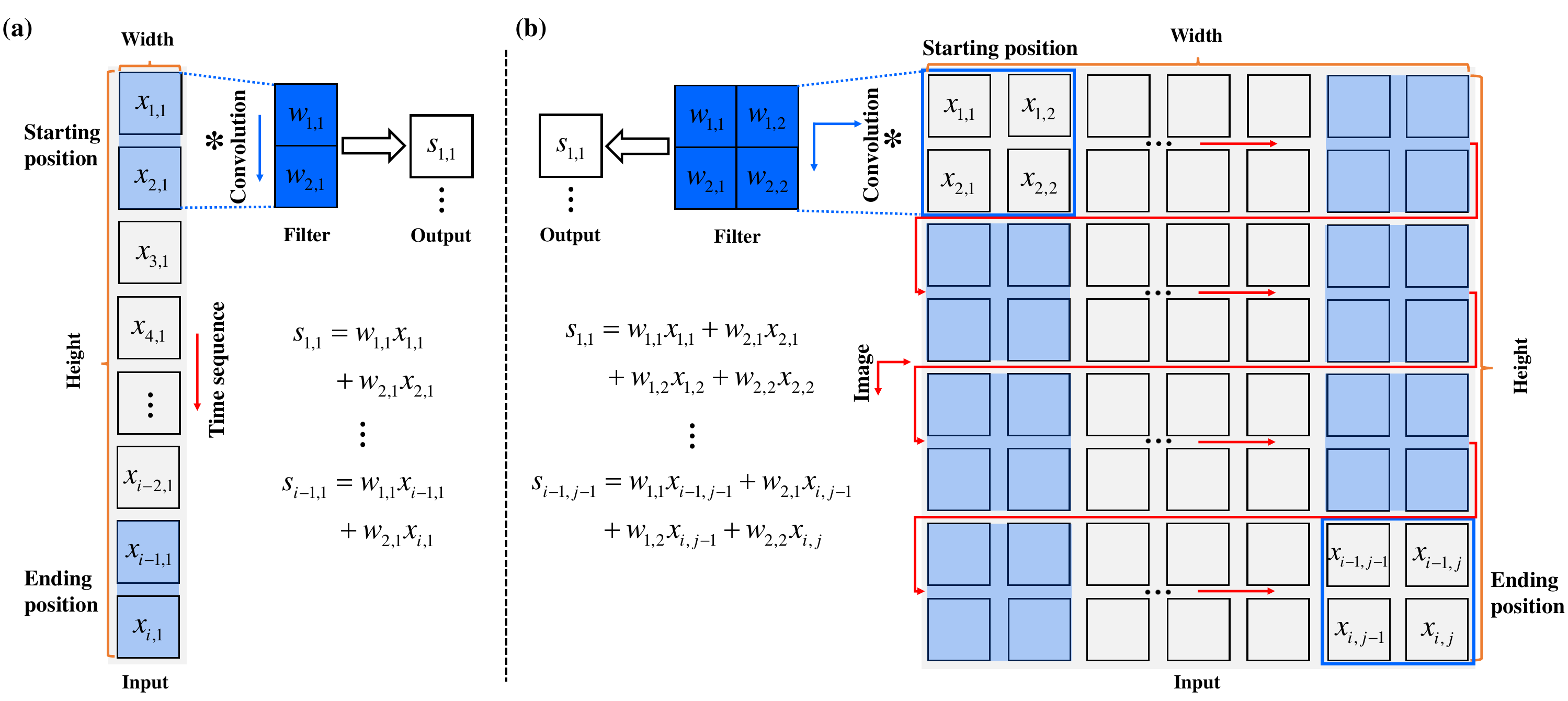}
    \caption{Schematic diagram of convolution operation. (a) and (b) show the working principle of 1D and 2D convolution, respectively.}
    \label{fig: Conv}
\end{figure*}
\begin{equation} 
\label{eq: cnn conv} 
x_{ik}^h = \sigma( \sum\limits_j {x_j^{h-1} \cdot w_{jk}^h + b_k^h}), 
\end{equation}
where $x_{ik}^h $ represents the output of the $k$-th feature map in the $h$-th layer, $x_j^{h-1}$ is the $j$-th input feature map from the $(h-1)$-th layer, $\sigma( \cdot )$ is the activation function (e.g. ReLU, Sigmoid, Tanh, etc.), $w_{jk}^h$ and $b_k^h$ are the convolution kernel and bias corresponding to input channel $j$ and output channel $k$, respectively. Pooling layers are used to reduce the spatial or temporal resolution of the feature maps and enhance invariance to translation and noise. For example, the max pooling can be formulated as:
\begin{equation} 
\label{eq: max pooling} 
p_k^{(h)}(u) = \max_{s \in \Omega_u} (x_{ik}^h)(s), 
\end{equation}
where ${\Omega _u }$ denotes the pooling region centered at position $u$. After several convolution and pooling layers, the high level features are flattened and passed into fully connected layers. The predicted output of the $k$-th neuron in the output layer is computed as:
\begin{equation} 
\label{eq: cnn output} 
\hat{y}_{ik} = \sigma^{out} (\sum\limits_j {w_{jk}^{out} \cdot z_j + b_k^{out}}), 
\end{equation} 
where $z_j$ denotes the $j$-th neuron output in the fully connected layer, $\sigma^{out}( \cdot )$ is typically Sigmoid or Softmax function, $w_{jk}^{out}$ and $b_k^{out}$ are the weights and biases of the output layer, respectively. The objective of CNN is to minimize the between the predicted output and the ground truth:
\begin{equation} 
\label{eq: cnn objective} 
\min_{\theta} \sum_{i=1}^N \mathcal{L}(y_i, \hat{y}_i), 
\end{equation}
where $\theta$ represents the model parameters and are optimized using stochastic gradient descent (SGD) or its variants (Adam, RMSprop, etc.), with the update formulas:
\begin{equation} 
\label{eq: cnn update} 
\begin{array}{cc} 
w^* \leftarrow w - \eta \cdot \frac{\partial \mathcal{L}}{\partial w},\;\;\;\;\;\; b^* \leftarrow b - \eta \cdot \frac{\partial \mathcal{L}}{\partial b}, 
\end{array} 
\end{equation}
where $\eta$ is the learning rate, ${\partial \mathcal{L}}/{\partial w}$ and ${\partial \mathcal{L}}/{\partial b}$ are the gradients with respect to $w$ and $b$, respectively. In general, CNN can be classified into 1D CNN, 2D CNN and 3D CNN. Based on the above, researchers have developed various architectures, such as LeNet \cite{lecun2002gradient}, AlexNet \cite{krizhevsky2012imagenet}, VGGNet \cite{simonyan2015very}, DenseNet \cite{huang2017densely}, ResNet \cite{he2016deep}, MobileNet \cite{howard2017mobilenets}, ShuffleNet \cite{zhang2018shufflenet}.

The application of CNN to CIR are systematically reviewed in Table \ref{tab: CNN Methods}. As shown in Table \ref{tab: CNN Methods}, CNN-based methods have been widely applied to CIR in various hydraulic machinery and fluid systems, such as pumps, turbines, pipes, valve and propellers. Among these, pumps are the most extensively studied, with both 1D CNN and 2D CNN architectures being applied. Specifically, 1D CNN is mainly used for processing time series signals (e.g. acoustic, vibration, prsssure, current, etc.), while 2D CNN often relies on time-frequency representations or high-speed images to capture spatial features of cavitation. Research on turbines, pipes, propellers and valves is relatively limited, which indicates potential for further expansion. Many CNN-based studies directly employ established network architectures, which makes it challenging to fully tailor the models to the unique characteristics of CIR. In general, CNN-based CIR studies show a trend toward integrating multiple signal types and network structures, though the maturity varies across different objects. 
\begin{table*}[htbp]
\centering
\caption{Summary of applications of CNN-based methods in cavitation intensity recognition.}
\label{tab: CNN Methods}
\small
\setlength{\tabcolsep}{5mm}{
\begin{tabular}{lclc}
\toprule 
Objects                  & Types                                       & References & Signals                 \\ 
\midrule
\multirow{7}{*}{pump}    
& \multirow{3}{*}{1D CNN}                     & \cite{tong2023cavitation},
                                                \cite{chao2020cavitation},
                                                \cite{liu2021convolution},
                                                \cite{tiwari2021blockage},
                                                \cite{han2022comparative},
                                                \cite{li2022application},
                                                \cite{han2025use},
                                                \cite{song2025cavitation},
                                                \cite{lu2024research},

                                                & \multirow{8}{*}{\makecell{Acoustic, \\Vibration, \\ Pressure,\\ Current, \\ High-speed images}} \\

&                                             & 
                                                \cite{he2024status},
                                                \cite{turunen2023deep},
                                                \cite{wang2025cavitation},
                                                 
                                                \cite{feng2024experimental},
                                                \cite{vasiliev2023pump},
                                                \cite{pan2018improved},
                                                \cite{zheng2025intensity},
                                                \cite{e2024development},
                                                \cite{ugli2023automatic},

                                                &                         \\
                                            && 
                                                
                                                \cite{li2020adaptive},
                                               \cite{zheng2023hydrodynamic},
                                               \cite{guo2022fault}

                                            \\
\cmidrule{2-3}
& \multirow{4}{*}{2D CNN}                     & 
                                                \cite{kumar2020improved}, 
                                                \cite{chao2020identification},
                                                \cite{Wei2021cavitation},
                                                \cite{zhu2021intelligent},
                                                \cite{hajnayeb2021cavitation},
                                                \cite{chao2022improving},
                                                \cite{liu2023intelligent},
                                                \cite{chennai2023deep},
                                                \cite{tang2024light},

                                              &                         \\
&                                             & 
                                                \cite{zhu2023data},
                                                \cite{ullah2023intelligent},
                                                \cite{tang2022adaptive},
                                                \cite{prasshanth2024deep},
                                                \cite{prasshanth2024fault},
                                                \cite{chao2023cavitation},
                                                \cite{sun2025cavitation},
                                                \cite{lee2025deep},
                                                \cite{chang2022centrifugal},

                                                &                         \\
                                                && 
                                                   \cite{huh2020new},
                                                   \cite{li2024fault},
                                                   \cite{chao2022fault},
                                                   \cite{tang2020intelligent},
                                                   \cite{zaman2023centrifugal},
                                                   \cite{kumar2025inceptionv3},
                                                   \cite{sunal2024centrifugal},
                                                   \cite{ma2023systematic},
                                                   \cite{anvar2023novel},
                                                
                                                \\
                                                && 
                                                  \cite{wang2019novel},
                                                  \cite{dai2023cavitation},
                                                  \cite{li2024vmd}

                                                \\
\midrule

\multirow{3}{*}{Turbine}    
& \multirow{2}{*}{1D CNN}                     & \cite{zheng2025intensity},
                                                \cite{zheng2023hydrodynamic},
                                                \cite{liu2025data},
                                                \cite{singh2025hybrid},
                                                \cite{dao2024wear},
                                                \cite{qian2022cnn},
                                                \cite{chen2025fault},
                                                \cite{jin2019eemd},
                                                \cite{wan2025fault},

                                                & \multirow{3}{*}{\makecell{Acoustic, Vibration, \\Pressure, Current}} \\
&                                             & \cite{lin2023application}          &                         \\
\cmidrule{2-3}
& \multirow{1}{*}{2D CNN}                     & \cite{kirschner2023cavitation},
                                                \cite{look2018building},
                                                \cite{tang2022intelligent},
                                                \cite{he2023sgst},
                                                \cite{wang2021high},
                                                \cite{liu2021method},
                                                \cite{xu2023cost}

&                         \\

\midrule

\multirow{2}{*}{Pipe}    
& \multirow{1}{*}{1D CNN}                     & \cite{you2024cavitation},
                                                \cite{sha2022regional},
                                                \cite{sha2022multi},
                                                \cite{gou2024hierarchical},
                                                \cite{wang2022deeppipe},
                                                \cite{li2022rolling}

                                                                               & \multirow{2}{*}{Acoustic} \\

\cmidrule{2-3}
& \multirow{1}{*}{2D CNN}                     & \cite{sandhu2024comparative},
                                                \cite{sha2024hierarchical},
                                                \cite{liu2022automatic}
                                              &                         \\

\midrule  

\multirow{2}{*}{Propeller}    
& \multirow{1}{*}{1D CNN}                     & \cite{tsai2021multi},
                                                \cite{jamal2024passive},
                                                \cite{kim2022study}

                                                                                     & \multirow{2}{*}{\makecell{Acoustic, Current, \\High-speed images}} \\

\cmidrule{2-3}
& \multirow{1}{*}{2D CNN}                     & \cite{miglianti2020predicting},
                                                \cite{bach2021classification},
                                                \cite{bach2021enhancing},
                                                \cite{liu2025feature},
                                                \cite{yilmaz2025deep},
                                                \cite{chen2024ship},
                                                \cite{shen2022quantitative},
                                                \cite{sheikh2025marine},
                                                \cite{sheikh2025ensemble}
                                                
                                                &                         \\

\midrule  
\multirow{2}{*}{Valve}    
& \multirow{1}{*}{1D CNN}                     & \cite{sha2022regional},
                                                \cite{sha2022multi},
                                                \cite{gou2024hierarchical}     
                                                
                                                                                     & \multirow{2}{*}{\makecell{Acoustic, \\Vibration}} \\

\cmidrule{2-3}
& \multirow{1}{*}{2D CNN}                     &  \cite{sha2024hierarchical},
                                                \cite{wei2023cavitation}
                                               
                                                    &                         \\

\bottomrule
\end{tabular}}
\end{table*}

Compared with DBNs, CNNs offer advantages in CIR, including automatic extraction of local features, reduced reliance on manual feature design and higher computational efficiency. In addition, CNNs also demonstrate higher performance and robustness when processing images and time-frequency two-dimensional features. However, CNNs have limited capability in capturing strong nonlinear and long-term dependency features and their performance is sensitive to the quantity and quality of training data.

\subsubsection{RNN-based Approaches}
\label{subsubsec: RNN-based Approaches}
\begin{figure}[htbp]
    \centering
    \includegraphics[width=0.45\textwidth,height=35mm]{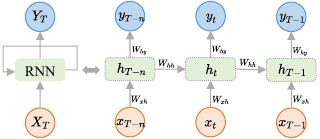}
    \caption{Schematic diagram of a RNN.}
    \label{fig: RNN}
\end{figure}

Recurrent neural network (RNN) is class of neural network models designed to capture temporal dependencies and sequential correlations in data \cite{zaremba2014recurrent}. Unlike CNNs, RNN introduce recurrent connections in hidden layer, allowing information from previous time steps to influence the current output, as shown in Figure \ref{fig: RNN}. The basic operation of a vanilla RNN for a sequence $\left \{x_1,x_2,   \cdots ,x_T\right \}$ can be expressed as:
\begin{equation} 
\label{eq: rnn hidden} 
h_t = \sigma(W_{xh} x_t + W_{hh} h_{t-1} + b_h), 
\end{equation}
\begin{equation} 
\label{eq: rnn output} y
_t = \sigma^{out}(W_{hy} h_t + b_y), 
\end{equation} 
where $h_t$ is the hidden state at time $t$, $W_{xh}$, $W_{hh}$, $W_{hy}$ are the input-to-hidden, hidden-to-hidden and hidden-to-output weight matrices, respectively. $b_h$ and $b_y$ are biases, $\sigma(\cdot)$ is typically a nonlinear activation function and $\sigma^{out}(\cdot)$ is the output activation function. The objective function is to minimize the sequence prediction loss over all time steps:
\begin{equation} 
\label{eq: rnn objective} 
\min_{\theta} \sum_{i=1}^{N} \sum_{t=1}^T \mathcal{L}(y_t^{(i)}, \hat{y}_t^{(i)}), 
\end{equation}
where $\theta =\left \{W_{xh},W_{hh},W_{hy},b_h,b_y\right \}$ and are updated through backpropagation through time. To address the vanishing and exploding gradient issues in vanilla RNN, researchers have developed various architectures, such as LSTM \cite{hochreiter1997long}, BiLSTM \cite{graves2005framewise} and GRU \cite{cho2014learning}.

RNN-based approaches have been effectively applied to CIR. For instance, some research \cite{ehemann2023ai,tan2023fault,al2024hydropower} directly applied LSTM to CIR of valves and pumps, respectively. Lee et al. \cite{lee2025deep} and Zhu et al. \cite{zhu2023novel} proposed combining CNN and RNN methods to achieve CIR in pumps using high-speed images and vibration time-frequency images, respectively. Zhang et al. \cite{zhang2024cavitation} developed a multi-dimensional feature fusion method based on a convolutional gated recurrent unit to achieve CIR in centrifugal pumps. Liu et al. \cite{liu2024failure} designed a GRU-based method using vibration data to a realize CIR of firefighting pumps. Zhao et al. \cite{zhao2025optimizing} presented a VMD-dung bettle optimization-GRU-attention model using pressure data to achieve CIR in pumps and turbines.

Compared to DBNs and CNNs, RNN-based models have superior capabilities in modeling temporal dependencies and capturing highly time-correlated signal patterns. However, their training process is often time-consuming and they are more susceptible to gradient vanishing and exploding issues. Moreover, their performance can degrade when data are insufficient or noisy.

\subsubsection{Transformer-based Approaches}
\label{subsubsec: Transformer-based Approaches}
\begin{figure}[htbp]
    \centering
    \includegraphics[width=0.45\textwidth,height=100mm]{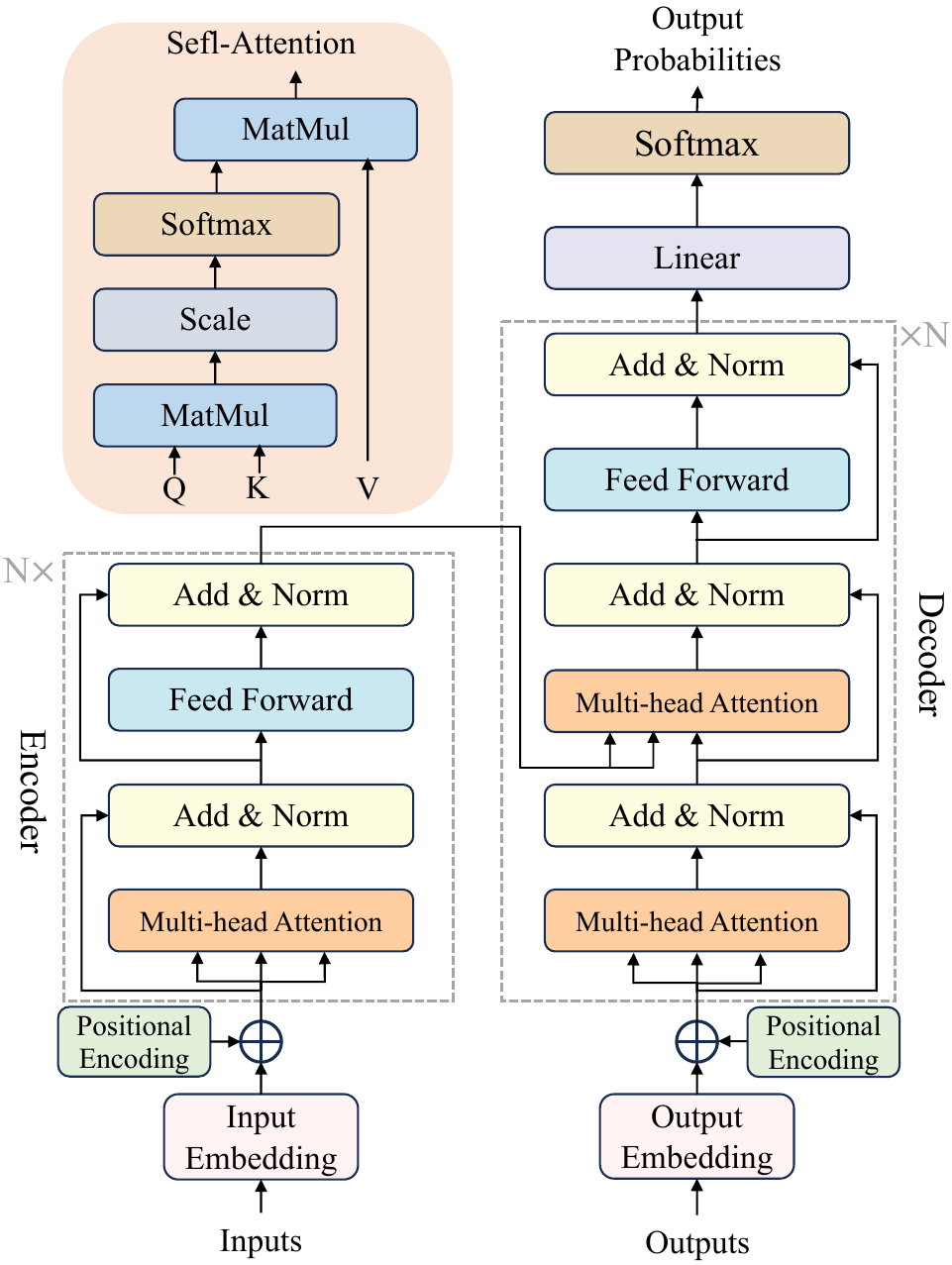}
    \caption{Schematic diagram of a Transformer.}
    \label{fig: Transformer}
\end{figure}

Transformer is a deep learning architecture entirely based on the attention mechanism, which has become the infrastructure for large language models \cite{vaswani2017attention,zeng2025futuresightdrive}. Unlike CNNs and RNNS, the Transformer discards recurrent and convolutional structures, instead of employing a fully parallelizable multi-head self-attention mechanism to model global dependencies, significantly improving training speed and long sequence modeling capability \cite{zeng2024driving,lin2024multi}. A standard Transformer consists of an encoder and a decoder stack. The encoder comprises multi-head self-attention layers, positional encoding, feed forward network, residual connections and layer normalization, see Figure \ref{fig: Transformer}. The decoder adds masked self-attention and encoder-decoder attention mechanisms to the encoder. Given an input sequence $\mathcal{X} = [{x_1},{x_{2,}} \cdots ,{x_N}]$, linear projections are applied to obtain the Query, Key and Value matrices:
\begin{equation} 
\label{eq: QKV}
Q = XW_Q, \quad K = XW_K, \quad V = XW_V, 
\end{equation}
where $W_Q$, $W_K$, $W_V$ are learnable weight matrices. The self-attention computation is defined as:
\begin{equation} 
\label{eq: self attention}
\text{Attention}(Q, K, V) = \text{softmax}  \left( \frac{QK^{\top}}{\sqrt{d_k}}\right) V, 
\end{equation}
where $d_k$ is the dimension of the key vectors. The multi-head self-attention mechanism computes $h$ parallel attention heads and concatenates results:
\begin{equation} 
\text{MultiHead}(Q,K,V) = \text{Concat}(\text{h}_1, \dots, \text{h}_h)W_O, 
\end{equation}
\begin{equation} 
\text{h}_i = \text{Attention}(QW_Q^{i},KW_K^{i},VW_V^{i}), 
\end{equation}
where $W_O$ is the output projection matrix. Based on the above theories, researchers present a variety of improved architectures for different tasks, such as ViT \cite{dosovitskiy2020image}, SwinT \cite{liu2021swin}, BERT \cite{devlin2019bert}, Linformer \cite{wang2020linformer} and CLIP \cite{radford2021learning}. 

Transformer as an emerging deep learning architecture that has also has been introduced into the field of CIR. For example, He et al. \cite{he2023sgst} proposed a model that combines Short-time Fourier Transform, generative adversarial network and Swin-Transformer to achieve CIR using vibration signal spectrograms. Jiang et al. \cite{jiang2024deep} presented a combination of continuous wavelet transform and Swin Transformer method for CIR in centrifugal pumps. Shang et al. \cite{shang2025lightweight} developed a lightweight vision transformer model combining adaptive convolution and mobile vision transformer blocks to achieve early-stage CIR under complex and noisy conditions. Zhang et al. \cite{zhang2025data} designed a deep learning approach using mel-spectrogram representations combined with self-attention mechanisms to realise CIR from hydrofoil acoustic signals. Yu et al. \cite{sha2024hierarchical} proposed a hierarchical knowledge-guided acoustic CIR method based on graph convolutional networks and reweighted hierarchical knowledge correlation matrices using various Transformer architectures.

Compared to CNNs, the application of Transformer is constrained by multiple factors, including the requirement for large-scale and high-quality labeled datasets, higher computational cost due to complex self-attention operations and the challenge of effectively modeling domain-specific features when applied to relatively small or noisy datasets. Transformer offers strong capabilities in capturing long-term dependencies and global contextual information, making it well suited for tasks with strong spatial and temporal correlations of signals. In addition, it has high flexibility in addressing multi-modal data. However, Transformer typically requires more computational resources and careful parameter tuning, which can make it less efficient for on-site or real-time monitoring scenarios. Furthermore, it is prone to overfitting and may struggle to generalize well compared to more lightweight structures in the case of insufficient training data.

\section{Data-Knowledge Driven Deep Learning}
\label{sec: 4_DKDL}
\begin{figure*}
    \centering
    \includegraphics[width=\textwidth,height=70mm]{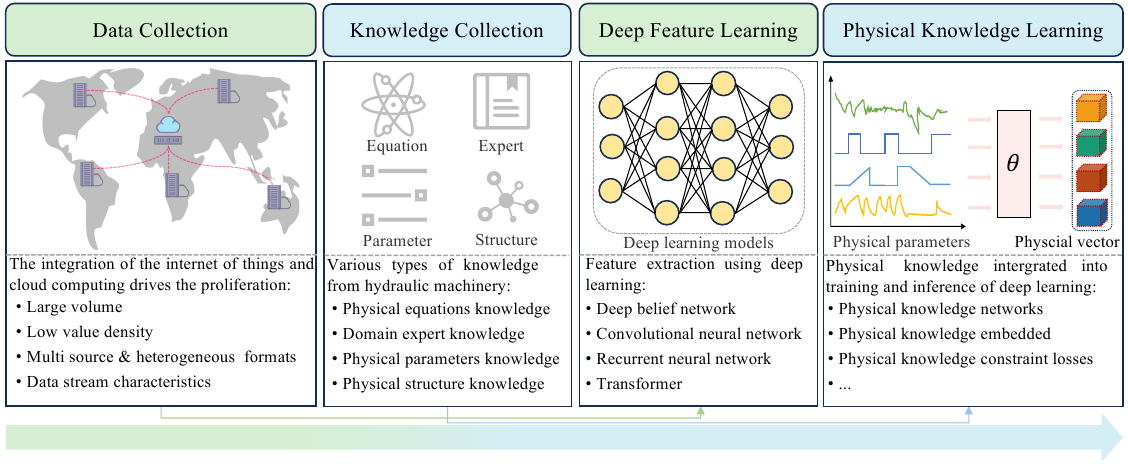}
    \caption{Diagnosis procedure of cavitation intensity recognition using data and knowledge driven deep learning methods, which consists of data acquisition, knowledge collection, deep feature learning and physical knowledge learning.}
    \label{fig: DKDL}
\end{figure*}

In contrast to data-driven deep learning approaches, data-knowledge driven deep learning (DKDL) methods have attracted growing attention in CIR because they integrate physical knowledge with the strong feature learning capability of deep models. By embedding domain-specific physics, governing equations or expert heuristics into the learning process, DKDL not only can enhance model interpretability, but also ensure the predictions comply with fundamental physical principle. This hybrid paradigm effectively bridges the gap between data-driven inference and physics-based modeling. A typical DKDL pipeline for CIR generally includes data and knowledge collection, deep feature learning and physical knowledge learning, as shown in Figure \ref{fig: DKDL}, which will be detailed in the following subsections.

\subsection{Data and Knowledge Collection}
\label{subsec: Data-Knowledge Collection}
For DKDL framework, the data component utilises various signals acquired from different sensors described previously, which characterize the operating states of hydraulic machinery under various work conditions. On this basis, the model incorporates multiple types of physical knowledge from the hydraulic machinery domain, which are structured and formalized to support feature learning and recognition tasks \cite{kumar2023knowledge}. These knowledge types include:
\begin{itemize}
    \item Physical equations knowledge \cite{brennen2014cavitation}: Includes fluid mechanics governing equations (e.g. Navier-Stokes equations), bubble dynamics equations (e.g. Rayleigh-Plesset equation) and cavitation inception criteria (e.g. cavitation coefficient), which describe the fundamental principle of liquid flow and cavitation formation.
    \item Domain expert knowledge \cite{alhashmi2005detection}: Derived from the long-term practical experience and diagnostic heuristics of engineers (e.g. abnormal vibration patterns, acoustic characteristics, pressure variation rules, historical cases, fault mechanism analysis, determination rules, etc.), provides diagnostic basis and domain background information.
    \item Physical parameters knowledge \cite{roy1999cavitation}: Key physical quantities obtained during the operation or experimental monitoring of hydraulic machinery (e.g. flow rate, rotational speed, pressure values, temperature, etc.). These parameters reflect the operational state and cavitation occurrence conditions of the equipment, providing highly relevant physical information for models.
    \item Physical structure knowledge \cite{folden2023classification}: Covers the hierarchical and dependency relationships among target classes, the mutual exclusion or inclusion constraints between classes, the coupling and constraint relationships among physical variables and the structural layout and functional dependencies among hydraulic machinery components. This type of knowledge enables the model to incorporate multi-level correlation information during the learning process.
\end{itemize}

These knowledge can be used to construct physical constraints, enhance feature representations or directly construct physical knowledge networks, which effectively guides the model to learn discriminative patterns matching practical operating conditions and physical principle. The DKDL framework not only retains the automatic feature learning advantages of data-driven methods, but also significantly improves the physical consistency and generalization capability of recognition results.

\subsection{Deep Feature Learning}
\label{subsec: Deep Feature Learning}
Deep feature learning aims to utilize deep neural networks to automatically extract high-level feature representations from raw multi-source sensor data, supporting subsequent fusion with physical knowledge. The input data typically includes acoustic signals, vibration signals, pressure pulsation signals, current signals and cavitation flow field images captured by high-speed imaging, which are preprocessed and directly fed into the deep learning model to perform end-to-end feature extraction. The deep learning architectures employed are consistent with data-driven deep learning approaches.

\subsection{Physical Knowledge Learning}
\label{subsec: Physical Knowledge Learning}
Physical knowledge learning aims to seamlessly integrate fundamental physical principle, domain experience, key physical parameters and structural relationships information into both the training and inference processes of deep models, which can enhance the physical consistency, cross-condition generalization capability and interpretability of CIR results. In view of current research progress, three representative physical knowledge learning approaches are briefly introduced in the following sections.

\subsubsection{Physical Knowledge Network}
\begin{figure}[htbp]
    \centering
    \includegraphics[width=0.4\textwidth,height=65mm]{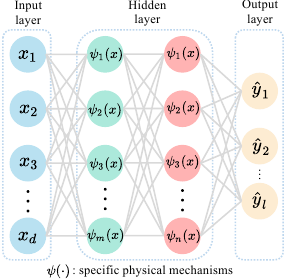}
    \caption{Schematic diagram of a PKN with two physical mechanism hidden layers.}
    \label{fig: PKN}
\end{figure}

The physical knowledge network (PKN) is an essential component of the DKDL framework. Its core principle involves explicitly incorporating the physical principle and characteristics of hydraulic machinery through customized network architectures, enhancing the physical consistency and interpretability of CIR. Unlike generic data-driven deep learning architectures, the PKNs incorporate specific physical mechanisms into their hierarchical design, connection patterns and kernel construction, enabling a better capture of intrinsic relationships among multidimensional physical variables, equipment structures and signal features, as shown in Figure \ref{fig: PKN}. In current research, typical PKN architectures include:
\begin{itemize}
    \item Hierarchical neural networks \cite{gou2024hierarchical,mavrovouniotis1992hierarchical}: Design network architectures based on hierarchical relationships among cavitation intensity labels or physical variables, enhancing the ability to discriminate multi-level cavitation states.
    \item GCN-based hybrid networks \cite{sha2024hierarchical,li2025graph}: Leverage graph convolutional netwotk (GCN) to model the correlation relationships between physical variables or cavitation intensity labels and integrate them with other feature learning networks (e.g. CNNs, RNNs, Transformer, etc.), enabling effective guidance and constraint of deep features.
    \item Transformation-domain neural works \cite{guo2022review,uteuliyeva2020fourier}: Directly replace or modify certain components of traditional data-driven neural networks (e.g. activation functions, kernel function, inter-layer weights, etc.) with signal transformation methods (e.g. wavelet transform, Fourier transform, etc.), enabling the model to acquire time-frequency domain or frequency domain feature extraction capabilities at the structural level.
\end{itemize}

Compared to data-driven deep learning approaches for CIR, the application of PKNs in this field remains in its infancy and the current research is relatively limited. For hierarchical neural networks, Gou et al. \cite{gou2024hierarchical} proposed a two-stage sub-master transition hierarchical network for CIR using acoustic signals from valve and pipeline systems. Wen et al. \cite{wen2019new} presented a hierarchical convolutional neural network method to achieve bearing fault diagnosis using vibration signals. Guo et al. \cite{guo2016hierarchical} developed a hierarchical learning rate adaptive deep convolutional neural network to implement bearing fault diagnosis using vibration signals. For GCN-based hybrid networks, Sha et al. \cite{sha2024hierarchical} proposed a hierarchical knowledge-guided acoustic CIR method based on graph convolutional networks and reweighted hierarchical knowledge correlation matrices. Li et al. \cite{li2025graph} presented a graph isomorphic network with a spatio-temporal attention mechanism to achieve fault diagnosis of hydraulic axial piston pumps using weighted graph data constructed from univariate signals. Hua et al. \cite{hua4977997fault} designed a graph convolutional transformer network method to perform fault diagnosis of slurry circulating equipment using frequency-domain graph data constructed from high-frequency vibration signals. For transformation-domain networks, Gao \cite{gao2000novel} presented a neural-wavelet network to achieve process diagnostics based on magnetic flowmeter data and power load demand prediction using electrical load signals. Won et al. \cite{won2004neural} proposed a method combining wavelet coefficient variance features with neural networks to realise cavitation detection in pumps using pressure signals. Wang et al. \cite{wang2009intelligent} suggested a method combining wavelet transform, rough sets and a partially linearized neural network to achieve centrifugal pump system fault diagnosis. Rani et al. \cite{rani2023fault} introduced a probabilistic Fourier neural operator method to conduct process fault detection using multivariate historical process data. Tan \cite{tan2006fourier} proposed a Fourier neural network and generalized single hidden layer network method to perform multi-fault diagnosis. Yu et al. \cite{yu2024method} developed a fast Fourier convolutional gated current unit method to perform fault prediction using remaining useful life data.

Compared to data-driven deep learning approaches, the PHN exhibits higher physical consistency and interpretability in CIR and demonstrate better generalization under small-sample or noisy conditions. However, the design of PKNs relies heavily on domain expert knowledge and involves complex construction. In addition, the training process requires simultaneous optimization of data features and knowledge constraints, resulting in relatively high computational costs.

\subsubsection{Physical Knowledge Embedded}
\begin{figure}[htbp]
    \centering
    \includegraphics[width=0.5\textwidth,height=35mm]{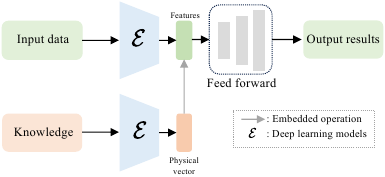}
    \caption{Schematic diagram of a PKE.}
    \label{fig: PKE}
\end{figure}

The core concept of physical knowledge embedded (PKE) is to explicitly integrate domain-specific knowledge into data-driven deep learning models, enabling feature extraction and physical constraints to operate simultaneously. Unlike purely data-driven feature learning, this approach injects physical information directly into the structural design or feature stream of the model, which balances data representation capability with physical consistency and enhances generalization and interpretability in CIR \cite{chi2022knowledge}, as shown in Figure \ref{fig: PKE}. Given a dataset $\{ \mathcal{X},\mathcal{K}\}  = \{({x_i},{k_i})\}_{i=1}^{N}$ with ${x_i}$ represents $i$-th sample and ${k_i} $ represents the corresponding physical knowledge, the embedding process can be expressed as:
\begin{equation}
\label{eq: knowledge embedd}
{F_{\text{embed}}} = \Phi ({f_{\text{data}}}(\mathcal{X}),{f_{\text{know}}}(\mathcal{K})),
\end{equation}
where ${f_{\text{data}}}( \cdot )$ is the operator for data-driven feature extraction, ${f_{\text{know}}}( \cdot )$ is the mapping operator for physical knowledge and $\Phi ( \cdot , \cdot )$ denotes the fusion strategy. The common forms of physical knowledge embedding include:
\begin{itemize}
    \item Feature-level fusion \cite{smirnov2019knowledge}: At the model input layer, raw sensor data and physical parameters are concatenated or fused, and uniformly encoded into the network so that both data features and physical information are provided from the outset.
    \item Intermediate embedding \cite{dai2020survey}: Domain specific prior knowledge calculated from physical principle are introduced into intermediate layers as additional channels, allowing alongside data-driven features toward subsequent layers. 
    \item Knowledge-guided attention \cite{cui2023knowledge}: Physical knowledge is generated as attention weights or masked areas to guide the model to focus on physically significant regions or variables during feature learning, enhancing discriminative and physical interpretability of features.
\end{itemize}

In contrast to data-driven deep learning approaches for CIR, the use of PKEs in this domain is still at an early developmental stage and the current research is relatively limited. For example, Sha et al. \cite{sha2024hierarchical} developed target classes hierarchical knowledge embedding representation learning based on convolutional neural network and word embedding technique for CIR. Miglianti \cite{miglianti2020modelling} proposed a method combining semi-empirical approaches and hybrid machine learning models to perform propeller cavitation noise prediction. Chen et al. \cite{chen2024ship} presented a hierarchical transformer method to achieve ship fault recognition using acoustic noise signals. Qiao et al. \cite{qiao2024prior} designed a prior knowledge embedding contrastive attention learning network method to conduct rolling bearing fault diagnosis using vibration signals. Du et al. \cite{du2023knowledge} suggested a knowledge-embedded deep belief network to achieve fault diagnosis of building chillers.

Compared to PHNs, PKEs directly embeds domain physical knowledge into the input layer or intermediate feature layers, offering flexible modification and low implementation cost, which allows for rapid application to existing deep neural networks. However, the physical knowledge embedded by PKE primarily consists of local parameters or prior features, making it difficult to construct complex global physical relationships. When the knowledge is incomplete or biased, it is susceptible to noise and variations in distribution. Therefore, PKE is more suitable for providing lightweight physical information enhancement to existing models, whereas PHN is better suited for tasks requiring in-depth characterization of multi-level physical mechanisms.

\subsubsection{Physical Knowledge Constraint Loss}
\begin{figure}[htbp]
    \centering
    \includegraphics[width=0.5\textwidth,height=35mm]{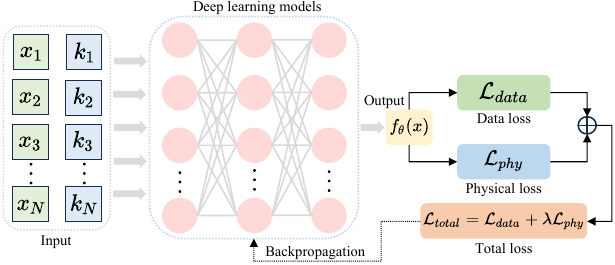}
    \caption{Schematic diagram of a PKCL.}
    \label{fig: PKCL}
\end{figure}

Physical knowledge constraint loss (PKCL) aims to explicitly incorporate physical principle, empirical rules or operational boundaries into the optimization objective of data-driven deep learning models \cite{shen2021physics}, as shown in Figure \ref{fig: PKCL}. The PKCL can guide the neural network to produce outputs consistent with know physical mechanisms and domain constraints by penalizing predictions violating physical feasibility, improving the reliability and interpretability of the deep learning model \cite{karniadakis2021physics}. Given a dataset $\{\mathcal{X},\mathcal{Y},\mathcal{K}\}  = \{({x_i},{y_i},{k_i})\}_{i=1}^{N}$, where $x_i$ denotes the $i$-th input sample, $y_i$ is the ground truth label and $k_i$ represents the corresponding physical knowledge. The total loss function can be expressed as:
\begin{equation}
\label{eq: physical constraint loss}
\mathcal{L}_{\text{total}}=\mathcal{L}_{\text{data}}(f_\theta(\mathcal{X}),\mathcal{Y} )+\lambda \mathcal{L}_{\text{phy}}(f_\theta(\mathcal{X}),\mathcal{K}),
\end{equation}
where $\mathcal{L}_{data}$ is the primary task data loss, $\mathcal{L}_{phy}$ is the physical knowledge constraint loss to measure the degree of violation of physical principle or constraints, $\lambda$ is a weighting factor controlling the balance between task performance and physical consistency, $f_\theta(\cdot)$ denotes the deep model represented by the parameter $\theta$. The common forms of PKCL include:
\begin{itemize}
    \item Boundary condition penalty: Restrict the output prediction values within predefined physical ranges.
    \item Physical principle penalty: Measure the deviation between predicted values and calculated values derived from physical equations or empirical models.
    \item Trend consistency penalty: Enforce the output of prediction results to remain consistent with the inherent properties of physical processes.
\end{itemize}

The PKCL has attracted significant attention and been widely applied in solving ordinary differential equations and other physical-related modeling tasks, but its research in CIR is still in the early developmental stage. For instance, Wang et al. \cite{wang2023physics} proposed a physical informed neural network method to achieve fault severity identification in axial piston pumps using pressure signals. Keshun et al. \cite{keshun2025novel} presented a dynamic physical with a data-driven quadratic neural network and bidirectional LSTM to conduct rolling bearing damage assessment using vibration signals. Keshun et al. \cite{keshun2025sound} developed a sound-vibration physical-information fusion constraint-guided deep learning method to perform bearing fault diagnosis using acoustic and vibration signals. Pan et al. \cite{pan2024interpretable} designed a physics-guided neural network to deliver condenser fault diagnosis using key physical indicator signals. Li et al. \cite{li2025pgmtkd} proposed a physical-guided multi-teacher knowledge distillation network method to achieve external gear pump fault diagnosis using multimodal signals.

Compared with PKE, the PKCL introduces physical constraints into the loss function, synchronizing the data-driven learning process with physical consistency optimization. Its advantages lie in the ability to continuously impose the supervision from physical principle on the model outputs during the training phase, which enhances the interpretability and physical credibility without modifying the network structure. However, PKCL is sensitive to the design of constraint terms and weight selection, which may lead to difficulties in model optimization when the physical knowledge is incomplete or noisy.

\section{Discussions}
\begin{figure*}
    \centering
    \includegraphics[width=\textwidth,height=190mm]{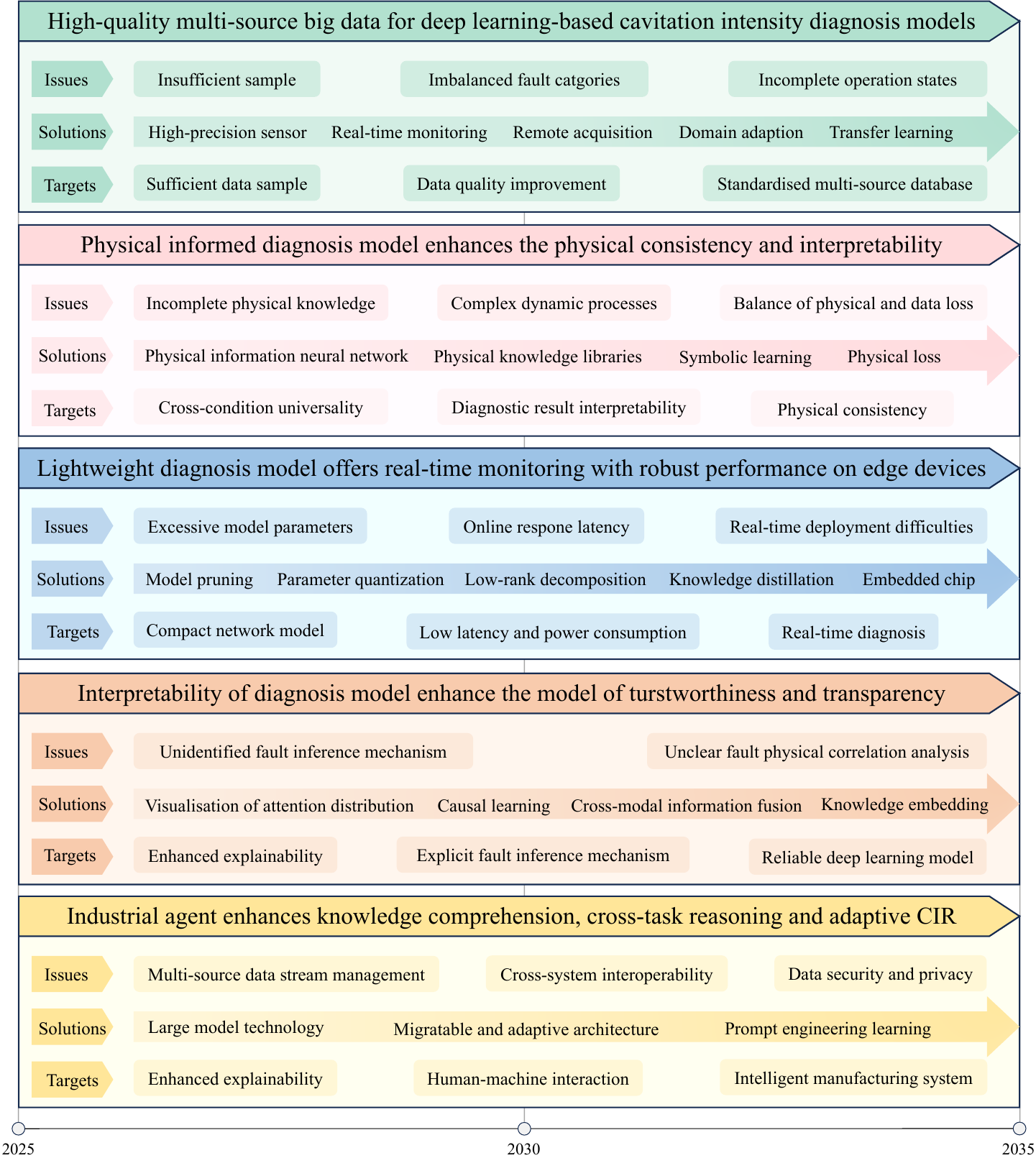}
    \caption{A roadmap outlining the application of machine learning to intelligent cavitation intensity recognition.}
    \label{fig: Discussions}
\end{figure*}

With the continuous development of cavitation intensity recognition (CIR) technology, research has evolved from traditional machine learning reliant on manually extracted features to deep learning with end-to-end feature extraction, which develops physics-guided diagnostic models incorporating physical knowledge. At the end of this review, we attempt to describe a roadmap and discuss challenges of CIR based on a systematic analysis of relevant literature from 2004 to 2025, as shown in Figure \ref{fig: Discussions}, aiming to inspire readers to understand potential research trends and directions in this field over the next five to ten years.

\label{sec: 5_Dissusion}
\subsection{High-Quality Multi-Source Data}
High-quality multi-source data is the foundation for reliable cavitation intensity recognition (CIR) and related industrial equipment fault diagnosis. Current datasets generally suffer from insufficient sample size, imbalanced fault categories, disproportionate distribution between normal and fault conditions and incomplete operational condition coverage. These issues not only hinder comprehensive feature learning during the training phase, but also reduce diagnostic performance and stability in practical applications. In addition, multi-source data (e.g. acoustic signals, vibration signals, pressure pulsations, current signals, high-speed imaging, etc.) are seldom acquired synchronously, which can lead to temporal misalignment and feature distortion during multimodal fusion.

In the future, the focus should be on establishing standardized multi-modal data collection protocols to ensure comparability and interoperability of data collected from different experimental platforms. At the same time, low-cost and high-accuracy sensing technologies should be explored to enhance the feasibility of data acquisition and large-scale cavitation status database should be established using real-time monitoring and remote acquisition technologies. Meanwhile, the integration of data augmentation techniques (e.g. signal simulation, noise modeling, signal mixing, etc.), domain adaptation and transfer learning can be employed to migrate knowledge from similar operating conditions or related equipment to expand the training sample space. In addition, hybrid modeling frameworks integrating simulated and experimental data can also effectively alleviate sample insufficiency and enhance the model robustness and cross-scenario adaptability.

\subsection{Physics Informed Diagnostic Model}
Physics informed diagnostic models can enhance the physical consistency and interpretability of fault diagnostic results by fusing data-driven learning with domain physical knowledge, which can strengthen the credibility of decision credibility and engineering applicability. Although physical information neural networks (PINNs), physical knowledge embedding (PKE) and physical knowledge constraint loss (PKCL) have achieved success in science and  engineering fields (e.g. fluid mechanics, structural health monitoring, energy systems, etc.), they are still in infancy in cavitation intensity recognition. In addition, the existing studies mainly focus on embedding individual physical laws or simplified models and lacks a unified deep integration framework. In practice, this class of methods mainly face challenges including the difficulty of translating multi-scale coupled dynamic equations of the cavitation process into directly usable constraints, the the incompleteness or noise in physical knowledge of industrial systems and the difficulty of balancing physical constraints with the flexibility of data-driven models.

Future research should explore automated and adaptive knowledge integration by leveraging machine-readable physical formula libraries, symbolic learning and graph-based knowledge representations to enable the automatic extraction and dynamic embedding of physical knowledge. In addition, physics-oriented interpretable tools should be developed to allow engineers to intuitively understand the model's decision process.

\subsection{Lightweight Diagnostic Model}
In industrial embedded monitoring systems, computational efficiency and hardware resource limitations create an urgent demand for lightweight diagnostic models, while requiring the model complexity be reduced without significantly compromising performance. However, many current deep learning models contain millions or even hundreds of millions of parameters, which not only increases storage and computational burdens but also makes real-time deployment difficult and limits adaptability to low-power edge devices. Furthermore, in multi-source data scenarios, models usually need to simultaneously process multimodal signals, resulting in traditional large-scale models are more prone to overfitting or response delays for on-line applications.

Future research should focus on designing task-specific compact network architectures for multi-source inputs, incorporating multimodal feature fusion and knowledge guidance into the architecture. In terms of compression techniques, model pruning, parameter quantization, low-rank decomposition and knowledge distillation can be explored. In particular, multi-teacher distillation strategies for handing missing modalities, which can retain critical feature extraction capabilities while reducing computational costs. In addition, hardware optimization algorithms should be integrated, such as operator acceleration for embedded chips and FPGA or ASIC deployment schemes. At the same time, exploration of neuromorphic computing and event-driven neural networks is also warranted to further lower latency and power consumption. Furthermore, lightweight models must also incorporate stability design in noisy environments, ensuring they maintain high reliability and performance in real-world industrial conditions.

\subsection{Interpretability of Diagnostic Model}
In safety-critical domains (e.g. aerospace, marine, energy, etc.), interpretability is a core factor in establishing trust in intelligent diagnostic systems. However, current deep learning models lack explicit reasoning about fault mechanisms and physical correlation analysis, making it difficult for maintenance personnel to fully trust the diagnostic conclusions of models. Existing studies aimed at improving interpretability primarily include visualizing attention weight distributions to reveal regions of interest for the model, conducting sensitivity analysis on multimodal inputs to evaluate feature contributions and embedding physical knowledge to link decision processes with fault evolution mechanisms. Nevertheless, there are still significant technical challenges in enhancing interpretability while maintaining diagnostic performance under complex multi-source data and non-linear feature scenarios.

Future research should focus on constructing diagnostic model architectures with interpretability as a core design principle, taking interpretability requirements into account during the network structure design stage. Causality-driven feature learning methods should be introduced to model the relationships between input features and fault causal chains, enhancing the scientific rigor and credibility of inferences. In addition, visualization and explanation tools based on domain expert decision logic should be developed, enabling model outputs to be presented in an intuitive manner. Moreover, research into explainability should incorporate cross-modal information fusion strategies, enabling complementary insights from different types of sensor data. This approach enhances trustworthiness and transparency while preserving the model's high performance and robustness in real-world industrial applications.

\subsection{Industrial Agents}
Industrial agents as an autonomous fault monitoring and decision control unit, which exhibit significant potential for intelligent cavitation intensity recognition in complex manufacturing systems. These agents integrated with industrial internet of things platforms, which can manage real-time multi-source data streams, perform distributed diagnostics and coordinate maintenance task. With the maturation of large model technology, its vertical applications have endowed industrial agents with enhanced knowledge comprehension and cross-task reasoning capabilities, enabling more precise feature extraction, complex pattern recognition and natural human-machine interaction in CIR. The industrial agent ensures interoperability across heterogeneous systems, safeguarding data exchange security and privacy during industrial data exchange. Furthermore, the transferable agent architecture adapts to diverse operational conditions. Future research should explore integrating large model prompt engineering with reinforcement learning, adaptive diagnostic strategies and cloud-edge-end collaborative frameworks, which can significantly enhance fault diagnosis performance interpretability and real-time performance in industrial operations.

\section{Conclusions}
\label{sec: conclusions}
In this paper, we systematically review of the development of cavitation intensity recognition (CIR), focusing on the evolution from traditional machine learning methods to physical-informed deep learning frameworks. In general, the development of CIR can be divided into three stages. In the past, CIR was implemented through step of data collection, manual feature extraction and intensity recognition. In this stage, traditional machine learning models were able to accomplish the recognition tasks, which are heavily depended on expert domain knowledge. With the rapid progression deep learning in recent years, end-to-end deep learning models have been extensively applied to automatically learn features from multi-modal cavitation data, significantly enhancing recognition performance and robustness. However, data-driven learning models cannot guarantee physical consistency, rendering them unsuitable for practical engineering applications. To address this limitation, physical-informed diagnostic models integrate domain knowledge into the deep learning process, enhancing the model interpretability and generalization capabilities under complex operating conditions. Finally, we discuss the challenges of CIR including multi-source data standardization, model lightweighting, interpretability and engineering deployment. In addition, we outline a roadmap for future studies. This review aims to systematically summarize the development of CIR technologies, providing valuable references for intelligent cavitation diagnosis in industrial applications.

\section*{Acknowledgements}
\label{sec:acknowledgements}
This research is supported by the China Postdoctoral Science Foundation under Grant No.\ 2025M781519 (Y. S.), the National Natural Science Foundation of China under Grant No.\ 92570117, the CUHK-Shenzhen University Development Fund under Grant Nos.\ UDF01003041 and UDF03003\-041, the Shenzhen Peacock Fund under Grant No.\ 2023TC0\-007 (Y. S, K. Z.), the AI grant at FIAS through SAMSON AG (K. Z.), the BMBF funded KISS consortium (05D23RI1) in the ErUM-Data action plan (K. Z.), SAMSON AG (D. V., A. W.), the Walter GreinerGesellschaft zur F\"orderung der physikalischen Grundla - genforschung e.V. through the Judah M. Eisenberg Lau-reatus Chair at Goethe Universit\"at Frankfurt am Main (H. S.) and the NVIDIA GPU grant through NVIDIA Corporation (K. Z.).

\bibliographystyle{unsrt}  
\bibliography{references}

@article{liu2023review,
  title={A review of pump cavitation fault detection methods based on different signals},
  author={Liu, Xiaohui and Mou, Jiegang and Xu, Xin and Qiu, Zhi and Dong, Buyu},
  journal={Processes},
  volume={11},
  number={7},
  pages={2007},
  year={2023},
  publisher={MDPI}
}

@article{mousmoulis2019review,
  title={A review of experimental detection methods of cavitation in centrifugal pumps and inducers},
  author={Mousmoulis, Georgios and Anagnostopoulos, John and Papantonis, Dimitrios},
  journal={International Journal of Fluid Machinery and Systems},
  volume={12},
  number={1},
  pages={71--88},
  year={2019},
  publisher={Turbomachinery Society of Japan, Korean Society for Fluid Machinery, Chinese~…}
}

@article{2017Cavitation,
  title={Cavitation damage: Theory and measurements – A review},
  author={ Sreedhar, B. K.  and  Albert, S. K.  and  Pandit, A. B. },
  journal={Wear},
  year={2017},
}

@article{kumar2010study,
  title={Study of cavitation in hydro turbines—A review},
  author={Kumar, Pardeep and Saini, RP},
  journal={Renewable and Sustainable Energy Reviews},
  volume={14},
  number={1},
  pages={374--383},
  year={2010},
  publisher={Elsevier}
}

@inproceedings{hasanpour2024pump,
  title={Pump Cavitation Detection with Machine Learning: A Comparative Study of SVM and Deep Learning},
  author={Hasanpour, Mohammad Amin and Engholm, Rasmus and Fafoutis, Xenofon},
  booktitle={2024 IEEE Annual Congress on Artificial Intelligence of Things (AIoT)},
  pages={219--225},
  year={2024},
  organization={IEEE}
}

@inproceedings{dutta2018centrifugal,
  title={Centrifugal pump cavitation detection using machine learning algorithm technique},
  author={Dutta, Nabanita and Umashankar, S and Shankar, VK Arun and Padmanaban, Sanjeevikumar and Leonowicz, Zbigniew and Wheeler, Patrick},
  booktitle={2018 IEEE International Conference on Environment and Electrical Engineering and 2018 IEEE Industrial and Commercial Power Systems Europe (EEEIC/I\&CPS Europe)},
  pages={1--6},
  year={2018},
  organization={IEEE}
}

@phdthesis{mahdavi2014performance,
  title={Performance characterisation and cavitation detection of variably angled V-shaped opening ball valves},
  author={Mahdavi, Rad Ali},
  year={2014}
}

@article{jazi2009detecting,
  title={Detecting cavitation in globe valves by two methods: Characteristic diagrams and acoustic analysis},
  author={Jazi, A Masjedian and Rahimzadeh, H},
  journal={Applied Acoustics},
  volume={70},
  number={11-12},
  pages={1440--1445},
  year={2009},
  publisher={Elsevier}
}

@article{wen2002time,
  title={Time frequency characteristics of the vibroacoustic signal of hydrodynamic cavitation},
  author={Wen, Yumei and Henry, Manus},
  journal={J. Vib. Acoust.},
  volume={124},
  number={4},
  pages={469--475},
  year={2002}
}

@article{de2015monitoring,
  title={Monitoring cavitation regime from pressure and optical sensors: Comparing methods using wavelet decomposition for signal processing},
  author={De Giorgi, Maria Grazia and Ficarella, Antonio and Lay-Ekuakille, Aime},
  journal={IEEE Sensors Journal},
  volume={15},
  number={8},
  pages={4684--4691},
  year={2015},
  publisher={IEEE}
}

@article{tash2009application,
  title={The Application of Laser Velocity Meter in Detecting Incipient Cavitation and Measurement its Intensity, Inside Axial Flow Pumps},
  author={Tash, HA and Sadeghi, M and Shervanitabar, MT and Ettefagh, MM},
  journal={Journal of Applied Sciences},
  volume={9},
  number={7},
  pages={1317--1323},
  year={2009}
}

@article{sakthivel2012automatic,
  title={Automatic rule learning using roughset for fuzzy classifier in fault categorization of mono-block centrifugal pump},
  author={Sakthivel, NR and Sugumaran, V and Nair, Binoy B},
  journal={Applied Soft Computing},
  volume={12},
  number={1},
  pages={196--203},
  year={2012},
  publisher={Elsevier}
}

@article{escaler2006detection,
  title={Detection of cavitation in hydraulic turbines},
  author={Escaler, Xavier and Egusquiza, Eduard and Farhat, Mohamed and Avellan, Francois and Coussirat, Miguel},
  journal={Mechanical systems and signal processing},
  volume={20},
  number={4},
  pages={983--1007},
  year={2006},
  publisher={Elsevier}
}

@inproceedings{escaler2004cavitation,
  title={Cavitation erosion prediction in hydro turbines from onboard vibrations},
  author={Escaler, Xavier and Egusquiza, Eduard and Farhat, Mohamed and Avellan, Fran{\c{c}}ois},
  booktitle={Proceedings of the 22nd IAHR Symposium on Hydraulic Machinery and Systems, Stockholm, Sweden},
  volume={1},
  pages={1--10},
  year={2004},
  organization={International Association For Hydraulic Research}
}

@article{jamalpassive,
  title={PASSIVE SONAR DETECTION AND CLASSIFICATION BASED ON DEMON-LOFAR ANALYSIS AND NEURAL NETWORK ALGORITHMS},
  author={Jamal, Said and Lakziz, Jawad and Benremdane, Yahya and Ouaskit, Said}
}

@inproceedings{grenie1990acoustic,
  title={Acoustic detection of propeller cavitation},
  author={Greni{\'e}, Michel},
  booktitle={International Conference on Acoustics, Speech, and Signal Processing},
  pages={2907--2910},
  year={1990},
  organization={IEEE}
}

@inproceedings{escaler2002field,
  title={Field assessment of cavitation detection methods in hydropower plants},
  author={Escaler, Xavier and Egusquiza, Eduard and Mebarki, Toufik and Avellan, Fran{\c{c}}ois and Farhat, Mohamed},
  booktitle={Proceedings of the 21st IAHR Symposium on Hydraulic Machinery and Systems, Lausanne, Switzerland},
  volume={1},
  pages={483--490},
  year={2002},
  organization={International Association For Hydraulic Research}
}

@article{kumar2021identification,
  title={Identification of inlet pipe blockage level in centrifugal pump over a range of speeds by deep learning algorithm using multi-source data},
  author={Kumar, Dhiraj and Dewangan, Aakash and Tiwari, Rajiv and Bordoloi, DJ},
  journal={Measurement},
  volume={186},
  pages={110146},
  year={2021},
  publisher={Elsevier}
}

@article{sha2022acoustic,
  title={An acoustic signal cavitation detection framework based on XGBoost with adaptive selection feature engineering},
  author={Sha, Yu and Faber, Johannes and Gou, Shuiping and Liu, Bo and Li, Wei and Schramm, Stefan and Stoecker, Horst and Steckenreiter, Thomas and Vnucec, Domagoj and Wetzstein, Nadine and others},
  journal={Measurement},
  volume={192},
  pages={110897},
  year={2022},
  publisher={Elsevier}
}

@inproceedings{zhang2015applicability,
  title={Applicability of Principle Components Analysis (PCA) to evaluate the dynamic characteristics of cavitation from captured images},
  author={Zhang, Shengzhuo and Xu, Hongguang and Li, Songjing},
  booktitle={2015 International Conference on Fluid Power and Mechatronics (FPM)},
  pages={307--311},
  year={2015},
  organization={IEEE}
}

@article{mahmoodi2009determination,
  title={Determination of Loss of Coolant Accident (LOCA) in Nuclear Power Plants Using Signal Processing Method},
  author={Mahmoodi, R and Shahriari, M and Zolfaghari, A},
  journal={Engineering Letters},
  volume={17},
  number={4},
  pages={305},
  year={2009}
}

@article{feng2024cavitation,
  title={Cavitation identification in a hydraulic bulb turbine based on vibration and pressure fluctuation measurements},
  author={Feng, Jianjun and Zhao, Nannan and Zhu, Guojun and Wu, Guangkuan and Li, Yunzhe and Luo, Xingqi},
  journal={Mechanical Systems and Signal Processing},
  volume={208},
  pages={111042},
  year={2024},
  publisher={Elsevier}
}

@article{kirschner2023cavitation,
  title={Cavitation detection in hydraulic machinery by analyzing acoustic emissions under strong domain shifts using neural networks},
  author={Kirschner, O and Riedelbauch, S and others},
  journal={Physics of Fluids},
  volume={35},
  number={2},
  year={2023},
  publisher={AIP Publishing}
}

@inproceedings{pereira2004experimental,
  title={Experimental investigation of a cavitating propeller in non-uniform inflow},
  author={Pereira, Francisco and Salvatore, Francesco and Di Felice, Fabio and Soave, Massimo},
  booktitle={Proceedings of the 25th Symposium on Naval Hydrodynamics, St. John’s, Canada},
  year={2004}
}

@article{tsai2021multi,
  title={Multi-sensor fault diagnosis of underwater thruster propeller based on deep learning},
  author={Tsai, Chia-Ming and Wang, Chiao-Sheng and Chung, Yu-Jen and Sun, Yung-Da and Perng, Jau-Woei},
  journal={Sensors},
  volume={21},
  number={21},
  pages={7187},
  year={2021},
  publisher={MDPI}
}

@article{samanipour2017cavitation,
  title={Cavitation detection in centrifugal pumps using pressure time-domain features},
  author={Samanipour, Pouya and Poshtan, Javad and Sadeghi, Hamed},
  journal={Turkish Journal of Electrical Engineering and Computer Sciences},
  volume={25},
  number={5},
  pages={4287--4298},
  year={2017}
}

@article{al2020detection,
  title={Detection of cavitation phenomenon within a centrifugal pump based on vibration analysis technique in both time and frequency domains},
  author={Al-Obaidi, Ahmed Ramadhan},
  journal={Experimental Techniques},
  volume={44},
  number={3},
  pages={329--347},
  year={2020},
  publisher={Springer}
}

@article{zhou2024multi,
  title={Multi-cavitation states identification of a sewage pump using CEEMDAN and BOA-SVM},
  author={Zhou, Peijian and Zeng, Weitao and Zhang, Wenwu and Zhou, Chengui and Yao, Zhifeng},
  journal={Journal of Water Process Engineering},
  volume={61},
  pages={105299},
  year={2024},
  publisher={Elsevier}
}

@article{shervani2018cavitation,
  title={Cavitation intensity monitoring in an axial flow pump based on vibration signals using multi-class support vector machine},
  author={Shervani-Tabar, Mohammad Taghi and Ettefagh, Mir Mohammad and Lotfan, Saeed and Safarzadeh, Hamed},
  journal={Proceedings of the Institution of Mechanical Engineers, Part C: Journal of Mechanical Engineering Science},
  volume={232},
  number={17},
  pages={3013--3026},
  year={2018},
  publisher={SAGE Publications Sage UK: London, England}
}

@article{panda2018prediction,
  title={Prediction of flow blockages and impending cavitation in centrifugal pumps using Support Vector Machine (SVM) algorithms based on vibration measurements},
  author={Panda, Asish Kumar and Rapur, Janani Shruti and Tiwari, Rajiv},
  journal={Measurement},
  volume={130},
  pages={44--56},
  year={2018},
  publisher={Elsevier}
}

@article{liu2024cavitation,
  title={Cavitation diagnosis method of centrifugal pump based on characteristic frequency and kurtosis},
  author={Liu, Yan and Wu, Denghao and Fei, Minghao and Deng, Jiaqi and Li, Qi and Wu, Zhenxing and Gu, Yunqing and Mou, Jiegang},
  journal={AIP Advances},
  volume={14},
  number={2},
  year={2024},
  publisher={AIP Publishing}
}

@article{li2024intelligent,
  title={Intelligent cavitation recognition of a canned motor pump based on a CEEMDAN-KPCA and PSO-SVM method},
  author={Li, Xiaojun and Yang, Haitao and Ge, Jie and Zhu, Shuai and Zhu, Zuchao},
  journal={IEEE Sensors Journal},
  volume={24},
  number={4},
  pages={5324--5334},
  year={2024},
  publisher={IEEE}
}

@article{kang2017analysis,
  title={Analysis of the incipient cavitation noise signal characteristics of hydroturbine},
  author={Kang, Ziyang and Feng, Chi and Liu, Zhiliang and Cang, Yan and Gao, Shan},
  journal={Applied Acoustics},
  volume={127},
  pages={118--125},
  year={2017},
  publisher={Elsevier}
}

@article{dong2019cavitation,
  title={Cavitation detection in centrifugal pump based on interior flow-borne noise using WPD-PCA-RBF},
  author={Dong, Liang and Wu, Kan and Zhu, Jian-cheng and Dai, Cui and Zhang, Li-xin and Guo, Jin-nan},
  journal={Shock and Vibration},
  volume={2019},
  number={1},
  pages={8768043},
  year={2019},
  publisher={Wiley Online Library}
}

@article{li2017feature,
  title={Feature selection: A data perspective},
  author={Li, Jundong and Cheng, Kewei and Wang, Suhang and Morstatter, Fred and Trevino, Robert P and Tang, Jiliang and Liu, Huan},
  journal={ACM computing surveys (CSUR)},
  volume={50},
  number={6},
  pages={1--45},
  year={2017},
  publisher={ACM New York, NY, USA}
}

@inproceedings{jovic2015review,
  title={A review of feature selection methods with applications},
  author={Jovi{\'c}, Alan and Brki{\'c}, Karla and Bogunovi{\'c}, Nikola},
  booktitle={2015 38th international convention on information and communication technology, electronics and microelectronics (MIPRO)},
  pages={1200--1205},
  year={2015},
  organization={Ieee}
}

@article{porkodi2014comparison,
  title={Comparison of filter based feature selection algorithms: an overview},
  author={Porkodi, R},
  journal={International journal of Innovative Research in Technology \& Science},
  volume={2},
  number={2},
  pages={108--113},
  year={2014}
}

@inproceedings{wang2010comparative,
  title={A comparative study of threshold-based feature selection techniques},
  author={Wang, Huanjing and Khoshgoftaar, Taghi M and Van Hulse, Jason},
  booktitle={2010 IEEE International Conference on Granular Computing},
  pages={499--504},
  year={2010},
  organization={IEEE}
}

@phdthesis{hall1999correlation,
  title={Correlation-based feature selection for machine learning},
  author={Hall, Mark A},
  year={1999},
  school={The University of Waikato}
}

@article{vergara2014review,
  title={A review of feature selection methods based on mutual information},
  author={Vergara, Jorge R and Est{\'e}vez, Pablo A},
  journal={Neural computing and applications},
  volume={24},
  pages={175--186},
  year={2014},
  publisher={Springer}
}

@article{patel2020euclidean,
  title={Euclidean distance based feature ranking and subset selection for bearing fault diagnosis},
  author={Patel, Sachin P and Upadhyay, Sanjay H},
  journal={Expert Systems with Applications},
  volume={154},
  pages={113400},
  year={2020},
  publisher={Elsevier}
}

@article{karegowda2010comparative,
  title={Comparative study of attribute selection using gain ratio and correlation based feature selection},
  author={Karegowda, Asha Gowda and Manjunath, AS and Jayaram, MA},
  journal={International Journal of Information Technology and Knowledge Management},
  volume={2},
  number={2},
  pages={271--277},
  year={2010}
}

@article{peng2005feature,
  title={Feature selection based on mutual information criteria of max-dependency, max-relevance, and min-redundancy},
  author={Peng, Hanchuan and Long, Fuhui and Ding, Chris},
  journal={IEEE Transactions on pattern analysis and machine intelligence},
  volume={27},
  number={8},
  pages={1226--1238},
  year={2005},
  publisher={IEEE}
}

@incollection{rudnicki2014all,
  title={All relevant feature selection methods and applications},
  author={Rudnicki, Witold R and Wrzesie{\'n}, Mariusz and Paja, Wies{\l}aw},
  booktitle={Feature selection for data and pattern recognition},
  pages={11--28},
  year={2014},
  publisher={Springer}
}

@article{mochammad2021stable,
  title={Stable hybrid feature selection method for compressor fault diagnosis},
  author={Mochammad, Solichin and Kang, Young-Jin and Noh, Yoojeong and Park, Sunhwa and Ahn, Byeongha},
  journal={IEEE Access},
  volume={9},
  pages={97415--97429},
  year={2021},
  publisher={IEEE}
}

@inproceedings{bermejo2009incremental,
  title={Incremental wrapper-based subset selection with replacement: An advantageous alternative to sequential forward selection},
  author={Bermejo, Pablo and Gamez, Jose A and Puerta, Jose M},
  booktitle={2009 IEEE symposium on computational intelligence and data mining},
  pages={367--374},
  year={2009},
  organization={IEEE}
}

@article{jeon2020hybrid,
  title={Hybrid-recursive feature elimination for efficient feature selection},
  author={Jeon, Hyelynn and Oh, Sejong},
  journal={Applied Sciences},
  volume={10},
  number={9},
  pages={3211},
  year={2020},
  publisher={MDPI}
}

@article{sadeghian2025review,
  title={A review of feature selection methods based on meta-heuristic algorithms},
  author={Sadeghian, Zohre and Akbari, Ebrahim and Nematzadeh, Hossein and Motameni, Homayun},
  journal={Journal of Experimental \& Theoretical Artificial Intelligence},
  volume={37},
  number={1},
  pages={1--51},
  year={2025},
  publisher={Taylor \& Francis}
}

@inproceedings{siham2021feature,
  title={Feature selection based on machine learning for credit scoring: An evaluation of filter and embedded methods},
  author={Siham, Akil and Sara, Sekkate and Abdellah, Adib},
  booktitle={2021 International conference on innovations in intelligent systems and applications (INISTA)},
  pages={1--6},
  year={2021},
  organization={IEEE}
}

@article{moslemi2023tutorial,
  title={A tutorial-based survey on feature selection: Recent advancements on feature selection},
  author={Moslemi, Amir},
  journal={Engineering Applications of Artificial Intelligence},
  volume={126},
  pages={107136},
  year={2023},
  publisher={Elsevier}
}

@article{manikandan2021feature,
  title={Feature selection is important: state-of-the-art methods and application domains of feature selection on high-dimensional data},
  author={Manikandan, Gopi and Abirami, S},
  journal={Applications in Ubiquitous Computing},
  pages={177--196},
  year={2021},
  publisher={Springer}
}

@inproceedings{masaeli2010transformation,
  title={From transformation-based dimensionality reduction to feature selection},
  author={Masaeli, Mahdokht and Dy, Jennifer G and Fung, Glenn M},
  booktitle={Proceedings of the 27th international conference on machine learning (ICML-10)},
  pages={751--758},
  year={2010}
}

@article{sumithra2015review,
  title={A review of various linear and non linear dimensionality reduction techniques},
  author={Sumithra, V and Surendran, Subu},
  journal={Int J Comput Sci Inf Technol},
  volume={6},
  number={3},
  pages={2354--60},
  year={2015},
  publisher={Citeseer}
}

@techreport{fodor2002survey,
  title={A survey of dimension reduction techniques},
  author={Fodor, Imola K},
  year={2002},
  institution={Lawrence Livermore National Lab.(LLNL), Livermore, CA (United States)}
}

@article{han2024off,
  title={Off-design operation and cavitation detection in centrifugal pumps using vibration and motor stator current analyses},
  author={Han, Yuejiang and Zou, Jiamin and Presas, Alexandre and Luo, Yin and Yuan, Jianping},
  journal={Sensors},
  volume={24},
  number={11},
  pages={3410},
  year={2024},
  publisher={MDPI}
}

@article{he2022intelligent,
  title={Intelligent Identification of Cavitation State of Centrifugal Pump Based on Support Vector Machine},
  author={He, Xiaoke and Song, Yu and Wu, Kaipeng and Ali, Asad and Shen, Chunhao and Si, Qiaorui},
  journal={Energies},
  volume={15},
  number={23},
  pages={8907},
  year={2022},
  publisher={MDPI}
}

@inproceedings{chen2022csa,
  title={CSA-SVM method for internal cavitation defects detection and its application of district heating pipes},
  author={Chen, Yanran and Ma, Shugen and Li, Longchuan and Li, Zhiqing and Yang, Yulin},
  booktitle={2022 IEEE/RSJ International Conference on Intelligent Robots and Systems (IROS)},
  pages={6242--6249},
  year={2022},
  organization={IEEE}
}

@article{gregg2018method,
  title={A Method for Automated Cavitation Detection with Adaptive Thresholds},
  author={Gregg, Seth W and Steele, John PH and Van Bossuyt, Douglas L},
  year={2018},
  publisher={Wiley}
}

@inproceedings{dutta2020comparative,
  title={Comparative study of cavitation problem detection in pumping system using SVM and K-nearest neighbour method},
  author={Dutta, Nabanita and Subramaniam, Umashankar and Sanjeevikumar, P and Bharadwaj, Sai Charan and Leonowicz, Zbigniew and Holm-Nielsen, Jens Bo},
  booktitle={2020 IEEE International Conference on Environment and Electrical Engineering and 2020 IEEE Industrial and Commercial Power Systems Europe (EEEIC/I\&CPS Europe)},
  pages={1--6},
  year={2020},
  organization={IEEE}
}

@article{rapur2019multifault,
  title={Multifault diagnosis of combined hydraulic and mechanical centrifugal pump faults using continuous wavelet transform and support vector machines},
  author={Rapur, Janani Shruti and Tiwari, Rajiv},
  journal={Journal of Dynamic Systems, Measurement, and Control},
  volume={141},
  number={11},
  pages={111013},
  year={2019},
  publisher={American Society of Mechanical Engineers}
}

@article{de2019intelligent,
  title={Intelligent incipient fault detection in wind turbines based on industrial IoT environment},
  author={De Sousa, Pedro H Feijo and Navar de Medeiros, M and Almeida, Jefferson S and Reboucas Filho, Pedro P and de Albuquerque, Victor Hugo C and others},
  journal={Journal of Artificial Intelligence and Systems},
  volume={1},
  number={1},
  pages={1--19},
  year={2019},
  publisher={Institute of Electronics and Computer}
}

@inproceedings{casoli2019multi,
  title={A multi-fault diagnostic method based on acceleration signal for a hydraulic axial piston pump},
  author={Casoli, Paolo and Pastori, Mirko and Scolari, Fabio},
  booktitle={AIP Conference Proceedings},
  volume={2191},
  number={1},
  pages={020037},
  year={2019},
  organization={AIP Publishing LLC}
}

@article{qiu2019experimental,
  title={Experimental investigation and multi-conditions identification method of centrifugal pump using Fisher discriminant ratio and support vector machine},
  author={Qiu, Guangqi and Huang, Si and Gu, Yingkui},
  journal={Advances in Mechanical Engineering},
  volume={11},
  number={9},
  pages={1687814019878041},
  year={2019},
  publisher={SAGE Publications Sage UK: London, England}
}

@article{de2015cavitation,
  title={Cavitation regime detection by LS-SVM and ANN with wavelet decomposition based on pressure sensor signals},
  author={De Giorgi, Maria Grazia and Ficarella, Antonio and Lay-Ekuakille, Aime},
  journal={IEEE Sensors Journal},
  volume={15},
  number={10},
  pages={5701--5708},
  year={2015},
  publisher={IEEE}
}

@inproceedings{liu2014underwater,
  title={Underwater target recognition based on line spectrum and support vector machine},
  author={Liu, Jian and He, Yang and Liu, Zhong and Xiong, Ying},
  booktitle={2014 International Conference on Mechatronics, Control and Electronic Engineering (MCE-14)},
  pages={79--84},
  year={2014},
  organization={Atlantis Press}
}

@article{jeong2012surface,
  title={Surface ship-wake detection using active sonar and one-class support vector machine},
  author={Jeong, Sungmoon and Ban, Sang-Woo and Choi, Sangmoon and Lee, Donghun and Lee, Minho},
  journal={IEEE Journal of Oceanic Engineering},
  volume={37},
  number={3},
  pages={456--466},
  year={2012},
  publisher={IEEE}
}

@inproceedings{tingfeng2011ls,
  title={LS-SVC based recognition method of the centrifugal pump cavitation intensity},
  author={Tingfeng, Ming and Yongsheng, Su},
  booktitle={2011 International Conference on Electric Information and Control Engineering},
  pages={3335--3338},
  year={2011},
  organization={IEEE}
}

@article{qian2025diffusion,
  title={A Diffusion-TGAN Framework for Spatio-Temporal Speed Imputation and Trajectory Reconstruction},
  author={Qian, Yu and Li, Xunhao and Zhang, Jian and Meng, Xiaolin and Li, Yongfu and Ding, Heng and Wang, Maoze},
  journal={IEEE Transactions on Intelligent Transportation Systems},
  year={2025},
  publisher={IEEE}
}

@article{kang2020stacked,
  title={Stacked sparse autoencoder in cavitation noise signal data classification of hydro turbine based on power spectrum},
  author={Kang, Ziyang and Feng, Chi and Wan, Xiaoqing and Liu, Zhiliang and Chen, Liwei},
  journal={Journal of Low Frequency Noise, Vibration and Active Control},
  volume={39},
  number={2},
  pages={233--245},
  year={2020},
  publisher={SAGE Publications Sage UK: London, England}
}

@article{gruberdetection,
  title={Detection of cavitating states (swirls) in a Francis test pump-turbine using ultrasonic and transient pressure measurements},
  author={Gruber, P and Agner, R and Deniz, S},
  journal={Proceedings of the 2018 12th International Group for Hydraulic Efficiency Measurements (IGHEM), Beijing, China},
  pages={10--13},
  year={2018}
}

@article{zhang2022anti,
  title={Anti-cavitation leading-edge profile design of centrifugal pump impeller blade based on genetic algorithm and decision tree},
  author={Zhang, Fangfang and Tao, Ran and Zhu, Di and Wu, Yanzhao and Jin, Faye and Xiao, Ruofu},
  journal={Journal of the Brazilian Society of Mechanical Sciences and Engineering},
  volume={44},
  number={7},
  pages={279},
  year={2022},
  publisher={Springer}
}

@inproceedings{gruber2014detection,
  title={The detection of cavitation in hydraulic machines by use of ultrasonic signal analysis},
  author={Gruber, P and Odermatt, P and Etterlin, M and Lerch, Th and Frei, M and Farhat, M},
  booktitle={IOP Conference Series: Earth and Environmental Science},
  volume={22},
  number={5},
  pages={052005},
  year={2014},
  organization={IOP Publishing}
}

@article{gatica2025classifying,
  title={Classifying acoustic cavitation with machine learning trained on multiple physical models},
  author={Gatica, Trinidad and Haqshenas, Reza and others},
  journal={Physics of Fluids},
  volume={37},
  number={3},
  year={2025},
  publisher={AIP Publishing}
}

@article{kamali2025characterization,
  title={Characterization of ventilated supercavitation regimes using Bayesian optimized Random Forest models},
  author={Kamali, HosseinAli and Erfanian, Mohammad-Reza},
  journal={Physics of Fluids},
  volume={37},
  number={1},
  year={2025},
  publisher={AIP Publishing}
}

@article{liu2025machine,
  title={Machine learning-based cavitation stage identification in an axisymmetric Venturi under bubbly shock dominance},
  author={Liu, Teng and You, Weibin and Manickam, Sivakumar and Wang, Wenlong and Wang, Benlong and Sun, Xun},
  journal={Physics of Fluids},
  volume={37},
  number={6},
  year={2025},
  publisher={AIP Publishing}
}

@article{orhan2024machine,
  title={Machine learning-based prediction of NPSH, noise, and vibration levels in radial pumps under cavitation conditions},
  author={Orhan, Nuri and Kurt, Mehmet and K{\i}r{\i}lmaz, Hasan and Ertu{\u{g}}rul, Murat},
  journal={Tekirda{\u{g}} Ziraat Fak{\"u}ltesi Dergisi},
  volume={21},
  number={2},
  pages={533--546},
  year={2024},
  publisher={Tekirdag Namik Kemal University}
}

@inproceedings{ehemann2023ai,
  title={AI-based Cavitation Detection in Process Valves},
  author={Ehemann, Marisa and Tr{\"a}nkle, Frank and Stache, Nicolaj C},
  booktitle={2023 IEEE 21st International Conference on Industrial Informatics (INDIN)},
  pages={1--6},
  year={2023},
  organization={IEEE}
}

@inproceedings{kang2022analysis,
  title={Analysis of optimal parameters for discriminating cavitation types by SSAE-RF},
  author={Kang, Ziyang and Liu, Zhiliang},
  booktitle={Journal of Physics: Conference Series},
  volume={2242},
  number={1},
  pages={012025},
  year={2022},
  organization={IOP Publishing}
}

@article{zhang2025data,
  title={Data-driven identification of cavitation regimes using acoustic signatures of hydrofoil},
  author={Zhang, Qi and Lin, Yuxing and el Moctar, Ould and Jiang, Changqing},
  journal={Physics of Fluids},
  volume={37},
  number={4},
  year={2025},
  publisher={AIP Publishing}
}

@article{karagiovanidis2023early,
  title={Early detection of cavitation in centrifugal pumps using low-cost vibration and sound sensors},
  author={Karagiovanidis, Marios and Pantazi, Xanthoula Eirini and Papamichail, Dimitrios and Fragos, Vassilios},
  journal={Agriculture},
  volume={13},
  number={8},
  pages={1544},
  year={2023},
  publisher={MDPI}
}

@article{sakthivel2010vibration,
  title={Vibration based fault diagnosis of monoblock centrifugal pump using decision tree},
  author={Sakthivel, NR and Sugumaran, V and Babudevasenapati, SJESwA},
  journal={Expert Systems with Applications},
  volume={37},
  number={6},
  pages={4040--4049},
  year={2010},
  publisher={Elsevier}
}

@article{xu2022leakage,
  title={Leakage identification in water pipes using explainable ensemble tree model of vibration signals},
  author={Xu, Weinan and Fan, Shidong and Wang, Chunping and Wu, Jie and Yao, Yunan and Wu, JunChen},
  journal={Measurement},
  volume={194},
  pages={110996},
  year={2022},
  publisher={Elsevier}
}

@article{muralidharan2013feature,
  title={Feature extraction using wavelets and classification through decision tree algorithm for fault diagnosis of mono-block centrifugal pump},
  author={Muralidharan, V and Sugumaran, V},
  journal={Measurement},
  volume={46},
  number={1},
  pages={353--359},
  year={2013},
  publisher={Elsevier}
}

@article{farokhzad2013fault,
  title={Fault classification of centrifugal water pump based on decision tree and regression model},
  author={Farokhzad, Saeid and Ahmadi, Hojjat and Jafary, Ali},
  journal={J. Sci. today’s world},
  volume={2},
  number={2},
  pages={170--176},
  year={2013}
}

@article{yakupov2023application,
  title={Application of machine learning to predict the acoustic cavitation threshold of fluids},
  author={Yakupov, Bulat and Smirnov, Ivan},
  journal={Fluids},
  volume={8},
  number={6},
  pages={168},
  year={2023},
  publisher={MDPI}
}

@article{stephen2025evaluation,
  title={Evaluation of supervised machine learning techniques for cavitation detection and diagnosis in a pump-as-turbine system},
  author={Stephen, Calvin and Basu, Biswajit and McNabola, Aonghus},
  journal={Expert Systems with Applications},
  pages={129167},
  year={2025},
  publisher={Elsevier}
}

@article{hu2025cavitation,
  title={Cavitation Identification Based on Feature Extraction},
  author={Hu, Siyuan and Dong, Liang and Li, Honggang and Hua, Runan and Liu, Houlin and Dai, Cui and others},
  journal={Nuclear Engineering and Technology},
  pages={103745},
  year={2025},
  publisher={Elsevier}
}

@article{bagherzadeh2025prediction,
  title={Prediction of cavitation damage using SVM model based on air--water two-phase flow over dam spillway},
  author={Bagherzadeh, Saghi and Ghaeini-Hessaroeyeh, Mahnaz and Fadaei-Kermani, Ehsan},
  journal={Applied Water Science},
  volume={15},
  number={4},
  pages={70},
  year={2025},
  publisher={Springer}
}

@article{ranawat2025blockage,
  title={Blockage detection in centrifugal pump using semi-supervised machine learning based on SVM and LSTM},
  author={Ranawat, Nagendra Singh and Miglani, Ankur and Kankar, Pavan Kumar},
  journal={Measurement Science and Technology},
  volume={36},
  number={3},
  pages={036215},
  year={2025},
  publisher={IOP Publishing}
}

@article{dias2025edge,
  title={Edge-based intelligent fault diagnosis for centrifugal pumps in microbreweries},
  author={Dias, Andre Luis and Buzoli, Marcio Rafael and da Silva, Vinicius Rodrigues and da Silva, Jean Carlos Rodrigues and Turcato, Afonso Celso and Sestito, Guilherme Serpa},
  journal={Flow Measurement and Instrumentation},
  volume={101},
  pages={102730},
  year={2025},
  publisher={Elsevier}
}

@article{hong2025improving,
  title={Improving Pipeline Leak Detection Accuracy by Constructing a Multistate Leakage Sounds Training Set},
  author={Hong, Zhou and Zhao, Dan and Dong, Liqiang and Liu, Shaogang and Qiu, Feng and Jin, Yanji},
  journal={Journal of Pipeline Systems Engineering and Practice},
  volume={16},
  number={4},
  pages={04025056},
  year={2025},
  publisher={American Society of Civil Engineers}
}

@article{dias2024soft,
  title={A soft sensor edge-based approach to fault diagnosis for piping systems},
  author={Dias, Andre Luis and Turcato, Afonso Celso and Sestito, Guilherme Serpa},
  journal={Flow Measurement and Instrumentation},
  volume={97},
  pages={102618},
  year={2024},
  publisher={Elsevier}
}

@article{araste2023fault,
  title={Fault diagnosis of a centrifugal pump using electrical signature analysis and support vector machine},
  author={Araste, Zahra and Sadighi, Ali and Jamimoghaddam, Mohammad},
  journal={Journal of Vibration Engineering \& Technologies},
  volume={11},
  number={5},
  pages={2057--2067},
  year={2023},
  publisher={Springer}
}

@inproceedings{huang2023condition,
  title={Condition Monitoring Of Centrifugal Pump In Nuclear Power Plant Based On Improved Vmd And Svm},
  author={Huang, Xueying and Xia, Hong and Liu, Yongkuo and Yin, Wenzhe and Ran, Wenhao},
  booktitle={The Proceedings of the International Conference on Nuclear Engineering (ICONE) 2023.30},
  pages={1747},
  year={2023},
  organization={The Japan Society of Mechanical Engineers}
}

@article{xiao2022research,
  title={Research on common fault diagnosis and classification method of centrifugal pump based on reliefF and SVM},
  author={Xiao, Xingxin and Chen, Hui and Dong, Liang and Liu, Houlin and Fan, Chuanhan},
  journal={International Journal of Fluid Machinery and Systems},
  volume={15},
  number={2},
  pages={287--296},
  year={2022}
}

@article{al4615504incipient,
  title={Incipient Fault Diagnosis of Centrifugal Pumps Using Support Vector Machine with Continuous Wavelet Transform},
  author={Al-Toubi, Soud and Alkali, Babakalli and Kumar, Amit and CV, Sudhir and Al-Tobi, Maamar and Manappallil, Varghese},
  journal={Available at SSRN 4615504}
}

@incollection{kamiel2022cavitation,
  title={Cavitation detection of centrifugal pumps using SVM and statistical features},
  author={Kamiel, Berli and Akbar, Taufiq and Sudarisman and Krisdiyanto},
  booktitle={Recent Advances in Mechanical Engineering: Select Proceedings of ICOME 2021},
  pages={1--9},
  year={2022},
  publisher={Springer}
}

@article{bordoloi2017identification,
  title={Identification of suction flow blockages and casing cavitations in centrifugal pumps by optimal support vector machine techniques},
  author={Bordoloi, DJ and Tiwari, Rajiv},
  journal={Journal of the Brazilian Society of Mechanical Sciences and Engineering},
  volume={39},
  number={8},
  pages={2957--2968},
  year={2017},
  publisher={Springer}
}

@article{dutta2023svm,
  title={SVM Algorithm for Vibration Fault Diagnosis in Centrifugal Pump.},
  author={Dutta, Nabanita and Kaliannan, Palanisamy and Shanmugam, Paramasivam},
  journal={Intelligent Automation \& Soft Computing},
  volume={35},
  number={3},
  year={2023}
}

@article{gypa2023propeller,
  title={Propeller optimization by interactive genetic algorithms and machine learning},
  author={Gypa, Ioli and Jansson, Marcus and Wolff, Krister and Bensow, Rickard},
  journal={Ship Technology Research},
  volume={70},
  number={1},
  pages={56--71},
  year={2023},
  publisher={Taylor \& Francis}
}

@inproceedings{araste2020support,
  title={Support vector machine-based fault diagnosis of a centrifugal pump using electrical signature analysis},
  author={Araste, Zahra and Sadighi, Ali and Moghaddam, Mohammad Jami},
  booktitle={2020 6th Iranian Conference on Signal Processing and Intelligent Systems (ICSPIS)},
  pages={1--5},
  year={2020},
  organization={IEEE}
}

@inproceedings{youlin2017wavelet,
  title={Wavelet analysis-application of PCA-SVM in state identification method of marine centrifugal pump},
  author={Youlin, Xu and Ling, Xiong and Zhigang, Yao and Jingjia, Guo},
  booktitle={2017 13th IEEE International Conference on Electronic Measurement \& Instruments (ICEMI)},
  pages={560--564},
  year={2017},
  organization={IEEE}
}

@article{xue2014intelligent,
  title={Intelligent diagnosis method for centrifugal pump system using vibration signal and support vector machine},
  author={Xue, Hongtao and Li, Zhongxing and Wang, Huaqing and Chen, Peng},
  journal={Shock and Vibration},
  volume={2014},
  number={1},
  pages={407570},
  year={2014},
  publisher={Wiley Online Library}
}

@article{wang2019hydraulic,
  title={Hydraulic system fault diagnosis method based on a multi-feature fusion support vector machine},
  author={Wang, Lihua and Wu, Xiao-qiang and Zhang, Chunyou and Shi, Hongyan},
  journal={The Journal of Engineering},
  volume={2019},
  number={13},
  pages={215--218},
  year={2019},
  publisher={Wiley Online Library}
}

@article{muralidharan2014fault,
  title={Fault diagnosis of monoblock centrifugal pump using SVM},
  author={Muralidharan, V and Sugumaran, V and Indira, V},
  journal={Engineering Science and Technology, an International Journal},
  volume={17},
  number={3},
  pages={152--157},
  year={2014},
  publisher={Elsevier}
}

@article{sakthivel2010application,
  title={Application of support vector machine (SVM) and proximal support vector machine (PSVM) for fault classification of monoblock centrifugal pump},
  author={Sakthivel, NR and Sugumaran, V and Nair, Binoy B},
  journal={International Journal of Data Analysis Techniques and Strategies},
  volume={2},
  number={1},
  pages={38--61},
  year={2010},
  publisher={Inderscience Publishers}
}

@article{shen2022tree,
  title={A tree-based machine learning method for pipeline leakage detection},
  author={Shen, Yongxin and Cheng, Weiping},
  journal={Water},
  volume={14},
  number={18},
  pages={2833},
  year={2022},
  publisher={MDPI}
}

@article{sakthivel2011use,
  title={Use of histogram features for decision tree-based fault diagnosis of monoblock centrifugal pump},
  author={Sakthivel, NR and Indira, V and Nair, Binoy B and Sugumaran, V},
  journal={International Journal of Granular Computing, Rough Sets and Intelligent Systems},
  volume={2},
  number={1},
  pages={23--36},
  year={2011},
  publisher={Inderscience Publishers}
}

@inproceedings{stephen2024prediction,
  title={Prediction of cavitation using machine learning techniques on centrifugal pump},
  author={Stephen, Christopher and Guguloth, Vivek and Sivasailam, Kumaraswamy and Gu, Yandong and Parmar, Richa and Banerjee, Chandan},
  booktitle={Journal of Physics: Conference Series},
  volume={2854},
  number={1},
  pages={012014},
  year={2024},
  organization={IOP Publishing}
}

@article{chang2024random,
  title={Random forest-based multi-faults classification modeling and analysis for intelligent centrifugal pump system},
  author={Chang, Kyuchang and Park, Seung Hwan},
  journal={Journal of Mechanical Science and Technology},
  volume={38},
  number={1},
  pages={11--20},
  year={2024},
  publisher={Springer}
}

@article{manikandan2023vibration,
  title={VIBRATION BASED PREDICTIVE FAULT ANALYSIS OF BEARING SEAL FAILURE AND CAVITATION ON INDUSTRIAL MONOBLOCK CENTRIFUGAL PUMP USING DEEP LEARNING ALGORITHM},
  author={Manikandan, S and Duraivelu, K},
  journal={Jurnal Teknologi (Sciences \& Engineering)},
  volume={85},
  number={5},
  pages={151--162},
  year={2023}
}

@inproceedings{amihai2018industrial,
  title={An industrial case study using vibration data and machine learning to predict asset health},
  author={Amihai, Ido and Gitzel, Ralf and Kotriwala, Arzam Muzaffar and Pareschi, Diego and Subbiah, Subanataranjan and Sosale, Guruprasad},
  booktitle={2018 IEEE 20th Conference on Business Informatics (CBI)},
  volume={1},
  pages={178--185},
  year={2018},
  organization={IEEE}
}

@article{saari2018selection,
  title={Selection of features for fault diagnosis on rotating machines using random forest and wavelet analysis},
  author={Saari, Juhamatti and Lundberg, Jan and Odelius, Johan and Rantatalo, Matti},
  journal={Insight-Non-Destructive Testing and Condition Monitoring},
  volume={60},
  number={8},
  pages={434--442},
  year={2018},
  publisher={The British Institute of Non-Destructive Testing}
}

@article{breiman2001random,
  title={Random forests},
  author={Breiman, Leo},
  journal={Machine learning},
  volume={45},
  number={1},
  pages={5--32},
  year={2001},
  publisher={Springer}
}

@article{friedman2001greedy,
  title={Greedy function approximation: a gradient boosting machine},
  author={Friedman, Jerome H},
  journal={Annals of statistics},
  pages={1189--1232},
  year={2001},
  publisher={JSTOR}
}

@inproceedings{chen2016xgboost,
  title={Xgboost: A scalable tree boosting system},
  author={Chen, Tianqi and Guestrin, Carlos},
  booktitle={Proceedings of the 22nd acm sigkdd international conference on knowledge discovery and data mining},
  pages={785--794},
  year={2016}
}

@article{ke2017lightgbm,
  title={Lightgbm: A highly efficient gradient boosting decision tree},
  author={Ke, Guolin and Meng, Qi and Finley, Thomas and Wang, Taifeng and Chen, Wei and Ma, Weidong and Ye, Qiwei and Liu, Tie-Yan},
  journal={Advances in neural information processing systems},
  volume={30},
  year={2017}
}

@article{geurts2006extremely,
  title={Extremely randomized trees},
  author={Geurts, Pierre and Ernst, Damien and Wehenkel, Louis},
  journal={Machine learning},
  volume={63},
  number={1},
  pages={3--42},
  year={2006},
  publisher={Springer}
}

@misc{orr1996introduction,
  title={Introduction to radial basis function networks},
  author={Orr, Mark JL and others},
  year={1996},
  publisher={Technical Report, center for cognitive science, University of Edinburgh~…}
}

@book{kohonen2012self,
  title={Self-organizing maps},
  author={Kohonen, Teuvo},
  volume={30},
  year={2012},
  publisher={Springer Science \& Business Media}
}

@article{hopfield1982neural,
  title={Neural networks and physical systems with emergent collective computational abilities.},
  author={Hopfield, John J},
  journal={Proceedings of the national academy of sciences},
  volume={79},
  number={8},
  pages={2554--2558},
  year={1982}
}

@book{haykin2009neural,
  title={Neural networks and learning machines, 3/E},
  author={Haykin, Simon},
  year={2009},
  publisher={Pearson Education India}
}

@inproceedings{lavretsky2002health,
  title={Health monitoring of an electro-hydraulic system using ordered neural networks},
  author={Lavretsky, Eugene and Chidambaram, Bala},
  booktitle={Proceedings of the 2002 International Joint Conference on Neural Networks. IJCNN'02 (Cat. No. 02CH37290)},
  volume={3},
  pages={2893--2898},
  year={2002},
  organization={IEEE}
}

@article{hovcevar2005prediction,
  title={Prediction of cavitation vortex dynamics in the draft tube of a francis turbine using radial basis neural networks},
  author={Ho{\v{c}}evar, Marko and {\v{S}}irok, Brane and Blagojevi{\v{c}}, Bogdan},
  journal={Neural Computing \& Applications},
  volume={14},
  number={3},
  pages={229--234},
  year={2005},
  publisher={Springer}
}

@inproceedings{yang2008leak,
  title={Leak acoustic detection in water distribution pipelines},
  author={Yang, Jin and Wen, Yumei and Li, Ping},
  booktitle={2008 7th World Congress on Intelligent Control and Automation},
  pages={3057--3061},
  year={2008},
  organization={IEEE}
}

@inproceedings{bernardini2009expert,
  title={An Expert System Based on Parametric Net to Support Motor Pump Multi-Failure Diagnostic},
  author={Bernardini, Flavia Cristina and Garcia, Ana Cristina Bicharra and Ferraz, Inha{\'u}ma Neves},
  booktitle={IFIP International Conference on Artificial Intelligence Applications and Innovations},
  pages={13--20},
  year={2009},
  organization={Springer}
}

@article{rathinasabapathydevelopment,
  title={Development of neural network model for classification of cavitation signals},
  author={Rathinasabapathy, Ramadevi and Balasubramaniam, Sheela Rani and Vasudevan, Prakash and Perumal, Kalyanasundaram}
}

@article{saeed20133d,
  title={3D fluid--structure modelling and vibration analysis for fault diagnosis of Francis turbine using multiple ANN and multiple ANFIS},
  author={Saeed, RA and Galybin, AN and Popov, V},
  journal={Mechanical Systems and Signal Processing},
  volume={34},
  number={1-2},
  pages={259--276},
  year={2013},
  publisher={Elsevier}
}

@article{kane2016application,
  title={Application of psychoacoustics for gear fault diagnosis using artificial neural network},
  author={Kane, PV and Andhare, AB},
  journal={Journal of Low Frequency Noise, Vibration and Active Control},
  volume={35},
  number={3},
  pages={207--220},
  year={2016},
  publisher={SAGE Publications Sage UK: London, England}
}

@phdthesis{hussain2016condition,
  title={Condition-based monitoring system for diagnostics and prognostics of centrifugal pumps},
  author={Hussain, Bilal Kamal},
  year={2016},
  school={University of Missouri--Columbia}
}

@article{babikir2019noise,
  title={Noise prediction of axial piston pump based on different valve materials using a modified artificial neural network model},
  author={Babikir, Hassan A and Abd Elaziz, Mohamed and Elsheikh, Ammar H and Showaib, Ezzat A and Elhadary, M and Wu, Defa and Liu, Yinshui},
  journal={Alexandria Engineering Journal},
  volume={58},
  number={3},
  pages={1077--1087},
  year={2019},
  publisher={Elsevier}
}

@inproceedings{salah2019automated,
  title={Automated valve fault detection based on acoustic emission parameters and artificial neural network},
  author={Salah, M Ali Al-Obaidi and Hui, KH and Hee, LM and Leong, M Salman and Mahdi, Ali Abdul-Hussain and Abdelrhman, Ahmed M and Ali, YH},
  booktitle={MATEC Web of Conferences},
  volume={255},
  pages={02013},
  year={2019},
  organization={EDP Sciences}
}

@article{gao2019cavitation,
  title={Cavitation damage prediction of stainless steels using an artificial neural network approach},
  author={Gao, Guiyan and Zhang, Zheng and Cai, Cheng and Zhang, Jianglong and Nie, Baohua},
  journal={Metals},
  volume={9},
  number={5},
  pages={506},
  year={2019},
  publisher={MDPI}
}

@article{maradey2020methodology,
  title={A methodology for detection of wear in hydraulic axial piston pumps},
  author={Maradey L{\'a}zaro, Jessica Gissella and Borr{\'a}s Pinilla, Carlos},
  journal={International Journal on Interactive Design and Manufacturing (IJIDeM)},
  volume={14},
  number={3},
  pages={1103--1119},
  year={2020},
  publisher={Springer}
}

@article{sayed2020classification,
  title={Classification model using neural network for centrifugal pump fault detection},
  author={Sayed, Eslam and Abdelsamee, Ahmed A and Ghazaly, Nouby M},
  journal={International Journal},
  volume={13},
  number={4},
  year={2020}
}

@article{adeodu2020adaptive,
  title={An adaptive Industrial Internet of things (IIOts) based technology for prediction and control of cavitation in centrifugal pumps},
  author={Adeodu, Adefemi and Daniyan, Ilesanmi and Omitola, Olusegun and Ejimuda, Chinoyelum and Agbor, Esoso and Akinola, Oluwole},
  journal={Procedia CIRP},
  volume={91},
  pages={927--934},
  year={2020},
  publisher={Elsevier}
}

@article{zhao2020use,
  title={On the use of artificial neural networks for condition monitoring of pump-turbines with extended operation},
  author={Zhao, Weiqiang and Egusquiza, M{\`o}nica and Valero, Carme and Valent{\'\i}n, David and Presas, Alexandre and Egusquiza, Eduard},
  journal={Measurement},
  volume={163},
  pages={107952},
  year={2020},
  publisher={Elsevier}
}

@article{matloobi2021identification,
  title={Identification of cavitation in centrifugal pump by artificial immune network},
  author={Matloobi, Seyed M and Riahi, Mohammad},
  journal={Proceedings of the Institution of Mechanical Engineers, Part E: Journal of Process Mechanical Engineering},
  volume={235},
  number={6},
  pages={2271--2280},
  year={2021},
  publisher={SAGE Publications Sage UK: London, England}
}

@article{souza2024application,
  title={Application of machine learning models in predictive maintenance of Francis hydraulic turbines},
  author={Souza, J{\'u}lio C{\'e}sar Silva de and Honorato J{\'u}nior, Oswaldo and Tiago Filho, Geraldo L{\'u}cio and Carpinteiro, Ot{\'a}vio Augusto Salgado and Biancardine J{\'u}nior, Hailton Silveira Domingues and Santos, Ivan Felipe Silva dos},
  journal={RBRH},
  volume={29},
  pages={e48},
  year={2024},
  publisher={SciELO Brasil}
}

@article{kang2022cavitation,
  title={Cavitation noise signal classification of hydroturbine based on improved multi-scale symbol dynamic entropy},
  author={Kang, Ziyang and Liu, Zhiliang and Guo, Xinnian and Liu, Liu},
  journal={Int. J. Acoust. Vib.},
  volume={27},
  number={4},
  pages={326--333},
  year={2022}
}

@article{lan2022experimental,
  title={Experimental investigation on cavitation and cavitation detection of axial piston pump based on MLP-Mixer},
  author={Lan, Yuan and Li, Zhijie and Liu, Shengzheng and Huang, Jiahai and Niu, Linkai and Xiong, Xiaoyan and Niu, Chenguang and Wu, Bing and Zhou, Xu and Yan, Jinbao and others},
  journal={Measurement},
  volume={200},
  pages={111582},
  year={2022},
  publisher={Elsevier}
}

@article{tong2023cavitation,
  title={Cavitation diagnosis for water distribution pumps: An early-stage approach combing vibration signal-based neural network with high-speed photography},
  author={Tong, Zheming and Liu, Hao and Cao, Xiangkun Elvis and Westerdahld, Dane and Jin, Xiaofeng},
  journal={Sustainable Energy Technologies and Assessments},
  volume={55},
  pages={102919},
  year={2023},
  publisher={Elsevier}
}

@article{matloobi2018identification,
  title={Identification of Cavitation Phenomenon in Centrifugal Pump by Artificial Immune Network Method},
  author={Matloobi, Seyed Mostafa and Riahi, Mohammad and Sadeghi, Hamed},
  journal={Amirkabir Journal of Mechanical Engineering},
  volume={52},
  number={3},
  pages={717--730},
  year={2018},
  publisher={Amirkabir University of Technology}
}

@inproceedings{nasiri2011vibration,
  title={Vibration signature analysis for detecting cavitation in centrifugal pumps using neural networks},
  author={Nasiri, MR and Mahjoob, MJ and Vahid-Alizadeh, H},
  booktitle={2011 IEEE International Conference on Mechatronics},
  pages={632--635},
  year={2011},
  organization={IEEE}
}

@inproceedings{de2012neural,
  title={A Neural Network Approach to Analyse Cavitating Flow Regime in an Internal Orifice},
  author={De Giorgi, Maria Grazia and Bello, Daniela and Ficarella, Antonio},
  booktitle={Engineering Systems Design and Analysis},
  volume={44854},
  pages={51--62},
  year={2012},
  organization={American Society of Mechanical Engineers}
}

@article{siano2018diagnostic,
  title={Diagnostic method by using vibration analysis for pump fault detection},
  author={Siano, D and Panza, MA},
  journal={Energy Procedia},
  volume={148},
  pages={10--17},
  year={2018},
  publisher={Elsevier}
}

@article{wu2023neural,
  title={Neural network-based analysis on the unusual peak of cavitation performance of a mixed flow pipeline pump},
  author={Wu, Yanzhao and Guan, Xin and Tao, Ran and Xiao, Ruofu},
  journal={Iranian Journal of Science and Technology, Transactions of Mechanical Engineering},
  volume={47},
  number={4},
  pages={1515--1533},
  year={2023},
  publisher={Springer}
}

@article{farokhzad2012897,
  title={Artificial neural network based classification of faults in centrifugal water pump},
  author={Farokhzad, Saeid and Ahmadi, Hojjat and Jaefari, Ali and Abad, MRAA and Kohan, Mohammad Ranjbar},
  journal={Journal of Vibroengineering},
  volume={14},
  number={4},
  pages={1734--1744},
  year={2012}
}

@inproceedings{schizas2002artificial,
  title={Artificial neural networks in estimating marine propeller cavitation},
  author={Schizas, CN},
  booktitle={Proceedings of the 2002 International Joint Conference on Neural Networks. IJCNN'02 (Cat. No. 02CH37290)},
  volume={2},
  pages={1848--1852},
  year={2002},
  organization={IEEE}
}

@inproceedings{arendra2020investigating,
  title={Investigating pump cavitation based on audio sound signature recognition using artificial neural network},
  author={Arendra, Anis and Akhmad, Sabarudin and Winarso, Kukuh and others},
  booktitle={Journal of Physics: Conference Series},
  volume={1569},
  number={3},
  pages={032044},
  year={2020},
  organization={IOP Publishing}
}

@article{you2024cavitation,
  title={Cavitation intensity prediction and optimization for a Venturi cavitation reactor using deep learning},
  author={You, Weibin and Liu, Teng and Manickam, Sivakumar and Wang, Jilai and Wang, Wenlong and Sun, Xun},
  journal={Physics of Fluids},
  volume={36},
  number={11},
  year={2024},
  publisher={AIP Publishing}
}

@article{sun2019prediction,
  title={Prediction and multi-objective optimization of tidal current turbines considering cavitation based on GA-ANN methods},
  author={Sun, Zhaocheng and Li, Zengliang and Fan, Menghao and Wang, Meng and Zhang, Le},
  journal={Energy Science \& Engineering},
  volume={7},
  number={5},
  pages={1896--1912},
  year={2019},
  publisher={Wiley Online Library}
}

@article{xu2021multi,
  title={Multi-objective optimization of jet pump based on RBF neural network model},
  author={Xu, Kai and Wang, Gang and Zhang, Luyao and Wang, Liquan and Yun, Feihong and Sun, Wenhao and Wang, Xiangyu and Chen, Xi},
  journal={Journal of Marine Science and Engineering},
  volume={9},
  number={2},
  pages={236},
  year={2021},
  publisher={Multidisciplinary Digital Publishing Institute}
}

@article{zhang2016effect,
  title={Effect of impeller inlet geometry on cavitation performance of centrifugal pumps based on radial basis function},
  author={Zhang, Shuwei and Zhang, Renhui and Zhang, Sidai and Yang, Junhu},
  journal={International Journal of Rotating Machinery},
  volume={2016},
  number={1},
  pages={6048263},
  year={2016},
  publisher={Wiley Online Library}
}

@article{al2022using,
  title={Using MLP-GABP and SVM with wavelet packet transform-based feature extraction for fault diagnosis of a centrifugal pump},
  author={Al Tobi, Maamar and Bevan, Geraint and Wallace, Peter and Harrison, David and Okedu, Kenneth Eloghene},
  journal={Energy Science \& Engineering},
  volume={10},
  number={6},
  pages={1826--1839},
  year={2022},
  publisher={Wiley Online Library}
}

@article{al2021faults,
  title={Faults diagnosis of a centrifugal pump using multilayer perceptron genetic algorithm back propagation and support vector machine with discrete wavelet transform-based feature extraction},
  author={Al Tobi, Maamar and Bevan, Geraint and Wallace, Peter and Harrison, David and Okedu, Kenneth Eloghene},
  journal={Computational Intelligence},
  volume={37},
  number={1},
  pages={21--46},
  year={2021},
  publisher={Wiley Online Library}
}

@article{zouari2004fault,
  title={Fault detection system for centrifugal pumps using neural networks and neuro-fuzzy techniques},
  author={Zouari, Rafik and Sieg-Zieba, Sophie and Sidahmed, Menad},
  journal={Surveillance},
  volume={5},
  pages={11--13},
  year={2004}
}

@article{altobi2019fault,
  title={Fault diagnosis of a centrifugal pump using MLP-GABP and SVM with CWT},
  author={ALTobi, Maamar Ali Saud and Bevan, Geraint and Wallace, Peter and Harrison, David and Ramachandran, KP},
  journal={Engineering Science and Technology, an International Journal},
  volume={22},
  number={3},
  pages={854--861},
  year={2019},
  publisher={Elsevier}
}

@article{tan2023fault,
  title={Fault diagnosis of a mixed-flow pump under cavitation condition based on deep learning techniques},
  author={Tan, Yangyang and Wu, Guoying and Qiu, Yanlin and Fan, Honggang and Wan, Jun},
  journal={Frontiers in Energy Research},
  volume={10},
  pages={1109214},
  year={2023},
  publisher={Frontiers Media SA}
}

@article{orru2020machine,
  title={Machine learning approach using MLP and SVM algorithms for the fault prediction of a centrifugal pump in the oil and gas industry},
  author={Orr{\`u}, Pier Francesco and Zoccheddu, Andrea and Sassu, Lorenzo and Mattia, Carmine and Cozza, Riccardo and Arena, Simone},
  journal={Sustainability},
  volume={12},
  number={11},
  pages={4776},
  year={2020},
  publisher={MDPI}
}

@article{wang2025optimization,
  title={Optimization design of centrifugal pump cavitation performance based on the improved BP neural network algorithm},
  author={Wang, Yuqin and Shao, Jiale and Yang, Fan and Zhu, Qingzhuo and Zuo, Mengqiang},
  journal={Measurement},
  volume={245},
  pages={116553},
  year={2025},
  publisher={Elsevier}
}

@article{azizi2017improving,
  title={Improving accuracy of cavitation severity detection in centrifugal pumps using a hybrid feature selection technique},
  author={Azizi, Raziyeh and Attaran, Behrooz and Hajnayeb, Ali and Ghanbarzadeh, Afshin and Changizian, Maziar},
  journal={Measurement},
  volume={108},
  pages={9--17},
  year={2017},
  publisher={Elsevier}
}

@inproceedings{benabdesselam2025multi,
  title={A multi-layer perceptron approach for indirect measurement of erosive cavitation in hydraulic turbines},
  author={Benabdesselam, Abderraouf and Pham, Quang Hung and Gagnon, Martin and Tahan, Antoine},
  booktitle={IET Conference Proceedings CP927},
  volume={2025},
  number={10},
  pages={29--34},
  year={2025},
  organization={IET}
}

@article{song2024centrifugal,
  title={Centrifugal Pump Cavitation Fault Diagnosis Based on Feature-Level Multi-Source Information Fusion},
  author={Song, Mengbin and Zhi, Yifan and An, Mengdong and Xu, Wei and Li, Guohui and Wang, Xiuli},
  journal={Processes},
  volume={12},
  number={1},
  pages={196},
  year={2024},
  publisher={MDPI}
}

@article{kamiel2015vibration,
  title={Vibration-based multi-fault diagnosis for centrifugal pumps},
  author={Kamiel, Berli Paripurna},
  journal={Thesis presented for the Degree of Doctor of Philosophy of Curtin University},
  pages={268},
  year={2015}
}

@article{fadaei2018cavitation,
  title={Cavitation damage prediction on dam spillways using Fuzzy-KNN modeling},
  author={Fadaei Kermani, E and Barani, GA and Ghaeini-Hessaroeyeh, M},
  journal={Journal of Applied Fluid Mechanics},
  volume={11},
  number={2},
  pages={323--329},
  year={2018}
}

@article{zeng2024multi,
  title={Multi-cavitation states diagnosis of the vortex pump using a combined DT-CWT-VMD and BO-LW-KNN based on motor current signals},
  author={Zeng, Weitao and Zhou, Peijian and Wu, Yanzhao and Wu, Denghao and Xu, Maosen},
  journal={IEEE Sensors Journal},
  year={2024},
  publisher={IEEE}
}

@article{farokhzad2013vibration,
  title={Vibration based fault detection of centrifugal pump by fast fourier transform and adaptive neuro-fuzzy inference system},
  author={Farokhzad, Saeid},
  journal={Journal of mechanical engineering and technology},
  volume={1},
  number={3},
  pages={82--87},
  year={2013}
}

@article{azadeh2010fuzzy,
  title={A fuzzy inference system for pump failure diagnosis to improve maintenance process: The case of a petrochemical industry},
  author={Azadeh, Ali and Ebrahimipour, Vahid and Bavar, P},
  journal={Expert Systems with Applications},
  volume={37},
  number={1},
  pages={627--639},
  year={2010},
  publisher={Elsevier}
}

@article{wang2017hydraulic,
  title={A hydraulic fault diagnosis method based on sliding-window spectrum feature and deep belief network},
  author={Wang, Xinqing and Huang, Jie and Ren, Guoting and Wang, Dong},
  journal={Journal of Vibroengineering},
  volume={19},
  number={6},
  pages={4272--4284},
  year={2017},
  publisher={JVE International Ltd.}
}

@article{huang2019analysis,
  title={Analysis of weak fault in hydraulic system based on multi-scale permutation entropy of fault-sensitive intrinsic mode function and deep belief network},
  author={Huang, Jie and Wang, Xinqing and Wang, Dong and Wang, Zhiwei and Hua, Xia},
  journal={Entropy},
  volume={21},
  number={4},
  pages={425},
  year={2019},
  publisher={MDPI}
}

@inproceedings{liu2023optimization,
  title={Optimization Application of Deep Belief Network in Gear Fault Diagnosis},
  author={Liu, Haotao and Wang, Jingjing},
  booktitle={Journal of Physics: Conference Series},
  volume={2437},
  number={1},
  pages={012077},
  year={2023},
  organization={IOP Publishing}
}

@article{yan2019fault,
  title={A fault diagnosis method for gas turbines based on improved data preprocessing and an optimization deep belief network},
  author={Yan, Li-Ping and Dong, Xue-Zhi and Wang, Tao and Gao, Qing and Tan, Chun-Qing and Zeng, De-Tang and Zhang, Hua-liang and Chen, Hai-sheng},
  journal={Measurement Science and Technology},
  volume={31},
  number={1},
  pages={015015},
  year={2019},
  publisher={IOP Publishing}
}

@book{mayer2013big,
  title={Big data: A revolution that will transform how we live, work, and think},
  author={Mayer-Sch{\"o}nberger, Viktor and Cukier, Kenneth},
  year={2013},
  publisher={Houghton Mifflin Harcourt}
}

@article{hoang2019survey,
  title={A survey on deep learning based bearing fault diagnosis},
  author={Hoang, Duy-Tang and Kang, Hee-Jun},
  journal={Neurocomputing},
  volume={335},
  pages={327--335},
  year={2019},
  publisher={Elsevier}
}

@article{hinton2006fast,
  title={A fast learning algorithm for deep belief nets},
  author={Hinton, Geoffrey E and Osindero, Simon and Teh, Yee-Whye},
  journal={Neural computation},
  volume={18},
  number={7},
  pages={1527--1554},
  year={2006},
  publisher={MIT Press}
}

@article{cortes1995support,
  title={Support-vector networks},
  author={Cortes, Corinna and Vapnik, Vladimir},
  journal={Machine learning},
  volume={20},
  number={3},
  pages={273--297},
  year={1995},
  publisher={Springer}
}

@article{quinlan1986induction,
  title={Induction of decision trees},
  author={Quinlan, J. Ross},
  journal={Machine learning},
  volume={1},
  number={1},
  pages={81--106},
  year={1986},
  publisher={Springer}
}

@article{rumelhart1986learning,
  title={Learning representations by back-propagating errors},
  author={Rumelhart, David E and Hinton, Geoffrey E and Williams, Ronald J},
  journal={nature},
  volume={323},
  number={6088},
  pages={533--536},
  year={1986},
  publisher={Nature Publishing Group UK London}
}

@book{fukunaga2013introduction,
  title={Introduction to statistical pattern recognition},
  author={Fukunaga, Keinosuke},
  year={2013},
  publisher={Elsevier}
}

@article{jackson1986introduction,
  title={Introduction to expert systems},
  author={Jackson, Peter},
  year={1986},
  publisher={Addison-Wesley Pub. Co., Reading, MA}
}

@article{jiao2024self,
  title={Self-Supervised, Non-Contact Heartbeat Detection Based on Ballistocardiograms Utilizing Physiological Information Guidance},
  author={Jiao, Changzhe and Yang, Aoyu and Zhao, Hantao and Yi, Ruhan and Gou, Shuiping and Sha, Yu and Wen, Wanshun and Jiao, Licheng and Skubic, Marjorie},
  journal={IEEE Journal of Biomedical and Health Informatics},
  year={2024},
  publisher={IEEE}
}

@article{li2022study,
  title={A study on small magnitude seismic phase identification using 1D deep residual neural network},
  author={Li, Wei and Chakraborty, Megha and Sha, Yu and Zhou, Kai and Faber, Johannes and R{\"u}mpker, Georg and St{\"o}cker, Horst and Srivastava, Nishtha},
  journal={Artificial Intelligence in Geosciences},
  volume={3},
  pages={115--122},
  year={2022},
  publisher={Elsevier}
}

@article{gou2025dynamic,
  title={Dynamic spatio-temporal pruning for efficient spiking neural networks},
  author={Gou, Shuiping and Fu, Jiahui and Sha, Yu and Cao, Zhen and Guo, Zhang and Eshraghian, Jason K and Li, Ruimin and Jiao, Licheng},
  journal={Frontiers in Neuroscience},
  volume={19},
  pages={1545583},
  year={2025},
  publisher={Frontiers Media SA}
}

@article{yuan2025prioritization,
  title={Prioritization Hindsight Experience Based on Spatial Position Attention for Robots},
  author={Yuan, Ye and Sha, Yu and Sun, Feixiang and Lu, Haofan and Gou, Shuiping and Luo, Jie},
  journal={Machine Intelligence Research},
  volume={22},
  number={1},
  pages={160--175},
  year={2025},
  publisher={Springer}
}

@article{pang2018equation,
  title={An equation-of-state-meter of quantum chromodynamics transition from deep learning},
  author={Pang, Long-Gang and Zhou, Kai and Su, Nan and Petersen, Hannah and St{\"o}cker, Horst and Wang, Xin-Nian},
  journal={Nature communications},
  volume={9},
  number={1},
  pages={210},
  year={2018},
  publisher={Nature Publishing Group UK London}
}

@article{wu2023towards,
  title={Towards automated 3D evaluation of water leakage on a tunnel face via improved GAN and self-attention DL model},
  author={Wu, Chen and Huang, Hongwei and Zhang, Le and Chen, Jiayao and Tong, Yue and Zhou, Mingliang},
  journal={Tunnelling and Underground Space Technology},
  volume={142},
  pages={105432},
  year={2023},
  publisher={Elsevier}
}

@inproceedings{lu2022understanding,
  title={Understanding the dynamics of dnns using graph modularity},
  author={Lu, Yao and Yang, Wen and Zhang, Yunzhe and Chen, Zuohui and Chen, Jinyin and Xuan, Qi and Wang, Zhen and Yang, Xiaoniu},
  booktitle={European Conference on Computer Vision},
  pages={225--242},
  year={2022},
  organization={Springer}
}

@article{lu2024generic,
  title={A generic layer pruning method for signal modulation recognition deep learning models},
  author={Lu, Yao and Zhu, Yutao and Li, Yuqi and Xu, Dongwei and Lin, Yun and Xuan, Qi and Yang, Xiaoniu},
  journal={IEEE Transactions on Cognitive Communications and Networking},
  year={2024},
  publisher={IEEE}
}

@inproceedings{li2025encoder,
  title={Encoder: Entity mining and modification relation binding for composed image retrieval},
  author={Li, Zixu and Chen, Zhiwei and Wen, Haokun and Fu, Zhiheng and Hu, Yupeng and Guan, Weili},
  booktitle={Proceedings of the AAAI Conference on Artificial Intelligence},
  volume={39},
  number={5},
  pages={5101--5109},
  year={2025}
}

@inproceedings{huang2025median,
  title={MEDIAN: Adaptive Intermediate-grained Aggregation Network for Composed Image Retrieval},
  author={Huang, Qinlei and Chen, Zhiwei and Li, Zixu and Wang, Chunxiao and Song, Xuemeng and Hu, Yupeng and Nie, Liqiang},
  booktitle={ICASSP 2025-2025 IEEE International Conference on Acoustics, Speech and Signal Processing (ICASSP)},
  pages={1--5},
  year={2025},
  organization={IEEE}
}

@inproceedings{fu2025pair,
  title={PAIR: Complementarity-guided Disentanglement for Composed Image Retrieval},
  author={Fu, Zhiheng and Li, Zixu and Chen, Zhiwei and Wang, Chunxiao and Song, Xuemeng and Hu, Yupeng and Nie, Liqiang},
  booktitle={ICASSP 2025-2025 IEEE International Conference on Acoustics, Speech and Signal Processing (ICASSP)},
  pages={1--5},
  year={2025},
  organization={IEEE}
}

@inproceedings{lavanya2021deep,
  title={Deep learning techniques on text classification using Natural language processing (NLP) in social healthcare network: A comprehensive survey},
  author={Lavanya, PM and Sasikala, E},
  booktitle={2021 3rd international conference on signal processing and communication (ICPSC)},
  pages={603--609},
  year={2021},
  organization={IEEE}
}

@article{wu2024novel,
  title={A novel Tree-augmented Bayesian network for predicting rock weathering degree using incomplete dataset},
  author={Wu, Chen and Huang, Hongwei and Chen, Jiayao and Zhou, Mingliang and Han, Shiju},
  journal={International Journal of Rock Mechanics and Mining Sciences},
  volume={183},
  pages={105933},
  year={2024},
  publisher={Elsevier}
}

@inproceedings{palaz2015analysis,
  title={Analysis of CNN-based speech recognition system using raw speech as input.},
  author={Palaz, Dimitri and Magimai-Doss, Mathew and Collobert, Ronan and others},
  booktitle={Interspeech},
  pages={11--15},
  year={2015},
  organization={Dresden, Germany}
}

@article{abdel2014convolutional,
  title={Convolutional neural networks for speech recognition},
  author={Abdel-Hamid, Ossama and Mohamed, Abdel-rahman and Jiang, Hui and Deng, Li and Penn, Gerald and Yu, Dong},
  journal={IEEE/ACM Transactions on audio, speech, and language processing},
  volume={22},
  number={10},
  pages={1533--1545},
  year={2014},
  publisher={IEEE}
}

@inproceedings{abdel2013exploring,
  title={Exploring convolutional neural network structures and optimization techniques for speech recognition.},
  author={Abdel-Hamid, Ossama and Deng, Li and Yu, Dong},
  booktitle={Interspeech},
  volume={2013},
  pages={1173--5},
  year={2013}
}

@article{huang2022feature,
  title={Feature map distillation of thin nets for low-resolution object recognition},
  author={Huang, Zhenhua and Yang, Shunzhi and Zhou, MengChu and Li, Zhetao and Gong, Zheng and Chen, Yunwen},
  journal={IEEE Transactions on Image Processing},
  volume={31},
  pages={1364--1379},
  year={2022},
  publisher={IEEE}
}

@article{bao2023lightweight,
  title={A lightweight block with information flow enhancement for convolutional neural networks},
  author={Bao, Zhiqiang and Yang, Shunzhi and Huang, Zhenhua and Zhou, MengChu and Chen, Yunwen},
  journal={IEEE Transactions on Circuits and Systems for Video Technology},
  volume={33},
  number={8},
  pages={3570--3584},
  year={2023},
  publisher={IEEE}
}

@article{lin2024multi,
  title={A multi-level relation-aware transformer model for occluded person re-identification},
  author={Lin, Guorong and Bao, Zhiqiang and Huang, Zhenhua and Li, Zuoyong and Zheng, Wei-shi and Chen, Yunwen},
  journal={Neural Networks},
  volume={177},
  pages={106382},
  year={2024},
  publisher={Elsevier}
}

@inproceedings{xiang2022phase,
  title={Phase Retrieval for Terahertz Holography with Physics-Informed Deep Learning},
  author={Xiang, Mingjun and Wang, Lingxiao and Sha, Yu and Yuan, Hui and Zhou, Kai and Roskos, Hartmut G},
  booktitle={Digital Holography and Three-Dimensional Imaging},
  pages={Tu4A--4},
  year={2022},
  organization={Optica Publishing Group}
}

@article{lecun2002gradient,
  title={Gradient-based learning applied to document recognition},
  author={LeCun, Yann and Bottou, L{\'e}on and Bengio, Yoshua and Haffner, Patrick},
  journal={Proceedings of the IEEE},
  volume={86},
  number={11},
  pages={2278--2324},
  year={2002},
  publisher={Ieee}
}

@article{krizhevsky2012imagenet,
  title={Imagenet classification with deep convolutional neural networks},
  author={Krizhevsky, Alex and Sutskever, Ilya and Hinton, Geoffrey E},
  journal={Advances in neural information processing systems},
  volume={25},
  year={2012}
}

@inproceedings{simonyan2015very,
  title={Very deep convolutional networks for large-scale image recognition},
  author={Simonyan, K and Zisserman, A},
  booktitle={3rd International Conference on Learning Representations (ICLR 2015)},
  year={2015},
  organization={Computational and Biological Learning Society}
}

@inproceedings{he2016deep,
  title={Deep residual learning for image recognition},
  author={He, Kaiming and Zhang, Xiangyu and Ren, Shaoqing and Sun, Jian},
  booktitle={Proceedings of the IEEE conference on computer vision and pattern recognition},
  pages={770--778},
  year={2016}
}

@inproceedings{huang2017densely,
  title={Densely connected convolutional networks},
  author={Huang, Gao and Liu, Zhuang and Van Der Maaten, Laurens and Weinberger, Kilian Q},
  booktitle={Proceedings of the IEEE conference on computer vision and pattern recognition},
  pages={4700--4708},
  year={2017}
}

@article{howard2017mobilenets,
  title={Mobilenets: Efficient convolutional neural networks for mobile vision applications},
  author={Howard, Andrew G and Zhu, Menglong and Chen, Bo and Kalenichenko, Dmitry and Wang, Weijun and Weyand, Tobias and Andreetto, Marco and Adam, Hartwig},
  journal={arXiv preprint arXiv:1704.04861},
  year={2017}
}

@inproceedings{zhang2018shufflenet,
  title={Shufflenet: An extremely efficient convolutional neural network for mobile devices},
  author={Zhang, Xiangyu and Zhou, Xinyu and Lin, Mengxiao and Sun, Jian},
  booktitle={Proceedings of the IEEE conference on computer vision and pattern recognition},
  pages={6848--6856},
  year={2018}
}

@article{zaremba2014recurrent,
  title={Recurrent neural network regularization},
  author={Zaremba, Wojciech and Sutskever, Ilya and Vinyals, Oriol},
  journal={arXiv preprint arXiv:1409.2329},
  year={2014}
}

@article{hochreiter1997long,
  title={Long short-term memory},
  author={Hochreiter, Sepp and Schmidhuber, J{\"u}rgen},
  journal={Neural computation},
  volume={9},
  number={8},
  pages={1735--1780},
  year={1997},
  publisher={MIT press}
}

@article{graves2005framewise,
  title={Framewise phoneme classification with bidirectional LSTM and other neural network architectures},
  author={Graves, Alex and Schmidhuber, J{\"u}rgen},
  journal={Neural networks},
  volume={18},
  number={5-6},
  pages={602--610},
  year={2005},
  publisher={Elsevier}
}

@inproceedings{cho2014learning,
  title={Learning phrase representations using RNN encoder-decoder for statistical machine translation},
  author={Cho, Kyunghyun and van Merrienboer, B and Gulcehre, Caglar and Bougares, F and Schwenk, H and Bengio, Yoshua},
  booktitle={Conference on Empirical Methods in Natural Language Processing (EMNLP 2014)},
  year={2014}
}

@article{vaswani2017attention,
  title={Attention is all you need},
  author={Vaswani, Ashish and Shazeer, Noam and Parmar, Niki and Uszkoreit, Jakob and Jones, Llion and Gomez, Aidan N and Kaiser, {\L}ukasz and Polosukhin, Illia},
  journal={Advances in neural information processing systems},
  volume={30},
  year={2017}
}

@article{zeng2025futuresightdrive,
  title={FutureSightDrive: Thinking Visually with Spatio-Temporal CoT for Autonomous Driving},
  author={Zeng, Shuang and Chang, Xinyuan and Xie, Mengwei and Liu, Xinran and Bai, Yifan and Pan, Zheng and Xu, Mu and Wei, Xing},
  journal={arXiv preprint arXiv:2505.17685},
  year={2025}
}

@article{zeng2024driving,
  title={Driving with prior maps: Unified vector prior encoding for autonomous vehicle mapping},
  author={Zeng, Shuang and Chang, Xinyuan and Liu, Xinran and Pan, Zheng and Wei, Xing},
  journal={arXiv preprint arXiv:2409.05352},
  year={2024}
}

@inproceedings{dosovitskiy2020image,
  title={An Image is Worth 16x16 Words: Transformers for Image Recognition at Scale},
  author={Dosovitskiy, Alexey and Beyer, Lucas and Kolesnikov, Alexander and Weissenborn, Dirk and Zhai, Xiaohua and Unterthiner, Thomas and Dehghani, Mostafa and Minderer, Matthias and Heigold, G and Gelly, S and others},
  booktitle={International Conference on Learning Representations},
  year={2020}
}

@inproceedings{liu2021swin,
  title={Swin transformer: Hierarchical vision transformer using shifted windows},
  author={Liu, Ze and Lin, Yutong and Cao, Yue and Hu, Han and Wei, Yixuan and Zhang, Zheng and Lin, Stephen and Guo, Baining},
  booktitle={Proceedings of the IEEE/CVF international conference on computer vision},
  pages={10012--10022},
  year={2021}
}

@inproceedings{devlin2019bert,
  title={Bert: Pre-training of deep bidirectional transformers for language understanding},
  author={Devlin, Jacob and Chang, Ming-Wei and Lee, Kenton and Toutanova, Kristina},
  booktitle={Proceedings of the 2019 conference of the North American chapter of the association for computational linguistics: human language technologies, volume 1 (long and short papers)},
  pages={4171--4186},
  year={2019}
}

@article{wang2020linformer,
  title={Linformer: Self-attention with linear complexity},
  author={Wang, Sinong and Li, Belinda Z and Khabsa, Madian and Fang, Han and Ma, Hao},
  journal={arXiv preprint arXiv:2006.04768},
  year={2020}
}

@inproceedings{radford2021learning,
  title={Learning transferable visual models from natural language supervision},
  author={Radford, Alec and Kim, Jong Wook and Hallacy, Chris and Ramesh, Aditya and Goh, Gabriel and Agarwal, Sandhini and Sastry, Girish and Askell, Amanda and Mishkin, Pamela and Clark, Jack and others},
  booktitle={International conference on machine learning},
  pages={8748--8763},
  year={2021},
  organization={PmLR}
}

@article{look2018building,
  title={Building Robust Classifiers with Generative Adversarial Networks for Detecting Cavitation in Hydraulic Turbines.},
  author={Look, Andreas and Kirschner, Oliver and Riedelbauch, Stefan},
  journal={ICPRAM},
  volume={2018},
  pages={456--462},
  year={2018}
}

@article{miglianti2020predicting,
  title={Predicting the cavitating marine propeller noise at design stage: A deep learning based approach},
  author={Miglianti, Leonardo and Cipollini, Francesca and Oneto, Luca and Tani, Giorgio and Gaggero, Stefano and Coraddu, Andrea and Viviani, Michele},
  journal={Ocean Engineering},
  volume={209},
  pages={107481},
  year={2020},
  publisher={Elsevier}
}

@article{kumar2020improved,
  title={Improved deep convolution neural network (CNN) for the identification of defects in the centrifugal pump using acoustic images},
  author={Kumar, Anil and Gandhi, CP and Zhou, Yuqing and Kumar, Rajesh and Xiang, Jiawei},
  journal={Applied Acoustics},
  volume={167},
  pages={107399},
  year={2020},
  publisher={Elsevier}
}

@article{chao2020identification,
  title={Identification of cavitation intensity for high-speed aviation hydraulic pumps using 2D convolutional neural networks with an input of RGB-based vibration data},
  author={Chao, Qun and Tao, Jianfeng and Wei, Xiaoliang and Liu, Chengliang},
  journal={Measurement Science and Technology},
  volume={31},
  number={10},
  pages={105102},
  year={2020},
  publisher={IOP Publishing}
}

@article{chao2020cavitation,
  title={Cavitation intensity recognition for high-speed axial piston pumps using 1-D convolutional neural networks with multi-channel inputs of vibration signals},
  author={Chao, Qun and Tao, Jianfeng and Wei, Xiaoliang and Wang, Yuanhang and Meng, Linghui and Liu, Chengliang},
  journal={Alexandria Engineering Journal},
  volume={59},
  number={6},
  pages={4463--4473},
  year={2020},
  publisher={Elsevier}
}

@article{Wei2021cavitation,
  title={Cavitation fault diagnosis of high-speed axial piston pump based on spectrum analysis and convolutional neural network},
  author={Wei, Xiaoliang and Chao, Qun and Tao, Jianfeng and Liu, Chengliang and Wang, Lixiao},
  journal={Chinese Hydraulics \& Pneumatics},
  volume={45},
  number={7},
  pages={7--13},
  year={2021}
}

@inproceedings{liu2021convolution,
  title={Convolution Diagnosis Model of Centrifugal Pump Based on Fractal Dimension},
  author={Liu, Xiaopeng and Sun, Bingxiang and Jiang, Jiuchun and Zhang, Weige and Zhao, Chunlu and Zhao, Yuzhang and Mao, Benqiang and Li, Jian and Wang, Zhisheng},
  booktitle={Journal of Physics: Conference Series},
  volume={2095},
  number={1},
  pages={012061},
  year={2021},
  organization={IOP Publishing}
}

@article{zhu2021intelligent,
  title={Intelligent fault diagnosis of hydraulic piston pump based on wavelet analysis and improved alexnet},
  author={Zhu, Yong and Li, Guangpeng and Wang, Rui and Tang, Shengnan and Su, Hong and Cao, Kai},
  journal={Sensors},
  volume={21},
  number={2},
  pages={549},
  year={2021},
  publisher={MDPI}
}

@article{tiwari2021blockage,
  title={Blockage and cavitation detection in centrifugal pumps from dynamic pressure signal using deep learning algorithm},
  author={Tiwari, Rajiv and Bordoloi, DJ and Dewangan, Aakash},
  journal={Measurement},
  volume={173},
  pages={108676},
  year={2021},
  publisher={Elsevier}
}

@article{hajnayeb2021cavitation,
  title={Cavitation analysis in centrifugal pumps based on vibration bispectrum and transfer learning},
  author={Hajnayeb, Ali},
  journal={Shock and Vibration},
  volume={2021},
  number={1},
  pages={6988949},
  year={2021},
  publisher={Wiley Online Library}
}

@article{bach2021classification,
  title={Classification of surface vehicle propeller cavitation noise using spectrogram processing in combination with convolution neural network},
  author={Bach, Nhat Hoang and Vu, Le Ha and Nguyen, Van Duc},
  journal={Sensors},
  volume={21},
  number={10},
  pages={3353},
  year={2021},
  publisher={MDPI}
}

@inproceedings{han2022comparative,
  title={Comparative study on deep and shallow feature fusion CNN for fault diagnosis},
  author={Han, Lei and Ran, Dongsheng and Zhang, Lishan},
  booktitle={2022 IEEE 17th Conference on Industrial Electronics and Applications (ICIEA)},
  pages={609--613},
  year={2022},
  organization={IEEE}
}

@article{li2022application,
  title={Application of Deep Learning to Predict Cavitation Flow in Centrifugal Pump},
  author={Li, Gaoyang and He, Jiachao and Ding, Xuhui and Zhu, Yonghong and Zhu, Wenkun and Qin, Caiyan and Zhang, Xuelan and Liu, Siwei and Sun, Haiyi and Yu, Wenjin and others},
  journal={Available at SSRN 4182265},
  year={2022}
}

@inproceedings{sha2022regional,
  title={Regional-local adversarially learned one-class classifier anomalous sound detection in global long-term space},
  author={Sha, Yu and Gou, Shuiping and Faber, Johannes and Liu, Bo and Li, Wei and Schramm, Stefan and Stoecker, Horst and Steckenreiter, Thomas and Vnucec, Domagoj and Wetzstein, Nadine and others},
  booktitle={Proceedings of the 28th ACM SIGKDD Conference on Knowledge Discovery and Data Mining},
  pages={3858--3868},
  year={2022}
}

@article{tang2022intelligent,
  title={Intelligent fault identification of hydraulic pump using deep adaptive normalized CNN and synchrosqueezed wavelet transform},
  author={Tang, Shengnan and Zhu, Yong and Yuan, Shouqi},
  journal={Reliability Engineering \& System Safety},
  volume={224},
  pages={108560},
  year={2022},
  publisher={Elsevier}
}

@article{chao2022improving,
  title={Improving accuracy of cavitation severity recognition in axial piston pumps by denoising time--frequency images},
  author={Chao, Qun and Wei, Xiaoliang and Lei, Junbo and Tao, Jianfeng and Liu, Chengliang},
  journal={Measurement Science and Technology},
  volume={33},
  number={5},
  pages={055116},
  year={2022},
  publisher={IOP Publishing}
}

@article{sha2022multi,
  title={A multi-task learning for cavitation detection and cavitation intensity recognition of valve acoustic signals},
  author={Sha, Yu and Faber, Johannes and Gou, Shuiping and Liu, Bo and Li, Wei and Schramm, Stefan and Stoecker, Horst and Steckenreiter, Thomas and Vnucec, Domagoj and Wetzstein, Nadine and others},
  journal={Engineering Applications of Artificial Intelligence},
  volume={113},
  pages={104904},
  year={2022},
  publisher={Elsevier}
}

@article{gou2024hierarchical,
  title={Hierarchical cavitation intensity recognition using Sub-Master Transition Network-based acoustic signals in pipeline systems},
  author={Gou, Shuiping and Sha, Yu and Liu, Bo and Liu, Ningtao and Fabe, Johannes and Schramm, Stefan and Stoecker, Horst and Steckenreiter, Thomas and Vnucec, Domagoj and Wetzstein, Nadine and others},
  journal={Expert Systems with Applications},
  volume={258},
  pages={125155},
  year={2024},
  publisher={Elsevier}
}

@inproceedings{liu2023intelligent,
  title={Intelligent Identification Method of Flow State in Nuclear Main Pump Based on Deep Learning Method},
  author={Liu, Ying-Yuan and Liu, Di and Zhang, Zhenjun and An, Kang},
  booktitle={Chinese Intelligent Automation Conference},
  pages={691--699},
  year={2023},
  organization={Springer}
}

@article{chennai2023deep,
  title={Deep learning for enhanced fault diagnosis of monoblock centrifugal pumps: Spectrogram-based analysis},
  author={Chennai Viswanathan, Prasshanth and Venkatesh, Sridharan Naveen and Dhanasekaran, Seshathiri and Mahanta, Tapan Kumar and Sugumaran, Vaithiyanathan and Lakshmaiya, Natrayan and Paramasivam, Prabhu and Nanjagoundenpalayam Ramasamy, Sakthivel},
  journal={Machines},
  volume={11},
  number={9},
  pages={874},
  year={2023},
  publisher={MDPI}
}

@article{wei2023cavitation,
  title={Cavitation state recognition for control valve using AlexNet-type neural networks with three-channel images transformed by time series},
  author={Wei, Jianqiu and Liu, Xiumei and Li, Beibei and Zhang, Yujia and Shang, Ximing},
  journal={Measurement Science and Technology},
  volume={34},
  number={5},
  pages={055301},
  year={2023},
  publisher={IOP Publishing}
}

@article{jamal2024passive,
  title={Passive SONAR detection and classification based on DEMON-LOFAR analysis and neural network algorithms},
  author={Jamal, Said and Lakziz, Jawad and Benremdane, Yahya and Ouaskit, Said},
  journal={International Journal of Artificial Intelligence and Applications},
  volume={15},
  number={1},
  pages={87--98},
  year={2024}
}

@article{han2025use,
  title={The use of model-based voltage and current analysis for torque oscillation detection and improved condition monitoring of centrifugal pumps},
  author={Han, Yuejiang and Zou, Jiamin and Gong, Bo and Luo, Yin and Wang, Longyan and Batllo, Alexandre Presas and Yuan, Jianping and Wang, Chao},
  journal={Mechanical Systems and Signal Processing},
  volume={222},
  pages={111781},
  year={2025},
  publisher={Elsevier}
}

@article{tang2024light,
  title={A light deep adaptive framework toward fault diagnosis of a hydraulic piston pump},
  author={Tang, Shengnan and Khoo, Boo Cheong and Zhu, Yong and Lim, Kian Meng and Yuan, Shouqi},
  journal={Applied Acoustics},
  volume={217},
  pages={109807},
  year={2024},
  publisher={Elsevier}
}

@article{song2025cavitation,
  title={Cavitation identification method of centrifugal pumps based on signal demodulation and EfficientNet},
  author={Song, Yongxing and Zhang, Tonghe and Liu, Qiang and Ge, Bingxin and Liu, Jingting and Zhang, Linhua},
  journal={Arabian Journal for Science and Engineering},
  volume={50},
  number={12},
  pages={8779--8793},
  year={2025},
  publisher={Springer}
}

@article{sandhu2024comparative,
  title={A comparative study on deep learning models for condition monitoring of advanced reactor piping systems},
  author={Sandhu, Harleen Kaur and Bodda, Saran Srikanth and Yan, Erin and Sabharwall, Piyush and Gupta, Abhinav},
  journal={Mechanical Systems and Signal Processing},
  volume={209},
  pages={111091},
  year={2024},
  publisher={Elsevier}
}

@article{zhu2023data,
  title={A data-driven diagnosis scheme based on deep learning toward fault identification of the hydraulic piston pump},
  author={Zhu, Yong and Zhou, Tao and Tang, Shengnan and Yuan, Shouqi},
  journal={Journal of Marine Science and Engineering},
  volume={11},
  number={7},
  pages={1273},
  year={2023},
  publisher={MDPI}
}

@article{ullah2023intelligent,
  title={An intelligent framework for fault diagnosis of centrifugal pump leveraging wavelet coherence analysis and deep learning},
  author={Ullah, Niamat and Ahmad, Zahoor and Siddique, Muhammad Farooq and Im, Kichang and Shon, Dong-Koo and Yoon, Tae-Hyun and Yoo, Dae-Seung and Kim, Jong-Myon},
  journal={Sensors},
  volume={23},
  number={21},
  pages={8850},
  year={2023},
  publisher={MDPI}
}

@article{tang2022adaptive,
  title={An adaptive deep learning model towards fault diagnosis of hydraulic piston pump using pressure signal},
  author={Tang, Shengnan and Zhu, Yong and Yuan, Shouqi},
  journal={Engineering Failure Analysis},
  volume={138},
  pages={106300},
  year={2022},
  publisher={Elsevier}
}

@article{lu2024research,
  title={Research into Prediction Method for Pressure Pulsations in a Centrifugal Pump Based on Variational Mode Decomposition--Particle Swarm Optimization and Hybrid Deep Learning Models},
  author={Lu, Jiaxing and Zhou, Yuzhuo and Ge, Yanlong and Liu, Jiahong and Zhang, Chuan},
  journal={Sensors},
  volume={24},
  number={13},
  pages={4196},
  year={2024},
  publisher={MDPI}
}

@article{prasshanth2024deep,
  title={Deep learning for fault diagnosis of monoblock centrifugal pumps: A Hilbert--Huang transform approach},
  author={Prasshanth, CV and Venkatesh, S Naveen and Mahanta, Tapan K and Sakthivel, NR and Sugumaran, V},
  journal={International Journal of System Assurance Engineering and Management},
  pages={1--14},
  year={2024},
  publisher={Springer}
}

@inproceedings{liu2024failure,
  title={A failure diagnosis approach for firefighting pump based deep learning GRU Methodology},
  author={Liu, Wen-Bin and Bui, Thuc-Minh and Doan, Quyet Nguyen and Le Thi, Huong and Thu, Trang Nguyen Thi and Cho, Ming-Yuan and Da, Thao Nguyen and Thanh, Phuong Nguyen},
  booktitle={2024 International Conference on System Science and Engineering (ICSSE)},
  pages={1--6},
  year={2024},
  organization={IEEE}
}

@article{jiang2024deep,
  title={Deep learning-based multilabel compound-fault diagnosis in centrifugal pumps},
  author={Jiang, Lizhe and Du, Hongze and Bu, Yufeng and Zhao, Chunyu and Lu, Hailong and Yan, Jun},
  journal={Ocean Engineering},
  volume={314},
  pages={119697},
  year={2024},
  publisher={Elsevier}
}

@article{kim2022study,
  title={A study on deep learning-based fault diagnosis and classification for marine engine system auxiliary equipment},
  author={Kim, Jeong-yeong and Lee, Tae-hyun and Lee, Song-ho and Lee, Jong-jik and Lee, Won-kyun and Kim, Yong-jin and Park, Jong-won},
  journal={Processes},
  volume={10},
  number={7},
  pages={1345},
  year={2022},
  publisher={MDPI}
}

@article{prasshanth2024fault,
  title={Fault diagnosis of monoblock centrifugal pumps using pre-trained deep learning models and scalogram images},
  author={Prasshanth, Chennai Viswanathan and Venkatesh, Sridharan Naveen and Mahanta, Tapan Kumar and Sakthivel, Nanjagoundenpalayam Ramasamy and Sugumaran, Vaithiyanathan},
  journal={Engineering Applications of Artificial Intelligence},
  volume={136},
  pages={109022},
  year={2024},
  publisher={Elsevier}
}

@inproceedings{turunen2023deep,
  title={Deep learning for centrifugal pump condition monitoring using data from variable frequency drive},
  author={Turunen, Topias and Miettinen, Jesse and H{\"a}m{\"a}l{\"a}inen, Aleksanteri and Karhinen, Aku and Viitala, Raine},
  booktitle={IFToMM World Congress on Mechanism and Machine Science},
  pages={905--914},
  year={2023},
  organization={Springer}
}

@inproceedings{bach2021enhancing,
  title={Enhancing the Capacity of Detecting and Classifying Cavitation Noise Generated from Propeller Using the Convolution Neural Network},
  author={Bach, Hoang Nhat and Van Nguyen, Duc and Le Vu, Ha},
  booktitle={International Conference on Industrial Networks and Intelligent Systems},
  pages={268--275},
  year={2021},
  organization={Springer}
}

@article{chao2023cavitation,
  title={Cavitation recognition of axial piston pumps in noisy environment based on Grad-CAM visualization technique},
  author={Chao, Qun and Wei, Xiaoliang and Tao, Jianfeng and Liu, Chengliang and Wang, Yuanhang},
  journal={CAAI Transactions on Intelligence Technology},
  volume={8},
  number={1},
  pages={206--218},
  year={2023},
  publisher={Wiley Online Library}
}

@inproceedings{he2023sgst,
  title={SGST: A novel approach based on machine learning for cavitation fault diagnosis},
  author={He, Jinglin and Hao, Jiasheng and Cao, Zhen},
  booktitle={IECON 2023-49th Annual Conference of the IEEE Industrial Electronics Society},
  pages={1--7},
  year={2023},
  organization={IEEE}
}

@article{sun2025cavitation,
  title={Cavitation state recognition of centrifugal pumps across various working conditions by Multi-Adversarial Attention Network},
  author={Sun, Bowen and Zhang, Zhengzhuang and Dong, Pan and Zeng, Yi and Wang, Lu and Yang, Shuai and Wu, Kelin and Wu, Dazhuan},
  journal={Mechanical Systems and Signal Processing},
  volume={239},
  pages={113336},
  year={2025},
  publisher={Elsevier}
}

@article{wang2025cavitation,
  title={Cavitation intensity recognition for axial piston pump based on transient flow rate measurement and improved transfer learning method},
  author={Wang, Renyuan and Xu, Yuanzhi and Mu, Wenkan and Chen, Yucan and Jiao, Zongxia},
  journal={Mechanical Systems and Signal Processing},
  volume={232},
  pages={112667},
  year={2025},
  publisher={Elsevier}
}

@article{zhang2024cavitation,
  title={Cavitation state recognition method of centrifugal pump based on multi-dimensional feature fusion and convolutional gate recurrent unit},
  author={Zhang, Tonghe and Song, Yongxing and Liu, Qiang and Ge, Yi and Zhang, Linhua and Liu, Jingting},
  journal={Physics of Fluids},
  volume={36},
  number={10},
  year={2024},
  publisher={AIP Publishing}
}

@inproceedings{lee2025deep,
  title={Deep Learning-Based Identification of Turbopump Inducer Cavitation Instability Using High-Speed Camera Images},
  author={Lee, Sanghyun and Yoon, Youngkuk and Song, Seung Jin},
  booktitle={AIAA SCITECH 2025 Forum},
  pages={0760},
  year={2025}
}

@article{liu2025feature,
  title={A feature extraction method for hydrofoil attached cavitation based on deep learning image semantic segmentation algorithm},
  author={Liu, Yingyuan and Wang, Yizhi and An, Kang},
  journal={Scientific Reports},
  volume={15},
  number={1},
  pages={4415},
  year={2025},
  publisher={Nature Publishing Group UK London}
}

@article{liu2025data,
  title={A Data-Driven Fault Diagnosis Method for Marine Steam Turbine Condensate System Based on Deep Transfer Learning},
  author={Liu, Yuhui and Chen, Liping and Shangguan, Duansen and Yu, Chengcheng},
  journal={Machines},
  volume={13},
  number={8},
  pages={708},
  year={2025},
  publisher={MDPI}
}

@inproceedings{chang2022centrifugal,
  title={Centrifugal pump fault diagnosis method based on EAS and stacked capsule autoencoder},
  author={Chang, Zihan and Yuan, Wei and Yu, Menghong},
  booktitle={Third International Conference on Electronics and Communication; Network and Computer Technology (ECNCT 2021)},
  volume={12167},
  pages={233--238},
  year={2022},
  organization={SPIE}
}

@article{yilmaz2025deep,
  title={Deep Learning-Based Propeller Sound Classification Using GoogLeNet},
  author={Yilmaz, Alican and Ozkat, Erkan Caner},
  booktitle={6th International Conference on Innovative Academic Studies},
  year={2025}
}

@inproceedings{sha2024hierarchical,
  title={Hierarchical knowledge guided fault intensity diagnosis of complex industrial systems},
  author={Sha, Yu and Gou, Shuiping and Liu, Bo and Faber, Johannes and Liu, Ningtao and Schramm, Stefan and Stoecker, Horst and Steckenreiter, Thomas and Vnucec, Domagoj and Wetzstein, Nadine and others},
  booktitle={Proceedings of the 30th ACM SIGKDD Conference on Knowledge Discovery and Data Mining},
  pages={5657--5668},
  year={2024}
}

@inproceedings{huh2020new,
  title={New way of detecting vibration of mechanical systems by explainable deep learning},
  author={Huh, Hyunsuk and Lee, Soo Young and Lee, Seungchul and Sun, Kyung Ho and Jung, Joon Ha},
  booktitle={INTER-NOISE and NOISE-CON Congress and Conference Proceedings},
  volume={261},
  number={1},
  pages={5646--5650},
  year={2020},
  organization={Institute of Noise Control Engineering}
}

@inproceedings{feng2024experimental,
  title={Experimental study on multiple characteristics of multi-source AE signals during cavitation process of waterjet propulsion pump},
  author={Feng, Chao and Zhu, Hualun and Tao, Jin and Yang, Mengzi},
  booktitle={Journal of Physics: Conference Series},
  volume={2707},
  number={1},
  pages={012118},
  year={2024},
  organization={IOP Publishing}
}

@article{singh2025hybrid,
  title={Hybrid Physics-Infused One-Dimensional Convolutional Neural Network-Based Ensemble Learning Framework for Diesel Engine Fault Diagnostics},
  author={Singh, Shubhendu Kumar and Khawale, Raj Pradip and Hazarika, Subhashis and Rai, Rahul},
  journal={Journal of Computing and Information Science in Engineering},
  volume={25},
  number={4},
  pages={041006},
  year={2025},
  publisher={American Society of Mechanical Engineers}
}

@article{he2024status,
  title={Status Recognition of Marine Centrifugal Pumps Based on a Stacked Sparse Auto-Encoder},
  author={He, Yi and Yao, Yunan and Ou, Hongsen},
  journal={Applied Sciences},
  volume={14},
  number={4},
  pages={1371},
  year={2024},
  publisher={MDPI}
}

@article{liu2022automatic,
  title={Automatic reservoir model identification method based on convolutional neural network},
  author={Liu, Xuliang and Zha, Wenshu and Qi, Zhankui and Li, Daolun and Xing, Yan and He, Lei},
  journal={Journal of Energy Resources Technology},
  volume={144},
  number={4},
  pages={043002},
  year={2022},
  publisher={American Society of Mechanical Engineers}
}

@article{li2024fault,
  title={Fault diagnosis of a marine power-generation diesel engine based on the Gramian angular field and a convolutional neural network},
  author={Li, Congyue and Hu, Yihuai and Jiang, Jiawei and Cui, Dexin},
  journal={Journal of Zhejiang University-Science A},
  volume={25},
  number={6},
  pages={470--482},
  year={2024},
  publisher={Springer}
}

@article{dao2024wear,
  title={Wear fault diagnosis in hydro-turbine via the incorporation of the IWSO algorithm optimized CNN-LSTM neural network},
  author={Dao, Fang and Zeng, Yun and Zou, Yidong and Qian, Jing},
  journal={Scientific Reports},
  volume={14},
  number={1},
  pages={25278},
  year={2024},
  publisher={Nature Publishing Group UK London}
}

@article{qian2022cnn,
  title={CNN-based feature fusion motor fault diagnosis},
  author={Qian, Long and Li, Binbin and Chen, Lijuan},
  journal={Electronics},
  volume={11},
  number={17},
  pages={2746},
  year={2022},
  publisher={MDPI}
}

@article{chen2024ship,
  title={A ship-radiated noise classification method based on domain knowledge embedding and attention mechanism},
  author={Chen, Lu and Luo, Xinwei and Zhou, Hanlu},
  journal={Engineering Applications of Artificial Intelligence},
  volume={127},
  pages={107320},
  year={2024},
  publisher={Elsevier}
}

@article{chen2025fault,
  title={Fault Diagnosis Method of Rotating Machinery Based on Acoustic Signals and Improved Convolutional Neural Network-Bidirectional Long-Short-Term Memory Model},
  author={Chen, Yujia and Ran, Yingbing and Zhang, Chonglin and Xu, Shuxian and Lv, Shunli and Zeng, Yun},
  journal={Energy Technology},
  pages={2500955},
  year={2025},
  publisher={Wiley Online Library}
}

@article{chao2022fault,
  title={Fault diagnosis of axial piston pumps with multi-sensor data and convolutional neural network},
  author={Chao, Qun and Gao, Haohan and Tao, Jianfeng and Liu, Chengliang and Wang, Yuanhang and Zhou, Jian},
  journal={Frontiers of Mechanical Engineering},
  volume={17},
  number={3},
  pages={36},
  year={2022},
  publisher={Springer}
}

@article{tang2020intelligent,
  title={Intelligent diagnosis towards hydraulic axial piston pump using a novel integrated CNN model},
  author={Tang, Shengnan and Zhu, Yong and Yuan, Shouqi and Li, Guangpeng},
  journal={Sensors},
  volume={20},
  number={24},
  pages={7152},
  year={2020},
  publisher={MDPI}
}

@inproceedings{shen2022quantitative,
  title={Quantitative Detection of Blade Crack Damage Based on Vibro-Acoustic Information and Multi-Dimensional Feature Fusion CNN},
  author={Shen, Junxian and Song, Di and Ma, Tianchi and Xu, Feiyun},
  booktitle={2022 Global Reliability and Prognostics and Health Management (PHM-Yantai)},
  pages={1--6},
  year={2022},
  organization={IEEE}
}

@inproceedings{jin2019eemd,
  title={An EEMD and convolutional neural network based fault diagnosis method in intelligent power plant},
  author={Jin, Hongwei and Wang, Huanming and Tian, Feng and Zhao, Chunhui},
  booktitle={2019 Chinese Automation Congress (CAC)},
  pages={5215--5220},
  year={2019},
  organization={IEEE}
}

@article{zaman2023centrifugal,
  title={Centrifugal pump fault diagnosis based on a novel SobelEdge scalogram and CNN},
  author={Zaman, Wasim and Ahmad, Zahoor and Siddique, Muhammad Farooq and Ullah, Niamat and Kim, Jong-Myon},
  journal={Sensors},
  volume={23},
  number={11},
  pages={5255},
  year={2023},
  publisher={MDPI}
}

@article{wang2021high,
  title={A high-stability diagnosis model based on a multiscale feature fusion convolutional neural network},
  author={Wang, Pengxin and Song, Liuyang and Guo, Xudong and Wang, Huaqing and Cui, Lingli},
  journal={IEEE Transactions on Instrumentation and Measurement},
  volume={70},
  pages={1--9},
  year={2021},
  publisher={IEEE}
}

@article{kumar2025inceptionv3,
  title={InceptionV3 based blockage fault diagnosis of centrifugal pump},
  author={Kumar, Deepak and Ranawat, Nagendra Singh and Kankar, Pavan Kumar and Miglani, Ankur},
  journal={Advanced Engineering Informatics},
  volume={65},
  pages={103181},
  year={2025},
  publisher={Elsevier}
}

@article{wang2022deeppipe,
  title={Deeppipe: Operating condition recognition of multiproduct pipeline based on KPCA-CNN},
  author={Wang, Chang and Zheng, Jianqin and Liang, Yongtu and Liao, Qi and Wang, Bohong and Zhang, Haoran},
  journal={Journal of Pipeline Systems Engineering and Practice},
  volume={13},
  number={2},
  pages={04022006},
  year={2022},
  publisher={American Society of Civil Engineers}
}

@inproceedings{vasiliev2023pump,
  title={Pump Fault Classification based on Autoencoding Convolutional Neural Network Residuum},
  author={Vasiliev, Iulian and Frangu, Laurentiu and Cristea, Mihai Lucian and Costea, Mihai Cristian},
  booktitle={2023 IEEE 28th International Conference on Emerging Technologies and Factory Automation (ETFA)},
  pages={1--7},
  year={2023},
  organization={IEEE}
}

@article{pan2018improved,
  title={An improved bearing fault diagnosis method using one-dimensional CNN and LSTM},
  author={Pan, Honghu and He, Xingxi and Tang, Sai and Meng, Fanming},
  journal={Strojniski Vestnik-Journal of Mechanical Engineering},
  volume={64},
  number={7-8},
  pages={443--453},
  year={2018},
  publisher={University of Ljubljana, Faculty of Mechanical Engineering}
}

@inproceedings{liu2021method,
  title={Method for detecting surface defects of runner blades of large hydraulic turbines based on improved real-time lightweight network},
  author={Liu, Cheng and Su, XinYi and Wu, Jialing and Zhou, Qun and Li, Tao and Tian, Tang and Li, XianYong and Hu, GuiChuan},
  booktitle={Journal of Physics: Conference Series},
  volume={1955},
  number={1},
  pages={012090},
  year={2021},
  organization={IOP Publishing}
}

@article{wan2025fault,
  title={Fault diagnosis of air conditioning compressor bearings using wavelet packet decomposition and improved 1D-CNN},
  author={Wan, Anping and Li, Pengchong and Khalil, AL-Bukhaiti and Cheng, Xiaomin and Ji, Xiaosheng and Wang, Jinglin and Shan, Tianmin},
  journal={Next Energy},
  volume={9},
  pages={100424},
  year={2025},
  publisher={Elsevier}
}

@article{zheng2025intensity,
  title={Intensity recognition of vortex ropes in draft tube of a prototype pump turbine using an optimized CNN-BiLSTM framework with multi-head self-attention mechanism},
  author={Zheng, Xianghao and Yang, Chenxin and Zeng, Lan and He, Yuanshuai and Tian, Yulong and Zhang, Yuning and Li, Jinwei},
  journal={Journal of Energy Storage},
  volume={106},
  pages={114910},
  year={2025},
  publisher={Elsevier}
}

@inproceedings{li2022rolling,
  title={Rolling Bearing Fault Diagnosis Based on Convolutional Neural Network and Multi-sensor Information Fusion},
  author={Li, Lin and Zhao, Xing and Liu, Xiaodong and Fei, Jiyou},
  booktitle={2022 IEEE 10th Joint International Information Technology and Artificial Intelligence Conference (ITAIC)},
  volume={10},
  pages={1396--1401},
  year={2022},
  organization={IEEE}
}

@article{e2024development,
  title={Development of a CNN-based fault detection system for a real water injection centrifugal pump},
  author={e Souza, Ana Cl{\'a}udia Oliveira and de Souza Jr, Maur{\'\i}cio B and da Silva, Fl{\'a}vio Vasconcelos},
  journal={Expert Systems with Applications},
  volume={244},
  pages={122947},
  year={2024},
  publisher={Elsevier}
}

@article{ugli2023automatic,
  title={Automatic optimization of one-dimensional CNN architecture for fault diagnosis of a hydraulic piston pump using genetic algorithm},
  author={Ugli, Oybek Eraliev Maripjon and Lee, Kwang-Hee and Lee, Chul-Hee},
  journal={IEEE Access},
  volume={11},
  pages={68462--68472},
  year={2023},
  publisher={IEEE}
}

@article{li2020adaptive,
  title={An adaptive data fusion strategy for fault diagnosis based on the convolutional neural network},
  author={Li, Shi and Wang, Huaqing and Song, Liuyang and Wang, Pengxin and Cui, Lingli and Lin, Tianjiao},
  journal={Measurement},
  volume={165},
  pages={108122},
  year={2020},
  publisher={Elsevier}
}

@article{sunal2024centrifugal,
  title={Centrifugal pump fault detection with convolutional neural network transfer learning},
  author={Sunal, Cem Ekin and Velisavljevic, Vladan and Dyo, Vladimir and Newton, Barry and Newton, Jake},
  journal={Sensors},
  volume={24},
  number={8},
  pages={2442},
  year={2024},
  publisher={MDPI}
}

@article{ma2023systematic,
  title={Systematic Comparison of Sensor Signals for Pump Operating Points Estimation Using Convolutional Neural Network},
  author={Ma, Hanbing and Kirschner, Oliver and Riedelbauch, Stefan},
  journal={International Journal of Turbomachinery, Propulsion and Power},
  volume={8},
  number={4},
  pages={39},
  year={2023},
  publisher={MDPI}
}

@article{anvar2023novel,
  title={A novel application of deep transfer learning with audio pre-trained models in pump audio fault detection},
  author={Anvar, Ali Akbar Taghizadeh and Mohammadi, Hossein},
  journal={Computers in Industry},
  volume={147},
  pages={103872},
  year={2023},
  publisher={Elsevier}
}

@article{zheng2023hydrodynamic,
  title={Hydrodynamic feature extraction and intelligent identification of flow regimes in vaneless space of a pump turbine using improved empirical wavelet transform and Bayesian optimized convolutional neural network},
  author={Zheng, Xianghao and Li, Hao and Zhang, Suqi and Zhang, Yuning and Li, Jinwei and Zhao, Weiqiang},
  journal={Energy},
  volume={282},
  pages={128705},
  year={2023},
  publisher={Elsevier}
}

@article{xu2023cost,
  title={A cost-effective CNN-BEM coupling framework for design optimization of horizontal axis tidal turbine blades},
  author={Xu, Jian and Wang, Longyan and Yuan, Jianping and Shi, Jiali and Wang, Zilu and Zhang, Bowen and Luo, Zhaohui and Tan, Andy CC},
  journal={Energy},
  volume={282},
  pages={128707},
  year={2023},
  publisher={Elsevier}
}

\end{document}